\documentclass[a4paper,11pt]{article}
\pdfoutput=1 % if your are submitting a pdflatex (i.e. if you have
\usepackage{Jiahep} 

\usepackage{bm,latexsym,galois,dsfont,euscript}

\usepackage[numbers,sort&compress]{natbib} %citation using numbers
\usepackage{url}
\usepackage{amsmath}
\usepackage{rotating}

\usepackage{multirow}

\usepackage[T1]{fontenc} % if needed

\usepackage{float, array, xspace, amscd, amsthm}
\usepackage{fancyhdr}
\usepackage{longtable}

\newtheorem{definition}{Definition}
\newtheorem{theorem}{Theorem}
\newtheorem{lemma}[theorem]{Lemma}
\newtheorem{proposition}[theorem]{Proposition}
\newtheorem{corollary}[theorem]{Corollary}
\newtheorem{thm-def}[theorem]{Theorem-Definition}

\allowdisplaybreaks

\usepackage[usenames,dvipsnames,svgnames,table,x11names]{xcolor}
\definecolor{MagentaXD}{RGB}{204, 48, 152}
\definecolor{MagentaXDdetail}{RGB}{150, 79, 126}
\definecolor{GreenMAF}{RGB}{28, 112, 46}
\definecolor{GreenMAFdetail}{RGB}{80, 117, 88}
\definecolor{detail}{RGB}{110,110,110}
\definecolor{quantumviolet}{HTML}{53257F} %Quantum violet
\definecolor{quantumgray}{HTML}{555555} %Quantum gray
\definecolor{quantumgreen}{HTML}{007474} %Quantum green
\definecolor{quantumblue}{HTML}{002366} %Quantum gray
\definecolor{quantumpurple}{HTML}{66023C} %Quantum purple
\definecolor{quantumdarkviolet}{HTML}{5D3954} %Quantum dark violet

\usepackage[bbgreekl]{mathbbol} %change mathbb style

\newcommand{\cone}{(1)}
\newcommand{\ctwo}{(2)}

\newcommand{\Cbb}{\mathbb{C}}

\newcommand{\Irr}{\operatorname{Irr}}
\newcommand{\Tr}{\operatorname{Tr}}
\newcommand{\End}{\operatorname{End}}
\newcommand{\TubeM}{\mathbf{Tube}({_{\EuScript{C}} }\EuScript{M}_{\EuScript{D}})}
\newcommand{\Tube}{\mathbf{Tube}}

\definecolor{nblue}{rgb}{0.2,0.2,0.7}
\definecolor{ngreen}{rgb}{0.2,0.6,0.2}
\definecolor{nred}{rgb}{0.7,0.2,0.2}
\definecolor{nblack}{rgb}{0,0,0}

\newcommand{\be}{\begin{equation}}
	\newcommand{\ee}{\end{equation}}
\def\bea#1\eea{\begin{align}#1\end{align}}

\usepackage{tikz}
\usetikzlibrary{positioning,intersections}
\usetikzlibrary{calc}
\usetikzlibrary{shapes.geometric}
\usetikzlibrary{tqft} %for TQFT graph
\usepackage{braids}
\usepackage{tikz-cd}
\usetikzlibrary{decorations.markings}

\tikzset{
    partial ellipse/.style args={#1:#2:#3}{
        insert path={+ (#1:#3) arc (#1:#2:#3)}
    }
}

\tikzset{->-/.style={decoration={
  markings,
  mark=at position #1 with {\arrow{>}}},postaction={decorate}}}
\tikzset{-<-/.style={decoration={
  markings,
  mark=at position #1 with {\arrow{<}}},postaction={decorate}}}

\newif\ifcomments
\commentstrue

\newif\ifdetails
\detailstrue
\usepackage{upgreek}
\usepackage[normalem]{ulem}
\usepackage{mathtools}
\usepackage{datetime}
\usepackage{cancel}

\usepackage[all,cmtip,knot]{xy}
\usepackage{xypic}

\usepackage{amsmath,amsfonts,amsthm}

\theoremstyle{definition}

\theoremstyle{remark}
\newtheorem{remark}{Remark}[section]

\theoremstyle{remark}

\newcommand\EB{\EuScript{B}}
\newcommand\EC{\EuScript{C}}
\newcommand\ED{\EuScript{D}}

\newcommand\EK{\EuScript{K}}
\newcommand\EM{\EuScript{M}}
\newcommand\EN{\EuScript{N}}
\newcommand\EP{\EuScript{P}}

\newcommand\EW{\EuScript{W}}
\newcommand\EX{\EuScript{X}}
\newcommand\EY{\EuScript{Y}}
\newcommand\EQ{\EuScript{Q}}

\newcommand\Fun{\mathsf{Fun}}
\newcommand\Vect{\mathsf{Vect}}

\newcommand\Rep {\mathsf{Rep}}

\newcommand\id {\mathrm{id}}
\newcommand\Hom {\mathrm{Hom}}
\newcommand{\one}{\mathbb{1}}

\newcommand\Mod{\mathsf{Mod}}

\newcommand{\FPdim}{\operatorname{FPdim}}

\newcommand\fk {\mathfrak{f}}
\newcommand\gk {\mathfrak{g}}
\newcommand\hk {\mathfrak{h}}

\usepackage{orcidlink}

\usepackage{mathtools,amssymb,scalerel}

\newcommand\biopencrossl{%
	\mathrel{\scalerel*{>\kern-.4\LMpt\joinrel\blacktriangleleft}{x}}}
\newcommand\biopencrossr{%
	\mathrel{\scalerel*{\blacktriangleright\joinrel\kern-.4\LMpt<}{x}}}
\newcommand\bicrosslr{%
	\mathrel{\scalerel*{\mathrel{\blacktriangleright}\joinrel\blacktriangleleft}{x}}}

\settimeformat{ampmtime}

\begin{document}

\title{Weak Hopf tube algebra for domain walls between 2d gapped phases of Turaev-Viro TQFTs}

\author[a,b,*]{Zhian Jia\orcidlink{0000-0001-8588-173X},}

\author[c,d,e,\dagger]{Sheng Tan\orcidlink{0009-0008-3318-9942}}

\affiliation[a]{Centre for Quantum Technologies, National University of Singapore, Singapore 117543, Singapore}
\affiliation[b]{Department of Physics, National University of Singapore, Singapore 117543, Singapore}
\affiliation[*]{Corresponding author, Email: \href{mailto:giannjia@foxmail.com}{giannjia@foxmail.com}}

\affiliation[c]{School of Mathematical Sciences, Capital Normal University, Beijing 100048, China}
\affiliation[d]{Beijing Institute of Mathematical Sciences and Applications, Beijing 101408, China}
\affiliation[e]{Yau Mathematical Sciences Center, Tsinghua University, Beijing 100084, China}
\affiliation[\dagger]{Corresponding author, Email: \href{mailto:tansheng@cnu.edu.cn}{tansheng@cnu.edu.cn}}

\abstract{
We investigate domain walls between $2d$ gapped phases of Turaev-Viro type topological quantum field theories (TQFTs) by constructing domain wall tube algebras. We begin by analyzing the domain wall tube algebra associated with bimodule categories, and then extend the construction to multimodule categories over $N$ base fusion categories. We prove that the resulting tube algebra is naturally equipped with a $C^*$ weak Hopf algebra structure. We show that topological excitations localized on domain walls are classified by representations of the corresponding domain wall tube algebra, in the sense that the functor category of bimodules admits a fusion-preserving embedding into the representation category of the domain wall tube algebra. We further establish the folding trick and Morita theory in this context.  Then, most crucially, we provide a rigorous construction of the Drinfeld quantum double from weak Hopf boundary tube algebras using a skew-pairing, and establish an isomorphism between domain wall tube algebra and Drinfeld quantum double of boundary tube algebras.  Motivated by the correspondence between the domain wall tube algebra and the quantum double of the boundary tube algebras, we introduce the notion of an $N$-tuple algebra and demonstrate how it arises in the multimodule domain wall setting. Finally, we consider defects between two domain walls, showing that such defects can be characterized by representations of a domain wall defect tube algebra. We briefly outline how these defects can be systematically treated within this representation-theoretic framework.
}


\maketitle

%\tableofcontents

\section{Introduction}

Topological phases of matter have garnered considerable interest in recent decades. These phases exhibit topological order that lies beyond the conventional Landau-Ginzburg paradigm of phase transitions~\cite{Wen2004,zeng2015quantum,fradkin2013field,simon2023topological}, necessitating a generalized framework to properly characterize their properties.
A defining feature of topological order is the emergence of anyonic excitations, which exhibit fractionalized quantum numbers and obey nontrivial braiding statistics. The ground state is characterized by long-range entanglement, and the topological excitations are intrinsically nonlocal—they cannot be created or annihilated by any local operator.
These phases are not only of fundamental theoretical interest but also hold significant promise for practical applications in quantum information science, particularly in the context of fault-tolerant quantum computation~\cite{Kitaev2003, freedman2002modular, Nayak2008, wang2010topological, pachos2012introduction}.

Topological quantum field theory (TQFT) provides a universal and mathematically rigorous framework for describing the low-energy, long-wavelength behavior of topologically ordered phases of matter~\cite{witten1988topological,dijkgraaf1990topological,turaev2017monoidal,turaev1992state,reshetikhin1991invariants}. 
Among various constructions, the Turaev-Viro TQFT stands out as a (2+1)D example built from a unitary spherical fusion category~\cite{turaev1992state,barrett1996invariants}, offering a combinatorial approach to modeling gapped topological phases, called Levin-Wen string-net model \cite{Levin2005,kirillov2011stringnet}.

The string-net model is equivalent to the quantum double model~\cite{Kitaev2003} when the input data of the string-net model is chosen to be the representation category $\Rep(W)$ of some quantum group $W$, and the input data of the quantum double model is taken as $W$~\cite{buerschaper2013electric,Buerschaper2013a,jia2023boundary,Jia2023weak,meusburger2017kitaev,chen2021ribbon,balsam2012kitaevs,chang2014kitaev,girelli2021semidual,jia2024weakTube}.
The topological excitations in both models are characterized by the Drinfeld center $\mathcal{Z}(\Rep(W))$, which is equivalent (as a unitary braided fusion category) to the representation category $\Rep(D(W))$ of the Drinfeld quantum double $D(W)$ of $W$. This means that the corresponding phases are all non-chiral phases.

For a non-chiral $(2+1)$D topological phase, its topological excitations are characterized by a unitary modular tensor category (UMTC) $\mathcal{Z}(\EC)$, which is the Drinfeld center of some unitary fusion category (UFC) $\EC$. 
It has been shown that these excitations can also be characterized by the \emph{tube algebra} constructed from $\EC$~\cite{ocneanu1994chirality,ocneanu2001operator,izumi2000structure,MUGER2003subfactor,evans1998quantum,Neshveyev2018tube,popa2015cohomology} (denoted as $\Tube_{\EC}$): the representation category of the tube algebra is equivalent to the original category describing the topological phase $\Rep(\Tube_{\EC}) \simeq \mathcal{Z}(\EC)$.
%$\Rep(\Tube_{\EC}) \simeq_{\otimes, \rm br} \mathcal{Z}(\EC)$.
Several variants of tube algebras have been introduced in the literature~\cite{jones2001annular,hoek2019drinfeld,lan2025tubecategory,bridgeman2020computing,bai2025weakhopf,lan2014topological,jia2024weakTube,Barter2019domainwall,Barter2022computing,liu2023alterfolds}, serving as algebraic tools for approximating various categories of interest.
The tube algebra (and its variants) also finds applications in various areas, for instance, in characterizing charges within the framework of symmetry topological field theory (SymTFT) and in the study of various topological excitations~\cite{choi2024generalized,cordova2025representation,Christian2023,Bullivant2019tube,Lin2023tube,Hu2018full}.

From the correspondence between the string-net model and the weak Hopf quantum double model, it is natural to seek a weak Hopf algebra whose representation category is equivalent to, or at least approximates, a given unitary fusion category (UFC) $\EC$. By the Tannaka–Krein reconstruction theorem~\cite{etingof2016tensor,etingof2005fusion}, such a weak Hopf algebra always exists. This raises a natural question: can we modify the tube algebra so that it becomes a weak Hopf algebra whose representation category approximates\footnote{Here we stress `approximates' since, in many situations, the resulting algebra does not yield exactly the category we want, but rather a larger one that contains it as a subcategory.} $\EC$—or, ideally, is equivalent to $\EC$? 
The problem of approximating a given category via some algebra has been a central topic in studying tensor categories~\cite{Johnson-Freyd2025homotopy,morrison2012blob}, since the language of algebras is often more tractable and algebraic models are typically easier to manipulate.

The traditional tube algebra is typically defined as an algebra without any coalgebra structure and hence does not form a weak Hopf algebra. In the study of topological boundary theories of string-net models, Ref.~\cite{Kitaev2012boundary} proposed a generalization of the tube algebra to the boundary setting. The resulting \emph{boundary tube algebra} $\Tube({_{\EC}\EM})$, constructed from a $\EC$-module category $\EM$, is a $C^*$ weak Hopf algebra (also referred to as the annular algebra in some of the literature). 
This generalized boundary tube algebra has garnered significant attention over the past decade~\cite{bridgeman2020computing,Bridgeman2017tensor,bai2025weakhopf,lan2014topological}, largely because its representation category serves as an approximation of the given fusion category—that is, it approximates the category of topological excitations on the boundary.

In Ref.~\cite{jia2024weakTube}, we proposed a generalization of the boundary tube algebra to bulk phase and topological domain walls by constructing $\Tube({_{\EC}\EM_{\ED}})$ from a $\EC|\ED$ bimodule category ${_{\EC}\EM_{\ED}}$, including cases where $\EC$ and $\ED$ are multifusion categories. The resulting \emph{domain wall tube algebra} is a weak Hopf algebra whose representation category provides an approximation to the topological excitations localized on the domain wall.
When the domain wall data is chosen to be trivial—namely, when $\EC$ is regarded as a $\EC|\EC$ bimodule category—the associated domain wall tube algebra approximates the bulk phase, which is described by the Drinfeld center $\mathcal{Z}(\EC)$. Ref.~\cite{jia2024weakTube} primarily focuses on this special case. The present work extends that investigation by systematically studying the domain wall tube algebra for general topological domain walls.

\subsection*{Summary of main results}

The first crucial observation of this paper is the existence of a deep connection between the boundary tube algebra and the domain wall tube algebra, as conjectured in Ref.~\cite{jia2024weakTube}: the domain wall tube algebra $\Tube({_{\EC}\EM_{\ED}})$ is isomorphic to a Drinfeld quantum double, given a suitably chosen pairing between the left and right boundary tube algebras $\Tube_{\rm bd}({_{\EC}\EM})$ and $\Tube_{\rm bd}({\EM}_{\ED})$.
In this work, we introduce the notion of a (generalized) Drinfeld quantum double based on a skew-pairing, which itself carries the structure of a weak Hopf algebra. We then apply this construction to the boundary tube algebras by defining a pairing function 
$p: \Tube_{\rm bd}({\EM}_{\ED}) \times \Tube_{\rm bd}({_{\EC}\EM}) \to \mathbb{C}$, and use it to form the Drinfeld quantum double 
\begin{equation}
    \Tube_{\rm bd}({\EM}_{\ED}) \Join_p \Tube_{\rm bd}({_{\EC}\EM}).
\end{equation}
We show that the following isomorphism of weak Hopf algebras holds:
\begin{equation}
    \Tube({_{\EC}\EM_{\ED}}) \simeq \Tube_{\rm bd}({\EM}_{\ED}) \Join_p \Tube_{\rm bd}({_{\EC}\EM}).
\end{equation}
The physics underlying this isomorphism can be understood as follows. From the correspondence between the string-net model and weak Hopf lattice gauge theory, the boundary tube algebra plays the role of a weak Hopf gauge symmetry, whose representation category corresponds to the input unitary fusion category (UFC) of the string-net model.
It is well known that topological excitations are characterized by the representations of the Drinfeld quantum double of the gauge symmetry—referred to here as the charge symmetry. On the other hand, within the string-net framework, topological excitations can also be described by the bulk tube algebra $\Tube({_{\EC}}\EC_{\EC})$, which is constructed from the trivial domain wall ${_{\EC}}\EC_{\EC}$.
Thus, both the domain wall tube algebra and the Drinfeld quantum double of the boundary tube algebra serve as weak Hopf charge symmetries. Consequently, they should be isomorphic as weak Hopf algebras. This can be generalized to the nontrivial domain wall straightforwardly.

Inspired by the above observation, we propose a generalization of the Drinfeld quantum double for two weak Hopf algebras to the case of $N$ weak Hopf algebras, which we refer to as the \emph{quantum $N$-tuple algebra}. Mathematically, this construction corresponds to the domain wall tube algebra associated with a multimodule category over $N$ UFCs. We show that the resulting object is also a $C^*$ weak Hopf algebra.
The physical setting underlying this construction is that of a domain wall connecting $N$ distinct $2d$ string-net phases. The bimodule domain wall tube algebra arises as a special case when $N=2$. The relation between the multimodule domain wall tube algebra and the boundary tube algebras is precisely captured by the $N$-tuple algebra structure: we can construct a weak Hopf algebra from $N$ given boundary tube algebras.
Furthermore, for any pair of boundary tube algebras among the $N$, their Drinfeld quantum double is naturally embedded within the $N$-tuple algebra. Likewise, for any triple, their corresponding \emph{quantum triple algebra} is also contained within this structure. This hierarchical embedding continues for higher-order combinations, revealing that the $N$-tuple algebra possesses a rich hierarchical structure encoding the interactions among all subsets of the $N$ weak Hopf algebras.

Moreover, we also study the representation theory of the domain wall tube algebra and demonstrate how topological excitations localized on the domain wall can be characterized within this framework. In particular, we rigorously establish the folding trick and its relation to Morita theory, providing a systematic understanding of how domain wall excitations are encoded in the representation category of the tube algebra.

The remainder of the paper is organized as follows.

In Section~\ref{sec:walltube}, we construct the domain wall tube algebra within the framework of the Levin-Wen string-net model, which describes gapped phases of the Turaev-Viro TQFT. We explicitly define its algebra, coalgebra, antipode, and $C^*$-structures, and demonstrate that they together form a $C^*$ weak Hopf algebra. The main result is presented in Theorem~\ref{thm:tube_alg_weak_hopf}.
Furthermore, we derive the Haar integral and the canonical grouplike element associated with the domain wall tube algebra. Several concrete examples are provided to illustrate the structure and properties of these algebras.

Section~\ref{sec:multi_tube} extends the results of the previous section to the setting of multimodule categories, in which multiple $2d$ topological phases are joined by a single domain wall characterized by a multimodule category over a collection of base fusion categories. We provide a precise definition of the multimodule domain wall tube algebra and demonstrate that it carries the structure of a $C^*$ weak Hopf algebra.
In addition, we explore the folding trick, which yields a multimodule generalization of the boundary tube algebra. We also develop the Morita theory for multimodule domain wall tube algebras, revealing structural correspondences between different characterization of excitations of gapped domain walls.
When the multimodule category involves exactly two base fusion categories, the corresponding domain wall tube algebra reduces to the bimodule tube algebra discussed in Section~\ref{sec:walltube}. All results developed in this section therefore apply to the weak Hopf tube algebra framework introduced in Section~\ref{sec:walltube}.

Section~\ref{sec:tubeDouble} provides a brief discussion on the generalization of the quantum double algebra to a quantum $N$-tuple algebra using the framework of multimodule tube algebras. As a special case, the domain wall tube algebra can be regarded as a ``tube'' quantum double of the boundary tube algebra.

In Section~\ref{sec:QDTube}, we introduce a generalized Drinfeld quantum double construction for weak Hopf algebras based on a skew-pairing. Applying this framework to the tube algebra, we define a skew-pairing between the left and right boundary tube algebras, $\Tube_{\rm bd}({_{\EC}\EM})$ and $\Tube_{\rm bd}({\EM}_{\ED})$, and construct the corresponding generalized quantum double of the boundary tube algebras. We then show that this quantum double is isomorphic, as a weak Hopf algebra, to the domain wall tube algebra.

Section~\ref{sec:RepTube} establishes the representation theory of the domain wall tube algebra. We show that topological excitations localized on the domain wall are characterized by the representations of the domain wall tube algebra. In particular, we construct a monoidal functor between the category $\Fun_{\EC|\ED}(\EM, \EM)$ of bimodule endofunctors and the representation category $\Rep(\Tube({_{\EC}\EM_{\ED}}))$.
By establishing a Schur orthogonality relation, we further demonstrate that simple objects in $\Fun_{\EC|\ED}(\EM, \EM)$ are mapped to simple objects in $\Rep(\Tube({_{\EC}\EM_{\ED}}))$. These results naturally extend to the case of multimodule domain wall tube algebras.

In Section~\ref{sec:DefectTube}, we briefly discuss the tube algebra framework for characterizing defects that occur between different domain walls.

In the final section, we present concluding remarks. The Appendix reviews foundational concepts related to fusion categories and weak Hopf algebras, which underpin the main results of this work, and provides further technical details.

%\subsection{Approximating a given tensor category}

%\subsection{Tube algebra}

%\subsection{Organization of the paper}

\section{Weak Hopf tube algebra for domain wall}
\label{sec:walltube}

The weak Hopf symmetry plays a crucial role in understanding $2d$ topological quantum spin liquids. All $2d$ non-chiral topological phases can be realized either by a weak Hopf quantum double model (lattice gauge theory)~\cite{Kitaev2003,Buerschaper2013a,chang2014kitaev,Jia2023weak}, or equivalently, by a string-net model~\cite{Levin2005}—more generally, a string-net model with input data given by a multifusion category~\cite{chang2015enriching,jia2024weakTube}. 
In Ref.~\cite{Kitaev2012boundary}, the boundary tube algebra for the string-net model was constructed. Inspired by this, Ref.~\cite{jia2024weakTube} provided an explicit construction of the bulk tube algebra for the string-net model and demonstrated that it forms a $C^*$ weak Hopf algebra. In this section, we extend these results to the case of domain walls, observing that bulk excitations can be viewed as a special case of domain wall excitations. The domain wall tube algebra exhibits a richer structure—it is isomorphic to the Drinfeld quantum double of the boundary weak Hopf tube algebra. Moreover, we compute the explicit expressions for the Haar integral and the canonical grouplike element.

\begin{figure}[t]
    \centering
    \includegraphics[width=0.7\linewidth]{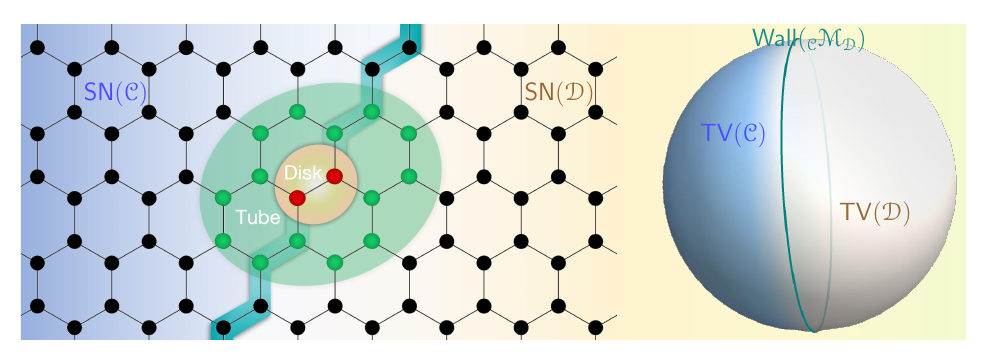}
    \caption{Depiction of a domain wall between two Levin-Wen string-net phases realizing two Turaev–Viro TQFTs. The input data are (multi)fusion categories $\EC$ and $\ED$, respectively, and the domain wall is specified by a $\EC|\ED$ bimodule category ${_{\EC}}\EM_{\ED}$. For a disk region on the domain wall hosting an excitation, one can consider a tubular neighborhood surrounding this region; the domain wall tube algebra is constructed from such a tubular region.}
    \label{fig:SNlattice}
\end{figure}

\subsection{Topological domain wall of string-net phases}
\label{sec:walltheory}

Consider two string-net phases whose input data are (multi)fusion categories \(\EC\) and \(\ED\), respectively. The domain wall between these two phases is characterized by a \(\EC|\ED\) bimodule category \({_{\EC}}\EM_{\ED}\)\footnote{In this work, we primarily focus on unitary fusion categories (UFCs) to simplify the discussion. The generalization to the multifusion case is straightforward, following the method outlined in Ref.~\cite{jia2024weakTube}.}, see Figure \ref{fig:SNlattice} for a depiction. 

In this work, unless otherwise specified, we assume that the module category is an abelian category that is indecomposable, unitary, and finitely semisimple. For foundational material on category theory and weak Hopf algebras, see Appendix~\ref{sec:app_fusion}; see also~\cite{etingof2016tensor,bakalov2001lectures,turaev2016quantum}.

In the vicinity of the domain wall, it is necessary to introduce the vertex and face operators using the module category data.

The Hilbert space $\mathcal{H}_{\rm wall}$ of the domain wall is the tensor product of morphism spaces of $\EM$.
For a given vertex in the domain wall that connecting $\EC$ bulk edge, we define $\delta_{m\to a, n}=1$ if $\Hom_{\EM}( m,a\otimes n)\neq 0$ for $a\in \Irr(\EC)$, $m,n\in \Irr(\EM)$; otherwise, we set $\delta_{m\to a, n}=0$. From this, we introduce the vertex projector $Q_v$ as follows:
\begin{equation}
    {Q}_v\, \big{|}  \begin{aligned}
		\begin{tikzpicture}
			\draw[->,line width=.6pt,black] (0,-0.5) -- (0,-.1);
			\draw[line width=.6pt,black] (0,-0.4) -- (0,0);
			\draw[->,line width=.6pt,black](0,0)--(-0.3,0.3);
			\draw[line width=.6pt,black](-0.1,0.1)--(-0.4,0.4);
			\draw[->,line width=.6pt,black](0,0)--(0.3,0.3);
			\draw[line width=.6pt,black](0.1,0.1)--(0.4,0.4);
			\node[ line width=0.6pt, dashed, draw opacity=0.5] (a) at (-0.5,0.6){$a$};
			\node[ line width=0.6pt, dashed, draw opacity=0.5] (a) at (0.5,0.6){{\color{teal}$n$}};
			\node[ line width=0.6pt, dashed, draw opacity=0.5] (a) at (0,-0.7){{\color{teal}$m$}};
			\node[ line width=0.6pt, dashed, draw opacity=0.5] (a) at (0.3,-0.1){$\alpha$};
		\end{tikzpicture}
	\end{aligned} \big{\rangle} =\delta_{m\to a, n} \,\big{|}\begin{aligned}
		\begin{tikzpicture}
			\draw[->,line width=.6pt,black] (0,-0.5) -- (0,-.1);
			\draw[line width=.6pt,black] (0,-0.4) -- (0,0);
			\draw[->,line width=.6pt,black](0,0)--(-0.3,0.3);
			\draw[line width=.6pt,black](-0.1,0.1)--(-0.4,0.4);
			\draw[->,line width=.6pt,black](0,0)--(0.3,0.3);
			\draw[line width=.6pt,black](0.1,0.1)--(0.4,0.4);
			\node[ line width=0.6pt, dashed, draw opacity=0.5] (a) at (-0.5,0.6){$a$};
			\node[ line width=0.6pt, dashed, draw opacity=0.5] (a) at (0.5,0.6){{\color{teal}$n$}};
			\node[ line width=0.6pt, dashed, draw opacity=0.5] (a) at (0,-0.7){{\color{teal}$m$}};
			\node[ line width=0.6pt, dashed, draw opacity=0.5] (a) at (0.3,-0.1){$\alpha$};
		\end{tikzpicture}
	\end{aligned}\big{\rangle}\,.
\end{equation}
The condition ${Q}_v=1$ corresponds to satisfying Gauss's law at the vertex $v$, in analogy with lattice gauge theory.
Similarly, we can define vertex operators for vertices that connect $\ED$ bulk edges.

The face operator near the domain wall also needs to be modified. Consider the domain wall face \( f \) in the \(\EC\) bulk. The face operator \( {B}_f \) is defined as a linear combination of basic face operators \( {B}^k_f \):
\begin{equation}
   {B}_f = \sum_{k \in \Irr(\EC)} w_k {B}^k_f,
\end{equation}
where \(\displaystyle w_k = \frac{Y^{\bar{k} k}_{\one}}{\sum_{l \in \Irr(\EC)} d_l^2} \), which is usually fixed by choosing the gauge as \( Y^{ab}_c = \left( \frac{d_a d_b}{d_c} \right)^{1/2} \).
The face operator ${B}_f^k$ acts on the domain wall face by inserting a $k$-loop into the face $f$, which graphically looks like

\begin{gather}
    {B}_f^k\, \Big{|} 
    \begin{aligned}
        \resizebox{0.2\textwidth}{!}{ % Adjust width here (0.5\textwidth for half the column width)

        }
    \end{aligned}
    \Big{\rangle} \, .
\end{gather}
Using the F-move, loop move, and parallel move in a module category (to be defined later), we can derive the matrix expression for ${B}_f^k$. 
A similar approach applies to defining the face operator for a domain wall face in the $\ED$ bulk. 
The condition ${B}_f = 1$ ensures that no flux penetrates the face.

The vertex operators and face operators are all projectors and commute with one another. 
The domain wall Hamiltonian is given by
\begin{equation}
   H_{\rm wall}=-\sum_{v:\rm wall} {Q}_v-\sum_{f:\EC\text{-}\rm wall}{B}_f-\sum_{f:\ED\text{-}\rm wall}{B}_f.
\end{equation}
The domain wall excitations are characterized by $\EC|\ED$ bimodule functors $F: \EM \to \EM$. 
These functors form a fusion category, denoted $\Fun_{\EC|\ED}(\EM, \EM)$, where the fusion of two functors is given by their composition \cite{etingof2010fusion,etingof2016tensor,Kitaev2012boundary,Huston2023domainwall}.

When defining the domain wall Hamiltonian and tube algebras, we make extensive use of \emph{domain wall topological local moves}. These moves primarily involve the defining data of the domain wall bimodule category ${_{\EC}}\EM_{\ED}$. 
Unlike UFCs, the bimodule category $\EM$ generally lacks a monoidal structure and duals of objects. To ensure the validity of string diagram calculations, we introduce a multifusion category in which these operations become meaningful. This multifusion category is defined separately for the left and right module structures.

For the left module structure ${_{\EC}}\EM$, we utilize a well-known equivalence of abelian categories established in Ref.~\cite[Proposition 3.5]{etingof2010fusion}:
\begin{equation}
    \mathsf{Fun}_{\EC}(\EM,\EN) \simeq \EM^{\mathrm{op}} \boxtimes_{\EC} \EN,
\end{equation}
where $\mathsf{Fun}_{\EC}(\EM,\EN)$ denotes the category of $\EC$-module functors from $\EM$ to $\EN$, and $\EM^{\mathrm{op}}$ is the dual category of $\EM$, with the same objects but reversed morphisms.
To organize the various sectors, we define $\EM_0 = \EC$ and $\EM_1 = \EM$, and set $\EP_{i,j} := \mathsf{Fun}_{\EC}(\EM_i, \EM_j)$. This allows us to define a multifusion category:
\begin{equation}\label{eq:ModuleP}
   \EP = \bigoplus_{i,j=0,1} \EP_{i,j}, 
\end{equation}
with components given by
\[
\EP_{0,0} \simeq \EC,\quad \EP_{0,1} \simeq \EM,\quad \EP_{1,0} \simeq \EM^{\mathrm{op}},\quad \EP_{1,1} := \EC_{\EM}^{\vee} = \mathsf{Fun}_{\EC}(\EM, \EM).
\]
The fusion rules between these sectors are defined as follows:
\begin{itemize}
  \item[(i)] For $j \ne j'$, if $X_{i,j} \in \EP_{i,j}$ and $Y_{j',k} \in \EP_{j',k}$, then $X_{i,j} \otimes Y_{j',k} = 0$;
  \item[(ii)] For $j = j'$, if $X_{i,j} \in \EP_{i,j}$ and $Y_{j,k} \in \EP_{j,k}$, then $X_{i,j} \otimes Y_{j,k} \in \EP_{i,k}$.
\end{itemize}
This is a direct consequence of the fact that the composition of functors is well-defined only across  sectors $\Fun(\EM_i,\EM_j)$ and $\Fun(\EM_j,\EM_k)$.
For the $\EM_{\ED}$ structure, an analogous construction applies: we define $\EQ = \bigoplus_{i,j=0,1} \EQ_{i,j}$ in a similar fashion.

To clarify our discussion for the reader, in this section we will distinguish the quantum dimensions of objects in $\EM$ obtained from the left and right actions, and denote them by $d^L_m$ and $d^R_m$, respectively. 
We emphasize that the notion of dual objects is also dependent on the module structure. Specifically, for the left module structure, the dual $\bar{m}$ of an object $m$ lies in $\EP$, and we have the identity $d^L_m = d^L_{\bar{m}}$. Similarly, under the right module structure, $\bar{m} \in \EQ$ and satisfies $d^R_m = d^R_{\bar{m}}$.

In the context of string diagram calculus, it will be clear which multifusion category structure—$\EP$ or $\EQ$—is being used, depending on the module structure under consideration. See Refs.~\cite{Kitaev2012boundary,jia2024weakTube,etingof2016tensor} for further details.

\begin{definition}[Domain wall topological local move]
There are three types of elementary topological local moves that we will use to prove the weak Hopf algebra structure of the domain wall tube algebra:
\begin{enumerate}
    \item Domain wall loop move. All categories in this work are assumed to be dagger category, meaning that for the morphism space there is an antilinear dagger map $\dagger: \Hom(a,b) \to \Hom(b,a)$ such that $f^{\dagger \dagger}=f$ and $(f\circ g)^{\dagger}=g^{\dagger} \circ f^{\dagger}$.
    For $\alpha,\beta \in  \Hom(c,a\otimes b)$, we can endow it with isometry inner product via dagger structure as:
    \begin{equation}
        \langle \beta |\alpha\rangle \id_c= \beta^{\dagger}\circ \alpha \in \Hom(c,c)\cong \Cbb.
    \end{equation}
    For $a\in \EC,m,n\in \EM$, we can choose an orthogonal basis $\alpha\in \Hom(m,a\otimes n)$ and set the following normalization constraint $\langle \alpha|\alpha\rangle =\sqrt{\frac{d_{a}d_n^L}{d_m^L}}$. 
    Expressed diagrammatically, we obtain the so-called loop move\footnote{Here, for clarity, we explicitly denote the dagger as $\beta^{\dagger}$. In the subsequent sections, we will omit the dagger notation for simplicity. The reader should understand that the dagger is implicitly present. For the parallel move, a similar convention will be adopted.}:
  \begin{equation} 
  \text{loop move:}\quad	\begin{aligned}
		\begin{tikzpicture}
			\draw[->,line width=.6pt,black](0,0)--(0,0.4);
			\draw[line width=.6pt,black](0,0.1)--(0,0.5);
			\draw[line width=.6pt,black](0,-0.3) circle (0.3);
			\draw[->,line width=.6pt,black](-0.3,-0.3)--(-.3,-.2);
			\draw[->,line width=.6pt,black](0.3,-0.3)--(.3,-.2);
			\draw[->,line width=.6pt,black](0,-1.1)--(0,-0.7);
			\draw[line width=.6pt,black](0,-1.1)--(0,-0.6);
			\node[ line width=0.6pt, dashed, draw opacity=0.5] (a) at (0.6,-0.3){$n$};
			\node[ line width=0.6pt, dashed, draw opacity=0.5] (a) at (0,0.75){$m'$};
			\node[ line width=0.6pt, dashed, draw opacity=0.5] (a) at (0,-1.3){$m$};
			\node[ line width=0.6pt, dashed, draw opacity=0.5] (a) at (-0.6,-0.3){$a$};
			\node[ line width=0.6pt, dashed, draw opacity=0.5] (a) at (0.3,0.2){$\beta^{\dagger}$};
			\node[ line width=0.6pt, dashed, draw opacity=0.5] (a) at (-0.3,-0.8){$\alpha$};
		\end{tikzpicture}
	\end{aligned}
	=
	\delta_{m,m'}\delta_{\alpha,\beta} {\sqrt{\frac{d_ad_n^L}{d_m^L}}}
	\begin{aligned}
		\begin{tikzpicture}
			\draw[line width=.6pt,black](0,-1.1)--(0,0.5);
			\draw[->,line width=.6pt,black](0,-0.3)--(0,0);
			\node[ line width=0.6pt, dashed, draw opacity=0.5] (a) at (0,0.7){$m$};
		\end{tikzpicture}
	\end{aligned}. \label{eq:loopev}
\end{equation}
The dimensions here refer to the quantum dimensions of objects in the category. Note that for objects in $\EM$, the dimension can also be defined through the module structure \cite{etingof2005fusion}.
For the right module structure $\EM_{\ED}$, a similar loop move applies.

\item  Domain wall parallel move. For $a \in \EC$ and $m \in \EM$, we have $a \otimes m = \bigoplus_{n \in \Irr(\EM)} N_{am}^n n$, which induces the isomorphism $\Hom(a \otimes m, a \otimes m) \cong \Hom\left(\bigoplus_{n \in \Irr(\EM)} N_{am}^n n, a \otimes m \right) \cong \Hom \left(a \otimes m, \bigoplus_{n \in \Irr(\EM)} N_{am}^n n\right)$. This leads to a decomposition of the identity map $\id_{a\otimes m}$ as follows:
\begin{equation} \label{eq:paraev}
	\text{parallel move:}\quad\begin{aligned}
		\begin{tikzpicture}
			\draw[line width=.6pt,black](0,-1.1)--(0,0.5);
			\draw[->,line width=.6pt,black](0,-0.3)--(0,0);
			\node[ line width=0.6pt, dashed, draw opacity=0.5] (a) at (0,0.7){$a$};
			\draw[line width=.6pt,black](0.6,-1.1)--(0.6,0.5);
			\draw[->,line width=.6pt,black](0.6,-0.3)--(0.6,0);
			\node[ line width=0.6pt, dashed, draw opacity=0.5] (a) at (0.6,0.7){$m$};
		\end{tikzpicture}
	\end{aligned}
	=
	\sum_{n,\alpha}\sqrt{\frac{d_n^L}{d_ad_m^L}}
	\begin{aligned}
		\begin{tikzpicture}
			\draw[->,line width=.6pt,black] (0.4,-0.4) -- (0.2,-0.2);
			\draw[line width=.6pt,black] (0.3,-0.3) -- (0,0);
			\draw[->,line width=.6pt,black] (-0.4,-0.4) -- (-0.2,-0.2);
			\draw[line width=.6pt,black] (-0.3,-0.3) -- (0,0);
			\draw[->,line width=.6pt,black](0,0)--(0,0.45);
			\draw[line width=.6pt,black](0,0.1)--(0,0.7);
			\draw[line width=.6pt,black](-0.4,1.1)--(0,0.7);
			\draw[line width=.6pt,black](0.4,1.1)--(0,0.7);
			\draw[->,line width=.6pt,black](0,0.7)--(0.3,1);
			\draw[line width=.6pt,black](0.4,1.1)--(0,0.7);
			\draw[->,line width=.6pt,black](0,0.7)--(-0.3,1);
			\node[ line width=0.6pt, dashed, draw opacity=0.5] (a) at (0.4,-0.6){$m$};
			\node[ line width=0.6pt, dashed, draw opacity=0.5] (a) at (-0.4,-0.6){$a$};
			\node[ line width=0.6pt, dashed, draw opacity=0.5] (a) at (0.3,0.1){$\alpha^{\dagger}$};
			\node[ line width=0.6pt, dashed, draw opacity=0.5] (a) at (0.3,0.6){$\alpha$};
			\node[ line width=0.6pt, dashed, draw opacity=0.5] (a) at (-0.25,0.3){$n$};
			\node[ line width=0.6pt, dashed, draw opacity=0.5] (a) at (0.4,1.3){$m$};
			\node[ line width=0.6pt, dashed, draw opacity=0.5] (a) at (-0.4,1.3){$a$};
		\end{tikzpicture}
	\end{aligned}\,.
\end{equation}
For the right module structure $\EM_{\ED}$, a similar parallel move applies.

\item Domain wall F-move. For a domain wall, the bimodule structure of ${_{\EC}}\EM_{\ED}$ gives rise to three types of F-moves (corresponding to the left and right module associators, as well as the middle associator), which are given in Appendix~\ref{sec:app_fusion}.
\end{enumerate}
\end{definition}

In addition to the loop moves involving the module category $\EM$, the fusion categories $\EC$ and $\ED$ are both unitary fusion categories (UFCs), each equipped with its own set of topological local moves~\cite{jia2024weakTube}. It is also important to note that the three types of domain wall topological local moves described above are equivalent to the Pachner moves, which are fundamental in establishing the topological invariance of the model.

\begin{remark}
For a bimodule category ${_{\EC}}\EM_{\ED}$, we say that it is \emph{pseudo-unitary} if the Frobenius–Perron dimensions of objects in $\EM$, as defined from the left action, coincide with the quantum dimensions defined via the multifusion category $\EP$ in Eq.~\eqref{eq:ModuleP}; similarly, for the right action, the Frobenius–Perron dimensions agree with the quantum dimensions defined via the multifusion category $\EQ$.
In this case, let us define
\[
X = \bigoplus_{E \in \Irr(\EC)} E, \qquad Y = \bigoplus_{K \in \Irr(\ED)} K,
\]
and consider the fusion matrices defined by $[N_X]^N_M := N_{X, M}^N$ and $[N_Y]^N_M := N_{M, Y}^N$. By the axioms of a bimodule category, it follows that $N_X N_Y = N_Y N_X$. Therefore, $N_X$ and $N_Y$ are simultaneously diagonalizable and share a common eigenvector, whose entries correspond to the Frobenius–Perron dimensions of simple objects in $\EM$. It follows that $d^L_m = d^R_m$ for all $m \in \EM$ in this case.
As a special example,  the trivial bimodule category ${_{\EC}}\EC_{\EC}$ is pseudo-unitary; thus the left and right quantum dimensions always coincide.
\end{remark}

\subsection{Weak Hopf domain wall tube algebra}

In this section, we extend the results of Ref.~\cite{jia2024weakTube} to the setting of domain walls and provide a detailed discussion of the Haar integral and the canonical grouplike element, which have not been addressed elsewhere in the literature. For readers unfamiliar with weak Hopf algebras, we refer to Ref.~\cite[Section 2]{Jia2023weak} for background material, including fundamental results, notation, and conventions used throughout this work. See also Appendix~\ref{sec:app_fusion} for a brief overview of the relevant algebraic concepts.

Consider an excited disk on the domain wall of the string-net model (cf.~Figure \ref{fig:SNlattice}). The surrounding tube region can be transformed into a `minimal tube' via topological local move, based on which the domain wall tube algebra is defined as follows:

\begin{definition}
    For a gapped domain wall ${_{\EC}}\EM_{\ED}$,  the domain wall tube algebra $\mathbf{Tube}({_{\EC}}\EM_{\ED})$ is the vector space spanned by the following basis:
\begin{equation} \label{eq:tube_basis}
 \left\{  
   \begin{aligned}
        \begin{tikzpicture}[scale=0.65]
        \filldraw[black!60, fill=gray!15, dotted, even odd rule] (0,0) circle[radius=0.5] (0,0) circle[radius=1.5];
             \draw[line width=.6pt,black] (0,0.5)--(0,1.5);
             \draw[line width=.6pt,black] (0,-0.5)--(0,-1.5);
             \draw[red] (0,0.8) arc[start angle=90, end angle=270, radius=0.8];
             \draw[blue] (0,1.3) arc[start angle=90, end angle=-90, radius=1.3];
            \node[ line width=0.6pt, dashed, draw opacity=0.5] (a) at (0,1.7){$\scriptstyle w$};
             \node[ line width=0.6pt, dashed, draw opacity=0.5] (a) at (0,-1.7){$\scriptstyle x$};
            \node[ line width=0.6pt, dashed, draw opacity=0.5] (a) at (-1,0){$\scriptstyle a$};
            \node[ line width=0.6pt, dashed, draw opacity=0.5] (a) at (1.1,0){$\scriptstyle b$};
            \node[ line width=0.6pt, dashed, draw opacity=0.5] (a) at (-0.2,-1){$\scriptstyle y$};
            \node[ line width=0.6pt, dashed, draw opacity=0.5] (a) at (-0.2,-1.35){$\scriptstyle \mu$};
            \node[ line width=0.6pt, dashed, draw opacity=0.5] (a) at (0.2,-0.8){$\scriptstyle \nu$};
            \node[ line width=0.6pt, dashed, draw opacity=0.5] (a) at (0,-0.3){$\scriptstyle z$};
            \node[ line width=0.6pt, dashed, draw opacity=0.5] (a) at (0,0.3){$\scriptstyle u$};
            \node[ line width=0.6pt, dashed, draw opacity=0.5] (a) at (-0.2,1){$\scriptstyle v$};
            \node[ line width=0.6pt, dashed, draw opacity=0.5] (a) at (-0.2,1.3){$\scriptstyle \gamma$};
            \node[ line width=0.6pt, dashed, draw opacity=0.5] (a) at (0.2,0.8){$\scriptstyle \zeta$};
        \end{tikzpicture}
    \end{aligned}
   \;:\; 
   \begin{aligned}
     & a\in \Irr(\EC),b\in \Irr(\ED), x,y,z,u,v,w\in \Irr(\EM), \\
     & \mu\in\Hom_{\EM}(x,y\otimes b)\neq 0,\nu\in\Hom_{\EM}(y,a\otimes z)\neq 0,\\
     & \zeta\in\Hom_{\EM}(a\otimes u,v)\neq 0,\gamma\in \Hom_{\EM}(v\otimes b,w)\neq 0  
   \end{aligned}
    \right\}.
\end{equation}
\end{definition}

As pointed out in Ref.~\cite{jia2024weakTube}, the construction is also applicable to multifusion categories. The main difference in the multifusion case is the need to consider the compatibility of the fusion structure imposed by the multifusion category. To simplify the discussion, we will focus on the fusion category case in this section. The main result is the following:

\begin{theorem} \label{thm:tube_alg_weak_hopf}
        The domain wall tube algebra $\mathbf{Tube}({_{\EC}}\EM_{\ED})$ is a $C^*$ weak Hopf algebra.  The structure morphisms will be provided by Eqs.~\eqref{eq:bi_tube_unit}, \eqref{eq:bi_tube_prod_1}, \eqref{eq:bi_tube_counit}, \eqref{eq:bi_tube_coprod}, \eqref{eq:bi_tube_antipode}, and \eqref{eq:bi_tube_star}. 
\end{theorem}

The proof will be illustrated in the subsequent sections. When considering the trivial bimodule category ${_\EC}\EC_\EC$, the domain wall tube algebra $\mathbf{Tube}({_{\EC}}\EC_{\EC})$ reduces to the bulk tube algebra in Ref.~\cite{jia2024weakTube}.

\begin{remark}[Local gauge transformation of tube algebra]
It is worth mentioning that the structure constants of the tube algebra are not unique. Different sets of structure constants are related by local unitary gauge transformations over the fusion/splitting vertex space.
Consider the fusion vertex $\Hom(a\otimes b,c)$, we introduce a unitary transformation $\Gamma_{ab}^c$:
\begin{equation}
    |ab \to c;\tilde{\mu}\rangle =\sum_{\mu}[\Gamma_{ab}^c]_{\tilde{\mu},\mu} |ab\to c;\mu\rangle,
\end{equation}
where $\{\tilde{\mu}\}$ and $\{\mu\}$ denote two sets of orthonormal bases for the fusion space.
Similarly, for splitting space, we can introduce unitary transformation $[\Gamma^{ab}_c]_{\tilde{\mu},\mu}$.
When applied to the the basis of tube algebra, we obtain of a new basis (red tube)
\begin{equation}
     \begin{aligned}

    \end{aligned}\,.
\end{equation}
These local unitary transformation will modify the structure constant in a way that the weak Hopf structure constant from red tube can be expressed as a unitary transformation of the structure constant obtained from the gray tube.
\end{remark}

% \begin{remark}
%     \begin{align}
%         \vcenter{\hbox{
%         %
%         }}
%     \end{align}
% \end{remark}

\subsubsection{The algebra structure}

The algebra structure of $\mathbf{Tube}({_{\EC}}\EM_{\ED})$ is presented as follows. First, the unit element is defined by 
\begin{equation} \label{eq:bi_tube_unit}
    1=\sum_{x,z\in\Irr(\EM)} \;\begin{aligned}
        %
    \end{aligned}\;\;. 
\end{equation}
The red dotted line is labeled by the tensor unit $\one_\EC$ of $\EC$, while the blue dotted line by the tensor unit $\one_\ED$ of $\ED$. In the multifusion case, the tensor unit might not be simple, hence the dotted labels should be understood as the simple components of the tensor units. 

Second, the multiplication map $\mu$ is defined as 
\begin{equation} \label{eq:bi_tube_prod_1}
      \mu\left( \begin{aligned}
        %
    \end{aligned}\;\;.
\end{equation}
Clearly, this is associative by direct inspection on diagrams. Actually we need to perform F-moves to the right-hand side in order to write it as a linear combination of the basis diagrams. This is realized as follows: 
\begin{align}
    \text{RHS} & = \delta_{u,w'}\delta_{z,x'} \sum_{i,\alpha,\beta}[F_{av'b'}^v]^{i\alpha\beta}_{u\gamma'\zeta}\sum_{j,\rho,\sigma}[F_{y}^{ay'b'}]^{j\rho\sigma}_{z\mu'\nu} \; \begin{aligned}
        %
    \end{aligned}  \nonumber \\
    & = \delta_{u,w'}\delta_{z,x'}\sum_{i,\alpha,\beta}[F_{av'b'}^v]^{i\alpha\beta}_{u\gamma'\zeta}\sum_{j,\rho,\sigma}[F_{y}^{ay'b'}]^{j\rho\sigma}_{z\mu'\nu} \sum_{k,\tau,\lambda} [F^i_{aa'u'}]^{k\tau\lambda}_{v'\zeta'\alpha}\sum_{\theta}[F_j^{aa'z'}]^{k\tau\theta}_{y'\nu'\rho} \nonumber \\
    & \quad \quad \times  \sum_{s,\delta,\phi}[F^w_{ib'b}]^{s\delta\phi}_{v\beta\gamma}\sum_{\xi}[F_x^{jb'b}]^{s\delta\xi}_{y\sigma\mu} \sqrt{\frac{d_ad_{a'}}{d_k}}\sqrt{\frac{d_bd_{b'}}{d_s}}\; \begin{aligned}

    \end{aligned}\;\;,  \label{eq:tube_prod_2}
\end{align}
where the last equality comes from loop moves 
\begin{equation*}
    \begin{aligned}
        %
    \end{aligned}\;\;. 
\end{equation*}

\subsubsection{The coalgebra structure}

The coalgebra structure of $\mathbf{Tube}({_{\EC}}\EM_{\ED})$ is provided by the following maps. The counit is defined as 
\begin{equation} \label{eq:bi_tube_counit}
     \varepsilon\left(\begin{aligned}%
\end{aligned}=\delta_{z,u}\delta_{x,w}\delta_{y,v}\delta_{\nu,\zeta}\delta_{\mu,\gamma} \sqrt{\frac{d_a d_z^L}{d_v^L}}\sqrt{\frac{d_v^R d_b}{d_x^R}}. 
\end{equation}
The second equality is from the loop moves. From now on, all duals of the form $\bar{x}$ in $\EM$ refer to the duals defined by the right module structure, unless stated otherwise. The comultiplication (or coproduct) is defined as 
\begin{equation} \label{eq:bi_tube_coprod}
     \Delta\left(\begin{aligned}%
\end{aligned}\;\;.
\end{equation}
It is immediate to see that this map is coassociative. Moreover, applying $\varepsilon\otimes \id$ to the right-hand side yields 
\begin{equation*}
    \sum_{i,j,k,\rho,\sigma} \sqrt{ \frac{d_j^L}{d_ad_i^L}}\sqrt{ \frac{d_k^R}{d_j^Rd_b}}\delta_{i,u}\delta_{k,w}\delta_{j,v}\delta_{\sigma,\zeta}\delta_{\rho,\gamma}\sqrt{\frac{d_ad_i^L}{d_j^L}}\sqrt{\frac{d_j^Rd_b}{d_k^R}}\;
    \begin{aligned}%
    \end{aligned}\;\;. 
\end{equation*}
Applying $\id\otimes\varepsilon$ will also have the same result. Hence $\varepsilon$ and $\Delta$ endow $\mathbf{Tube}({_{\EC}}\EM_{\ED})$ with a coalgebra structure, as claimed.  

Two brief comments are in order. The first one is a key observation for computing the counit. Taking $\varepsilon$ on the expression \eqref{eq:tube_prod_2} results in 
\begin{align*}
        & \delta_{u,w'}\delta_{z,x'}\sum_{i,\alpha,\beta}[F_{av'b'}^v]^{i\alpha\beta}_{u\gamma'\zeta}\sum_{j,\rho,\sigma}[F_{y}^{ay'b'}]^{j\rho\sigma}_{z\mu'\nu} \sum_{k,\tau,\lambda} [F^i_{aa'u'}]^{k\tau\lambda}_{v'\zeta'\alpha}\sum_{\theta}[F_j^{aa'z'}]^{k\tau\theta}_{y'\nu'\rho} \nonumber \\
    & \times  \sum_{s,\delta,\phi}[F^w_{ib'b}]^{s\delta\phi}_{v\beta\gamma}\sum_{\xi}[F_x^{jb'b}]^{s\delta\xi}_{y\sigma\mu} \sqrt{\frac{d_ad_{a'}}{d_k}}\sqrt{\frac{d_bd_{b'}}{d_s}}\delta_{z',u'}\delta_{x,w}\delta_{i,j}\delta_{\xi,\phi}\delta_{\theta,\lambda}\sqrt{\frac{d_kd_{z'}^L}{d_i^L}} \sqrt{\frac{d_i^Rd_s}{d_x^R}} \\ 
    =&\;\delta_{u,w'}\delta_{z,x'}\sum_{i,\alpha,\beta}[F_{av'b'}^v]^{i\alpha\beta}_{u\gamma'\zeta}\sum_{\rho,\sigma}[F_{y}^{ay'b'}]^{i\rho\sigma}_{z\mu'\nu} \underbrace{\sum_{k,\tau,\lambda} [F^i_{aa'z'}]^{k\tau\lambda}_{v'\zeta'\alpha}[F_i^{aa'z'}]^{k\tau\lambda}_{y'\nu'\rho}}_{\delta_{v',y'}\delta_{\zeta',\nu'}\delta_{\alpha,\rho}} \\
    & \times \underbrace{\sum_{s,\delta,\phi}[F^x_{ib'b}]^{s\delta\phi}_{v\beta\gamma}[F_x^{ib'b}]^{s\delta\phi}_{y\sigma\mu}}_{\delta_{v,y}\delta_{\beta,\sigma}\delta_{\gamma,\mu}} \sqrt{\frac{d_ad_{a'}}{d_k}}\sqrt{\frac{d_bd_{b'}}{d_s}} \delta_{z',u'}\delta_{x,w}\sqrt{\frac{d_kd_{z'}^L}{d_i^L}} \sqrt{\frac{d_i^Rd_s}{d_x^R}} \\
    =&\;\delta_{u,w'}\delta_{z,x'}\underbrace{\sum_{i,\alpha,\beta}[F_{av'b'}^v]^{i\alpha\beta}_{u\gamma'\zeta}[F_{v}^{av'b'}]^{i\alpha\beta}_{z\mu'\nu}\sqrt{\frac{d_i^R}{d_i^L}}}_{\delta_{u,z}\delta_{\gamma',\mu'}\delta_{\zeta,\nu}\sqrt{\frac{d_{v'}^Rd_u^Ld_v^R}{d_{v'}^Ld_u^Rd_v^L}}} \delta_{v',y'}\delta_{\zeta',\nu'} \delta_{v,y}\delta_{\gamma,\mu} \delta_{z',u'}\delta_{x,w}\sqrt{\frac{d_ad_{a'}d_{z'}^Ld_bd_{b'}}{d_x^R}} \\
    =&\; \delta_{u,w'}\delta_{z,x'}\delta_{u,z}\delta_{\gamma',\mu'}\delta_{\zeta,\nu}\delta_{v',y'}\delta_{\zeta',\nu'} \delta_{v,y}\delta_{\gamma,\mu} \delta_{z',u'}\delta_{x,w}\sqrt{\frac{d_{a'}d_{z'}^L}{d_{v'}^L}}\sqrt{\frac{d_{v'}^Rd_{b'}}{d_u^R}}\sqrt{\frac{d_ad_u^L}{d_v^L}}\sqrt{\frac{d_v^Rd_b}{d_x^R}}\;,
\end{align*}
which is equal to the value of taking $\varepsilon$ directly to the diagram on the right-hand side of \eqref{eq:bi_tube_prod_1}, namely the value of the following diagram
\begin{equation*}
    \frac{\delta_{u,w'}\delta_{z,x'}}{d_x^R}\; \begin{aligned}

    \end{aligned}\;\;.
\end{equation*}
This observation applies to product of arbitrary number of basis diagrams. 
See Proposition~\ref{prop:F-symbols_property} for the properties of F-symbols. 

The second one is how to compute the comultiplication. We insert a wall labeled by $\sum_{i\in\Irr(\EM)} i$, and then apply the parallel moves; the result gives the components in the comultiplication. For example, performing the parallel moves twice, 
\begin{equation}
    \sum_{i\in\Irr(\EM)}
    \begin{aligned}
        %
    \end{aligned}
\end{equation}
gives the coproduct in \eqref{eq:bi_tube_coprod}. This method can be used to compute the coproduct of a product of any number of basis diagrams. 

One computes that 
\begin{equation} \label{eq:coprod_unit}
    \Delta(1) = \sum_{x,y,z\in\Irr(\EM)}
    \;\begin{aligned}
        %
    \end{aligned}\;\;. 
\end{equation}
Therefore, the left and right counital maps are 
\begin{align}
    \varepsilon_L\left(\begin{aligned}%
    \end{aligned}\;\;. \label{eq:right_counit_tube}
\end{align}

\subsubsection{The weak bialgebra structure}

In order to demonstrate the previous maps $(1, \mu, \varepsilon, \Delta)$ give $\mathbf{Tube}({_{\EC}}\EM_{\ED})$ a weak bialgebra structure, we need to verify the following identities
\begin{gather}
    \Delta(XX') = \Delta(X)\Delta(X'), \\ 
    (\Delta\otimes\id)\comp \Delta(1) = (\Delta(1)\otimes 1)(1\otimes \Delta(1)) = (1\otimes \Delta(1))(\Delta(1)\otimes 1), \\
    % \sum 1^{(1)}\otimes 1^{(2)}1'^{(1)}\otimes 1'^{(2)} = \sum 1^{(1)}\otimes 1'^{(1)}1^{(2)}\otimes 1'^{(2)}, \\
    \varepsilon(XX'X'') = \sum \varepsilon(XX'^{(1)})\varepsilon(X'^{(2)}X'') = \sum \varepsilon(XX'^{(2)})\varepsilon(X'^{(1)}X''), 
\end{gather}
for $X, X', X'' \in \mathbf{Tube}({_{\EC}}\EM_{\ED})$. Suppose the labels of $X'$ are $\{a', b', x', \mu',\cdots\}$, and similarly for $X''$. 

To show the first identity, we use the method provided in the last subsection to compute $\Delta(XX')$. That is,  
    \begin{align*}
    \Delta(XX')=&\sum_{i}\sum_{j,\rho}\sum_{k,\sigma}\sum_{s,\tau}\sum_{t,\theta} \sqrt{\frac{d_j^L}{d_{a'}d_i^L}}\sqrt{\frac{d_k^R}{d_{b'}d_j^R}}\sqrt{\frac{d_s^L}{d_{a}d_k^L}}\sqrt{\frac{d_t^R}{d_{b}d_s^R}} \delta_{u,w'}\delta_{z,x'}\\
    & \quad \quad  \begin{aligned}

    \end{aligned}\right) \\
    =&\; \Delta(X)\Delta(X'). 
    \end{align*}
The second identity is immediate: 
\begin{equation*}
    \begin{aligned}
        (\Delta(1)\otimes 1)(1\otimes \Delta(1)) & = \sum_{w,x,y,z\in\Irr(\EM)}
    \;\begin{aligned}
        %
    \end{aligned} \\
    & = (1\otimes \Delta(1))(\Delta(1)\otimes 1) = (\Delta\otimes\id)\comp \Delta(1). 
    \end{aligned}
\end{equation*}
Let us prove the third identity. To compute $\varepsilon(XX'X'')$, we first stack three diagrams according to the product rule, and then take $\varepsilon$ directly. The result is 
\begin{equation*}
    \begin{aligned}
        \varepsilon(XX'X'')  =&\; \delta_{u,w'}\delta_{z,x'}\delta_{u',w''}\delta_{z',x''}\delta_{u'',z''}\delta_{x,w}\\
        & \times \delta_{y'',v''}\delta_{\nu'',\zeta''}\sqrt{\frac{d_{a''}d_{z''}^L}{d_{y''}^L}}\delta_{z',u'}\delta_{\mu'',\gamma''}\sqrt{\frac{d_{y''}^Rd_{b''}}{d_{z'}^R}} \\
        & \times \delta_{y',v'}\delta_{\nu',\zeta'}\sqrt{\frac{d_{a'}d_{z'}^L}{d_{y'}^L}}\delta_{z,u}\delta_{\mu',\gamma'}\sqrt{\frac{d_{y'}^Rd_{b'}}{d_z^R}}\\
        &\times \delta_{y,v}\delta_{\nu,\zeta}\sqrt{\frac{d_ad_z^L}{d_y^L}} \delta_{\mu,\gamma}\sqrt{\frac{d_y^Rd_b}{d_x^R}}\;.
    \end{aligned}
\end{equation*}
On the other hand, one has 
\begin{equation*}
    \begin{aligned}
        \sum \varepsilon(X&X'^{(1)})\varepsilon(X'^{(2)}X'') \\
        =&\; \sum_{i,j,k,\rho,\sigma} \sqrt{\frac{d_j^L}{d_{a'}d_i^L}} \sqrt{\frac{d_k^R}{d_{b'}d_j^R}} \delta_{u,w'}\delta_{z,k} \delta_{x,w}\delta_{i,u'} \\
        & \quad \times \delta_{j,v'}\delta_{\sigma,\zeta'}\sqrt{\frac{d_{a'}d_{u'}^L}{d_{v'}^L}} \delta_{z,u}\delta_{\rho,\gamma'} \sqrt{\frac{d_{v'}^Rd_{b'}}{d_z^R}} \delta_{y,v}\delta_{\nu,\zeta} \sqrt{\frac{d_ad_z^L}{d_y^L}} \delta_{\mu,\gamma} \sqrt{\frac{d_y^Rd_b}{d_x^R}}   \\
        &\quad \times\delta_{i,w''}\delta_{z',x''}\delta_{x',k}\delta_{z'',u''}\delta_{y'',v''}\delta_{\nu'',\zeta''}\sqrt{\frac{d_{a''}d_{z''}^L}{d_{y''}^L}}\\
        &\quad \times \delta_{z',w''}\delta_{\mu'',\gamma''}\sqrt{\frac{d_{y''}^Rd_{b''}}{d_{z'}^R}} \delta_{y',j}\delta_{\nu',\sigma} \sqrt{\frac{d_{a'}d_{z'}^L}{d_{y'}^L}} \delta_{\mu',\rho} \sqrt{\frac{d_{y'}^Rd_{b'}}{d_{x'}^R}}\;, \\
        \sum \varepsilon(X&X'^{(2)})\varepsilon(X'^{(1)}X'') \\
        =&\; \sum_{i,j,k,\rho,\sigma} \sqrt{\frac{d_j^L}{d_{a'}d_i^L}} \sqrt{\frac{d_k^R}{d_{b'}d_j^R}} \delta_{u,k}\delta_{z,x'} \delta_{z',i}\delta_{x,w} \\
        & \quad \times \delta_{y',j}\delta_{\nu',\sigma}\sqrt{\frac{d_{a'}d_{z'}^L}{d_{y'}^L}} \delta_{z,u}\delta_{\mu',\rho} \sqrt{\frac{d_{y'}^Rd_{b'}}{d_z^R}} \delta_{y,v}\delta_{\nu,\zeta} \sqrt{\frac{d_ad_z^L}{d_y^L}} \delta_{\mu,\gamma} \sqrt{\frac{d_y^Rd_b}{d_x^R}}   \\
        &\quad \times\delta_{u',w''}\delta_{i,x''}\delta_{z'',u''}\delta_{k,w'}\delta_{y'',v''}\delta_{\nu'',\zeta''}\sqrt{\frac{d_{a''}d_{z''}^L}{d_{y''}^L}}\\
        &\quad \times \delta_{x'',u'}\delta_{\sigma,\gamma'}\sqrt{\frac{d_{y''}^Rd_{b''}}{d_{x''}^R}} \delta_{j,v'}\delta_{\sigma,\gamma'} \sqrt{\frac{d_{a'}d_{x''}^L}{d_{v'}^L}} \delta_{\rho,\zeta'} \sqrt{\frac{d_{v'}^Rd_{b'}}{d_{w'}^R}}\;,
    \end{aligned}
\end{equation*}
both of which are equal to $\varepsilon(XX'X'')$ by simple inspection.

\subsubsection{The antipode map and weak Hopf structure}

The antipode is defined as 
\begin{equation} \label{eq:bi_tube_antipode}
    S\left( 
          \begin{aligned}
    \end{aligned}\;\;. 
\end{equation}
Let us derive that $S$ satisfies the antipode axioms \eqref{eq:weak_antipode}, which read explicitly as 
\begin{gather}
    \sum X^{(1)}S(X^{(2)}) = \varepsilon_L(X),\\
    \sum S(X^{(1)})X^{(2)} = \varepsilon_R(X), \\ 
     \sum S(X^{(1)})X^{(2)}S(X^{(3)}) = S(X). 
\end{gather}
To verify the first identity, one computes  
\begin{align*}
\sum X^{(1)}S(X^{(2)}) &=
    \sum_{i,j,k,\rho,\sigma} \sqrt{ \frac{d_j^L}{d_ad_i^L}}\sqrt{ \frac{d_k^R}{d_j^Rd_b}} \;
    \begin{aligned}%
    \end{aligned}\; = \varepsilon_L(X).  
\end{align*}
The proof of the second identity is similar:
\begin{align*}
\sum S(X^{(1)})X^{(2)} &=
    \sum_{i,j,k,\rho,\sigma} \sqrt{ \frac{d_j^L}{d_ad_i^L}}\sqrt{ \frac{d_k^R}{d_j^Rd_b}} \; \frac{d_u^Ld_v^R}{d_v^Ld_w^R}\;\begin{aligned}%
    \end{aligned}\; = \varepsilon_R(X).  
\end{align*}
It remains to show the last identity. This immediately follows from $\sum S(X^{(1)})X^{(2)}S(X^{(3)}) = \sum \varepsilon_R(X^{(1)})S(X^{(2)})$, as one has
\begin{align*}
     \sum \varepsilon_R(X^{(1)})S(X^{(2)}) 
     =& \sum_{i,j,k,\rho,\sigma}\sqrt{ \frac{d_j^L}{d_ad_i^L}}\sqrt{ \frac{d_k^R}{d_j^Rd_b}} 
    \sum_{z'\in\Irr(\EM)} \delta_{i,u}\delta_{k,w}\delta_{\rho,\gamma}\delta_{j,v}\delta_{\sigma,\zeta} \sqrt{ \frac{d_ad_i^L}{d_j^L}}\sqrt{ \frac{d_j^Rd_b}{d_k^R}}  \; \\
    &\begin{aligned}
        %
    \end{aligned}\; = S(X)\;. 
\end{align*}

\begin{remark}
Suppose that $d_x^L=d_x^R=:d_x$ for all objects $x$ in ${_{\EC}}\EM_{\ED}$. One typical example is the domain wall ${_{\EC}}\EC_{\EC}$. Under this condition, the antipode reduces to 
\begin{equation} 
    S\left( 
          \begin{aligned}%
    \end{aligned}\;\;. 
\end{equation}
Using F-move, the diagram on the right-hand side is equal to 
\begin{align*}
    \begin{aligned}%
    \end{aligned} \;\;. 
\end{align*}
Applying $S$ again to the above expression, one gets 
\begin{align*}
    S\left(\begin{aligned}%
    \end{aligned}  \\
     & =  \frac{d_x}{d_{z}} \sum_{y'',\mu'',\nu''}\underbrace{\sum_{y',\mu',\nu'}[F^z_{\bar{a}x\bar{b}}]_{y\mu\nu}^{y'\mu'\nu'}[(F_{\bar{a}x\bar{b}}^z)^{-1}]_{y'\mu'\nu'}^{y''\mu''\nu''}}_{\delta_{y,y''}\delta_{\mu,\mu''}\delta_{\nu,\nu''}} \\
     & \quad \times \sum_{v'',\zeta'',\gamma''}\underbrace{\sum_{v',\zeta',\gamma'}[(F_{aub}^w)^{-1}]_{v\zeta\gamma}^{v'\zeta'\gamma'}[F^w_{aub}]_{v'\zeta'\gamma'}^{v''\zeta''\gamma''}}_{\delta_{v,v''}\delta_{\zeta,\zeta''}\delta_{\gamma,\gamma''}}
    \; 
    \begin{aligned}
\end{aligned}\;\;.
\end{align*}
This means that the antipode commutes with the F-move. Also, it implies that 
\begin{equation} \label{eq:double_antipode}
    S^2\left( 
          \begin{aligned}%
\end{aligned}\;\;. 
\end{equation}
Hence, it follows that $S^2\neq \id$ as expected for a general weak Hopf algebra. 
\end{remark}

Up to this point, we have shown that the domain wall tube algebra $\mathbf{Tube}({_{\EC}}\EM_{\ED})$ is a weak Hopf algebra, with the structure morphisms $\eta,\mu,\varepsilon,\Delta,S$ given by \eqref{eq:bi_tube_unit}, \eqref{eq:bi_tube_prod_1}, \eqref{eq:bi_tube_counit}, \eqref{eq:bi_tube_coprod}, \eqref{eq:bi_tube_antipode}, respectively. Next we will demonstrate that it is also a $C^*$ weak Hopf algebra under some additional assumption, hence a semisimple algebra.

\subsubsection{The $C^*$ structure}

In the proof of the weak Hopf algebra structure, we did not impose the condition $d^L_m = d^R_m$ (for $m\in \EM$) in order to maintain generality. However, when considering a $C^*$-structure, it becomes necessary to impose the condition that $d^L_m = d^R_m=d_m$. We will adopt this assumption in this section.

The $*$-operation is defined as 
\begin{gather} \label{eq:bi_tube_star}
\left( \begin{aligned}
    \end{aligned}\;\;,
\end{gather}
extended antilinearly. To show this gives a $*$-structure, we need to verify the following conditions
\begin{equation}
    X^{**} = X,\quad (XX')^* = X'^*X^*,\quad \Delta(X^*) = \Delta(X)^{*}.
\end{equation}

First, using F-move, the diagram on the left-hand side of \eqref{eq:bi_tube_star} can be written as 
\begin{align*}
    \begin{aligned}%
    \end{aligned}\;. 
\end{align*}
Therefore, we have 
\begin{align*}
    \left(\begin{aligned}%
    \end{aligned} 
    \\
    & = \frac{d_x}{d_z}\sum_{v'',\gamma'',\zeta''} \underbrace{\sum_{v',\zeta',\gamma'} [(F_{\bar{a}w\bar{b}}^u)^{-1}]^{v\gamma\zeta}_{v'\gamma'\zeta'}[(F_u^{\bar{a}w\bar{b}})^{-1}]_{v'\gamma'\zeta'}^{v''\gamma''\zeta''}}_{\delta_{v,v''}\delta_{\gamma,\gamma''}\delta_{\zeta,\zeta''}} \\
    & \quad \times  \sum_{y'',\nu'',\mu''}\underbrace{\sum_{y',\mu',\nu'}[(F_z^{\bar{a}x\bar{b}})^{-1}]^{y\mu\nu}_{y'\mu'\nu'} [(F_{\bar{a}x\bar{b}}^z)^{-1}]_{y'\mu'\nu'}^{y''\mu''\nu''}}_{\delta_{y,y''}\delta_{\mu,\mu''}\delta_{\nu,\nu''}}\; 
    \begin{aligned}
    \end{aligned}\;, 
\end{align*}
where the second equality follows from $(F_{\bar{a}w\bar{b}}^u)^\dagger = (F_{\bar{a}w\bar{b}}^u)^{-1}$ and $(F^{\bar{a}x\bar{b}}_z)^\dagger = (F^{\bar{a}x\bar{b}}_z)^{-1}$ by unitarity. 
This shows that the $*$-operation commutes with the F-move. 
In particular, $X^{**}=X$. 

Second, to verify $(XX')^* = X'^*X^*$, it follows from the expression \eqref{eq:tube_prod_2} and unitarity that 
\begin{align*}
    (XX')^* &  = \delta_{u,w'}\delta_{z,x'}\sum_{i,\alpha,\beta}\overline{[F_{av'b'}^v]^{i\alpha\beta}_{u\gamma'\zeta}}\sum_{j,\rho,\sigma}\overline{[F_{y}^{ay'b'}]^{j\rho\sigma}_{z\mu'\nu}} \sum_{k,\tau,\lambda} \overline{[F^i_{aa'u'}]^{k\tau\lambda}_{v'\zeta'\alpha}}\sum_{\theta}\overline{[F_j^{aa'z'}]^{k\tau\theta}_{y'\nu'\rho}} \nonumber \\
    & \quad \quad \times  \sum_{s,\delta,\phi}\overline{[F^w_{ib'b}]^{s\delta\phi}_{v\beta\gamma}} \sum_{\xi}\overline{[F_x^{jb'b}]^{s\delta\xi}_{y\sigma\mu}} \sqrt{\frac{d_ad_{a'}}{d_k}}\sqrt{\frac{d_bd_{b'}}{d_s}}\frac{d_{z'}}{d_x} \;
\begin{aligned}\begin{tikzpicture}[scale=0.65]
\filldraw[black!60, fill=gray!15, dotted, even odd rule] (0,0) circle[radius=0.5] (0,0) circle[radius=1.5];
             \draw[line width=.6pt,black] (0,0.5)--(0,1.5);
             \draw[line width=.6pt,black] (0,-0.5)--(0,-1.5);
             \draw[red] (0,1.3) arc[start angle=90, end angle=270, radius=1.3];
             \draw[blue] (0,0.8) arc[start angle=90, end angle=-90, radius=0.8];
            \node[ line width=0.6pt, dashed, draw opacity=0.5] (a) at (0,1.7){$\scriptstyle u'$};
             \node[ line width=0.6pt, dashed, draw opacity=0.5] (a) at (0,-1.7){$\scriptstyle z'$};
            \node[ line width=0.6pt, dashed, draw opacity=0.5] (a) at (-1,0){$\scriptstyle \bar{k}$};
            \node[ line width=0.6pt, dashed, draw opacity=0.5] (a) at (1,0){$\scriptstyle \bar{s}$};
            \node[ line width=0.6pt, dashed, draw opacity=0.5] (a) at (-0.2,-1.05){$\scriptstyle j$};
            \node[ line width=0.6pt, dashed, draw opacity=0.5] (a) at (0.2,-1.3){$\scriptstyle \theta$};
            \node[ line width=0.6pt, dashed, draw opacity=0.5] (a) at (-0.2,-0.65){$\scriptstyle \xi$};
            \node[ line width=0.6pt, dashed, draw opacity=0.5] (a) at (0,-0.3){$\scriptstyle x$};
            \node[ line width=0.6pt, dashed, draw opacity=0.5] (a) at (0,0.3){$\scriptstyle w$};
            \node[ line width=0.6pt, dashed, draw opacity=0.5] (a) at (-0.2,1.1){$\scriptstyle i$};
            \node[ line width=0.6pt, dashed, draw opacity=0.5] (a) at (0.2,1.2){$\scriptstyle \lambda$};
            \node[ line width=0.6pt, dashed, draw opacity=0.5] (a) at (-0.2,0.7){$\scriptstyle \phi$};
        \end{tikzpicture}
    \end{aligned} \\
    &  = \delta_{u,w'}\delta_{z,x'}\sum_{i,\alpha,\beta}[(F_{av'b'}^v)^{-1}]_{i\alpha\beta}^{u\gamma'\zeta}\sum_{j,\rho,\sigma}[(F_{y}^{ay'b'})^{-1}]_{j\rho\sigma}^{z\mu'\nu} \sum_{k,\tau,\lambda} [(F^i_{aa'u'})^{-1}]_{k\tau\lambda}^{v'\zeta'\alpha}\sum_{\theta}[(F_j^{aa'z'})^{-1}]_{k\tau\theta}^{y'\nu'\rho} \nonumber \\
    & \quad \quad \times  \sum_{s,\delta,\phi}[(F^w_{ib'b})^{-1}]_{s\delta\phi}^{v\beta\gamma} \sum_{\xi}[(F_x^{jb'b})^{-1}]_{s\delta\xi}^{y\sigma\mu} \sqrt{\frac{d_ad_{a'}}{d_k}}\sqrt{\frac{d_bd_{b'}}{d_s}}\frac{d_{z'}}{d_x} \;
\begin{aligned}\begin{tikzpicture}[scale=0.65]
\filldraw[black!60, fill=gray!15, dotted, even odd rule] (0,0) circle[radius=0.5] (0,0) circle[radius=1.5];
             \draw[line width=.6pt,black] (0,0.5)--(0,1.5);
             \draw[line width=.6pt,black] (0,-0.5)--(0,-1.5);
             \draw[red] (0,1.3) arc[start angle=90, end angle=270, radius=1.3];
             \draw[blue] (0,0.8) arc[start angle=90, end angle=-90, radius=0.8];
            \node[ line width=0.6pt, dashed, draw opacity=0.5] (a) at (0,1.77){$\scriptstyle u'$};
             \node[ line width=0.6pt, dashed, draw opacity=0.5] (a) at (0,-1.7){$\scriptstyle z'$};
            \node[ line width=0.6pt, dashed, draw opacity=0.5] (a) at (-1,0){$\scriptstyle \bar{k}$};
            \node[ line width=0.6pt, dashed, draw opacity=0.5] (a) at (1,0){$\scriptstyle \bar{s}$};
            \node[ line width=0.6pt, dashed, draw opacity=0.5] (a) at (-0.2,-1.05){$\scriptstyle j$};
            \node[ line width=0.6pt, dashed, draw opacity=0.5] (a) at (0.2,-1.3){$\scriptstyle \theta$};
            \node[ line width=0.6pt, dashed, draw opacity=0.5] (a) at (-0.2,-0.65){$\scriptstyle \xi$};
            \node[ line width=0.6pt, dashed, draw opacity=0.5] (a) at (0,-0.3){$\scriptstyle x$};
            \node[ line width=0.6pt, dashed, draw opacity=0.5] (a) at (0,0.3){$\scriptstyle w$};
            \node[ line width=0.6pt, dashed, draw opacity=0.5] (a) at (-0.2,1.1){$\scriptstyle i$};
            \node[ line width=0.6pt, dashed, draw opacity=0.5] (a) at (0.2,1.2){$\scriptstyle \lambda$};
            \node[ line width=0.6pt, dashed, draw opacity=0.5] (a) at (-0.2,0.7){$\scriptstyle \phi$};
        \end{tikzpicture}
    \end{aligned}\;. 
\end{align*}
By a similar computation as in \eqref{eq:tube_prod_2}, one gets  
\begin{align*}
    X'^*X^* & = \frac{d_{z'}}{d_x}\delta_{u,w'}\delta_{z,x'}\sum_{i,\alpha,\beta}[F^{av'b'}_v]^{i\alpha\beta}_{u\gamma'\zeta}\sum_{j,\rho,\sigma}[F^{y}_{ay'b'}]^{j\rho\sigma}_{z\mu'\nu} \sum_{k,\tau,\lambda} [F_i^{aa'u'}]^{k\tau\lambda}_{v'\zeta'\alpha}\sum_{\theta}[F^j_{aa'z'}]^{k\tau\theta}_{y'\nu'\rho} \nonumber \\
    & \quad \quad \times  \sum_{s,\delta,\phi}[F_w^{ib'b}]^{s\delta\phi}_{v\beta\gamma} \sum_{\xi}[F^x_{jb'b}]^{s\delta\xi}_{y\sigma\mu} \sqrt{\frac{d_ad_{a'}}{d_k}}\sqrt{\frac{d_bd_{b'}}{d_s}} \;
\begin{aligned}\begin{tikzpicture}[scale=0.65]
\filldraw[black!60, fill=gray!15, dotted, even odd rule] (0,0) circle[radius=0.5] (0,0) circle[radius=1.5];
             \draw[line width=.6pt,black] (0,0.5)--(0,1.5);
             \draw[line width=.6pt,black] (0,-0.5)--(0,-1.5);
             \draw[red] (0,1.3) arc[start angle=90, end angle=270, radius=1.3];
             \draw[blue] (0,0.8) arc[start angle=90, end angle=-90, radius=0.8];
            \node[ line width=0.6pt, dashed, draw opacity=0.5] (a) at (0,1.76){$\scriptstyle u'$};
             \node[ line width=0.6pt, dashed, draw opacity=0.5] (a) at (0,-1.7){$\scriptstyle z'$};
            \node[ line width=0.6pt, dashed, draw opacity=0.5] (a) at (-1,0){$\scriptstyle \bar{k}$};
            \node[ line width=0.6pt, dashed, draw opacity=0.5] (a) at (1,0){$\scriptstyle \bar{s}$};
            \node[ line width=0.6pt, dashed, draw opacity=0.5] (a) at (-0.2,-1.05){$\scriptstyle j$};
            \node[ line width=0.6pt, dashed, draw opacity=0.5] (a) at (0.2,-1.3){$\scriptstyle \theta$};
            \node[ line width=0.6pt, dashed, draw opacity=0.5] (a) at (-0.2,-0.65){$\scriptstyle \xi$};
            \node[ line width=0.6pt, dashed, draw opacity=0.5] (a) at (0,-0.3){$\scriptstyle x$};
            \node[ line width=0.6pt, dashed, draw opacity=0.5] (a) at (0,0.3){$\scriptstyle w$};
            \node[ line width=0.6pt, dashed, draw opacity=0.5] (a) at (-0.2,1.1){$\scriptstyle i$};
            \node[ line width=0.6pt, dashed, draw opacity=0.5] (a) at (0.2,1.2){$\scriptstyle \lambda$};
            \node[ line width=0.6pt, dashed, draw opacity=0.5] (a) at (-0.2,0.7){$\scriptstyle \phi$};
        \end{tikzpicture}
    \end{aligned}\;. 
\end{align*} 
Then the identity $(XX')^* = X'^*X^*$ follows from the following properties on F-symbols: 
\begin{gather}
    [F^{av'b'}_v]^{i\alpha\beta}_{u\gamma'\zeta}= [(F_{av'b'}^v)^{-1}]_{i\alpha\beta}^{u\gamma'\zeta}, \quad [F^{y}_{ay'b'}]^{j\rho\sigma}_{z\mu'\nu} = [(F_{y}^{ay'b'})^{-1}]_{j\rho\sigma}^{z\mu'\nu},  \\
    [F_i^{aa'u'}]^{k\tau\lambda}_{v'\zeta'\alpha} = [(F^i_{aa'u'})^{-1}]_{k\tau\lambda}^{v'\zeta'\alpha},  \quad [F^j_{aa'z'}]^{k\tau\theta}_{y'\nu'\rho} = [(F_j^{aa'z'})^{-1}]_{k\tau\theta}^{y'\nu'\rho}, \\
    [F_w^{ib'b}]^{s\delta\phi}_{v\beta\gamma} = [(F^w_{ib'b})^{-1}]_{s\delta\phi}^{v\beta\gamma}, \quad 
    [F^x_{jb'b}]^{s\delta\xi}_{y\sigma\mu} = 
    [(F_x^{jb'b})^{-1}]_{s\delta\xi}^{y\sigma\mu}.
\end{gather}
Their proofs correspond to evaluating the following diagrams using different F-moves: 
\begin{equation*}
    \begin{aligned}

    \end{aligned}. 
\end{equation*}
Take the first one as an example, on the one hand, one has 
\begin{align*}
    \begin{aligned}
        %
    \end{aligned}; 
\end{align*}
on the other hand, one has 
\begin{align*}
    \begin{aligned}
        %
    \end{aligned}; 
\end{align*}
then the first identity comes from by comparing the above two expressions.

Lastly, the identity $\Delta(X^*) = \Delta(X)^*$ is established as follows: 
\begin{align*}
       \Delta(X)^*&= \sum_{i,j,k,\rho,\sigma} \sqrt{ \frac{d_k}{d_ad_id_b}}\frac{d_i}{d_k}\frac{d_z}{d_x}\;
\begin{aligned}%
    \end{aligned}=\Delta(X^*), 
    \end{align*}
where the last equality comes from the method of inserting a wall labeled by $\sum_{k\in\Irr(\EM)}k$, and then applying the parallel move.

\subsection{Haar integral and canonical grouplike element}

The Haar integral plays a central role in the representation theory of weak Hopf algebras, particularly in their applications to the construction of lattice models for topological phases~\cite{Buerschaper2013a,chen2021ribbon,jia2023boundary,Jia2023weak,jia2024generalized,jia2024weakhopfnoninvertible,jia2024quantumcluster}. 
In this section, we demonstrate that the domain wall tube algebra admits a Haar integral and provide its explicit expression.

Recall that for a weak Hopf algebra $W$, a \emph{Haar integral} is an element $\mathfrak{I} \in W$ satisfying the conditions
\[
w\mathfrak{I} = \varepsilon_L(w)\mathfrak{I}, \qquad \mathfrak{I}w = \mathfrak{I}\varepsilon_R(w), \qquad \text{and} \qquad \varepsilon_L(\mathfrak{I}) = \varepsilon_R(\mathfrak{I}) = 1
\]
for all $w \in W$. It has been demonstrated that a Haar integral exists uniquely in any $C^*$ weak Hopf algebra~\cite{BOHM1998weak}. Moreover, to verify that a given element $\mathfrak{I}$ is a Haar integral, it suffices to check the simplified conditions:
\[
w\mathfrak{I} = \varepsilon_L(w)\mathfrak{I} \quad \text{for all } w \in W, \qquad \text{and} \qquad \varepsilon_R(\mathfrak{I}) = 1.
\]
See \cite[Prop.~3.26]{BOHM1998weak}. 

\begin{proposition}
The Haar integral of the domain wall tube algebra $\Tube({_{\EC}}\EM_{\ED})$ is of the form
\begin{equation} \label{eq:haar_integral}
   \mathfrak{I} =\frac{1}{(\operatorname{rank} \EM)^2} \sum_{a,b,x,y,z,\mu,\nu} \frac{1}{d_y^Rd_z^L}\sqrt{\frac{d_a d_b}{d_y^Ld_z^Ld_x^Rd_y^R}} \;\;
    \begin{aligned}
    \end{aligned}\;,
\end{equation}
where $\operatorname{rank} \EM$ is the number of equivalence classes of simple objects in $\EM$.
\end{proposition}

\begin{proof}
    First we use topological moves to reformulate the right-hand side of \eqref{eq:haar_integral}. Indeed, since we have 
    \begin{align*}
        \sum_{a,b,\mu,\nu}\sqrt{\frac{d_a d_b}{d_y^Ld_z^Ld_x^Rd_y^R}}\;
    \begin{aligned}%
    \end{aligned} \;,
    \end{align*}
    we see that 
    \begin{equation} \label{eq:haar_integral_1}
        \mathfrak{I} = \frac{1}{\operatorname{rank} \EM} \sum_{x,z} \frac{1}{d_z^L}\;\; 
        \begin{aligned}%
    \end{aligned}\;.
    \end{equation}
    Here $\bar{z}$ is the left dual of $z$ defined by the left module structure.
    Now let us verify the condition $X\mathfrak{I} = \varepsilon_L(X)\mathfrak{I}$ for a general basis diagram 
    $$X= \begin{aligned}%
    \end{aligned}\;.$$
    In fact, one can compute 
    \begin{align*}
        X\mathfrak{I} & = \frac{1}{\operatorname{rank} \EM} \sum_{x,z} \frac{1}{d_z^L}\delta_{u,x}\delta_{z',x} \; 
        \begin{aligned}%
    \end{aligned} \;,
    \end{align*}
    which is equal to $\varepsilon_L(X)\mathfrak{I}$ by Eqs.~\eqref{eq:haar_integral_1} and \eqref{eq:left_counit_tube}. 

    It remains to verify that $\mathfrak{I}$ is right-normalized, i.e., $\varepsilon_R(\mathfrak{I}) = 1$. It follows from Eqs.~\eqref{eq:haar_integral} and \eqref{eq:right_counit_tube} that 
    \begin{align*}
        \varepsilon_R(\mathfrak{I}) & = \frac{1}{(\operatorname{rank} \EM)^2} \sum_{a,b,x,y,z} N_{az}^yN_{yb}^x\frac{1}{d_y^Rd_z^L}\sqrt{\frac{d_a d_b}{d_y^Ld_z^Ld_x^Rd_y^R}}
        \sum_{z'}\sqrt{\frac{d_a d_z^L}{d_y^L}}\sqrt{\frac{d_y^Rd_b}{d_x^R}} \; \begin{aligned}

    \end{aligned}\;,
    \end{align*}
    as expected. Here we used the fact that $N_{az}^y  = N_{z\bar{y}}^{\bar{a}}$ and $N_{yb}^x = N_{\bar{x}y}^{\bar{b}}$, which follows from $\Hom(a\otimes z,y) \simeq \Hom(z\otimes \bar{y},\bar{a})$ and $\Hom(y\otimes b,x) \simeq \Hom(\bar{x}\otimes y,\bar{b})$, and the formulas $d_z^Ld_{\bar{y}}^L = \sum_a N_{z\bar{y}}^ad_a$, $d_{\bar{x}}^Rd_y^R = \sum_b N_{\bar{x}y}^{\bar{b}}d_b$. 
\end{proof}

\begin{remark} \label{rmk:module_dual_obj}
    Note that we use $\bar{z}$ to denote the dual object of $z$ in $\EM$ for both the left and right module structures. Although the precise meaning of $\bar{z}$ differs depending on context, it is always clear from the usage. To avoid cluttering the equations, we do not distinguish between these cases notationally. We also emphasize that the notion of duality here is in the sense of the multifusion category formed by module categories; see Ref.~\cite{jia2024weakTube,Kitaev2012boundary} for details. 
  The following are some additional remarks regarding the subtleties involved in the proof of Eq.~\eqref{eq:haar_integral_1}:
   \begin{itemize}
       \item[(i)] For the $_{\EC}\EM$ module structure, to make the string diagram calculation work, we need to work in the multifusion category $\EuScript{P}=\oplus_{i,j=0,1}\EP_{i,j}$ where $\EP_{0,0}=\EC$, $\EP_{0,1}=\EM$, $\EP_{1,0}=\EM^{\rm op}$ and $\EP_{1,1}=\EC_{\EM}^{\vee}=\Fun_{\EC}(\EM,\EM)$. 
      For $z\in \EM$, the dual object $\bar{z}$ is in $\EP_{1,0}$.
       The tensor product $y \otimes \bar{z}$ is an object in $\EC$, so it is legitimate to have a morphism $\nu: a \to y \otimes \bar{z}$ with $a \in \EC$. However, $\bar{z} \otimes y$ lies in the category $\EC_{\EM}^{\vee}$, and hence there does not exist a morphism from $a \in \EC$ to $\bar{z} \otimes y$.    
       Since $y\otimes \bar{z} \in \EC$, the parallel move works as equality in Hom-space of $\EC$
       \[
       \sum_{a,\nu} \sqrt{\frac{d_a}{d_yd_{\bar{z}}}}
\begin{aligned}
.
	\end{aligned}
       \]   
    \item[(ii)] Similarly, for the $\EM_{\ED}$ module structure, $\bar{y} \otimes x \in \ED$, so it is valid to have morphism $\mu$ from $b \in \ED$ to $\bar{y} \otimes x$. On the other hand, $x \otimes \bar{y}$ lies in $\ED_{\EM}^{\vee}$, and thus there is no morphism from $b \in \ED$ to $x \otimes \bar{y}$.
    In the Hom-space of $\ED$, we have the parallel move
 \[
       \sum_{b,\mu} \sqrt{\frac{d_b}{d_{\bar{y}}d_x}}
\begin{aligned}
		%
	\end{aligned}.
       \]   
      \end{itemize} 
\end{remark}

For the pseudo-unitary bimodule category, where $d_m^L=d^R_m=d_m$ for all $m\in \EM$, the Haar integral of $\Tube({_{\EC}}\EM_{\ED})$ is of the form:
\begin{equation} \label{eq:haar_integral_2}
   \mathfrak{I} =\frac{1}{(\operatorname{rank} \EM)^2} \sum_{a,b,x,y,z,\mu,\nu} \frac{1}{d_y^2d_z}\sqrt{\frac{d_a d_b}{d_zd_x}} \;\;
    \begin{aligned}%
    \end{aligned}\;.
\end{equation}

Recall that a pivotal structure on a rigid monoidal category is an isomorphism between two functors: $\theta: \id \to (\bullet)^{\vee \vee}$. The representation category of weak Hopf algebra has a pivotal structure if it has a grouplike element. In this case, the weak Hopf algebra is pivotal. In this part, we will demonstrate that the domain wall tube algebra is pivotal by proving it has a canonical grouplike element.

By definition, this is a positive and invertible element $\xi\in W$ such that $S^2(w) = \xi w\xi^{-1}$ (for all $w\in W$), and $\mathfrak{I}^{(2)}\otimes \mathfrak{I}^{(1)} = \mathfrak{I}^{(1)}\otimes \xi \mathfrak{I}^{(2)}\xi$, where $\mathfrak{I}$ is the Haar integral. In a $C^*$ weak Hopf algebra, this element also exists uniquely. 

\begin{proposition}
    Assume that $d_m^L=d^R_m=d_m$ for all $m\in \EM$. Then the canonical grouplike element and its inverse of the domain wall tube algebra $\Tube({_{\EC}}\EM_{\ED})$ are of the form
\begin{equation} 
    \xi =\sum_{x,z\in\Irr(\EM)} \frac{d_x}{d_z} \;\begin{aligned}

    \end{aligned}\;\;.
\end{equation}
\end{proposition}

\begin{proof}
    The condition $S^2(X) = \xi X\xi^{-1}$ (for all $X\in \Tube({_{\EC}}\EM_{\ED})$) is easy to verify by Eq.~\eqref{eq:double_antipode}. Moreover, as 
    \begin{align} \label{eq:coprod_haar_int}
        \Delta(\mathfrak{I}) = \frac{1}{(\operatorname{rank} \EM)^2} \sum_{a,b,x,y,z,\mu,\nu} \frac{1}{d_y^2d_z}\sqrt{\frac{d_a d_b}{d_zd_x}} \sum_{i,j,k,\rho,\sigma}\sqrt{\frac{d_k}{d_ad_id_b}}\;\;
    \begin{aligned}%
    \end{aligned}\;,
    \end{align}
    we have that $\mathfrak{I}^{(1)}\otimes \xi \mathfrak{I}^{(2)}\xi$ is equal to 
    \begin{align*}
        &\quad \frac{1}{(\operatorname{rank} \EM)^2} \sum_{a,b,x,y,z,\mu,\nu} \frac{1}{d_y^2d_z}\sqrt{\frac{d_a d_b}{d_zd_x}} \sum_{i,j,k,\rho,\sigma}\sqrt{\frac{d_k}{d_ad_id_b}}\;\;
    \begin{aligned}%
    \end{aligned}
    \end{align*}
    which is just $\mathfrak{I}^{(2)}\otimes \mathfrak{I}^{(1)}$ by Eq.~\eqref{eq:coprod_haar_int}. Here we relabeled the indices as $i \leftrightarrow z$, $j \leftrightarrow y$, $k \leftrightarrow x$, $\sigma \leftrightarrow \nu$, $\rho \leftrightarrow \mu$.  
\end{proof}

\subsection{Examples of weak Hopf tube algebra of domain wall}

Before discussing the general properties of the domain wall tube algebra, we first provide several concrete examples in this section.
Throughout this section, to simplify notation, we use $1$ to label the string for the tensor unit $\one$ of the fusion category.

\subsubsection{Group tube algebra}
\label{sec:groupTube}

Given a finite group $G$, consider the category $\underline{G}$ whose objects are group elements in $G$, morphism spaces are $\Hom(g,h)=\delta_{g,h}\id_g$, the tensor product is given by
\begin{equation}
    g\otimes h:=gh,
\end{equation}
and the tensor unit is provided by the identity element $e\in G$. This is the categorification of the finite group $G$. We can represent all morphisms in $\underline{G}$ as 
\begin{equation*}
        \begin{tikzpicture}
            \draw [fill = black] (0,0) circle (1.3pt);
            \draw [black, ->-=1, line width=0.6pt] (0,0.4) [partial ellipse=245:-66:0.3 and 0.4];
            \node[ line width=0.6pt, dashed, draw opacity=0.5] (a) at (0,-0.2){$\scriptstyle g$};
            \draw [fill = black] (1.5,0) circle (1.3pt);
            \draw [black, ->-=1, line width=0.6pt] (1.5,0.4) [partial ellipse=245:-66:0.3 and 0.4];
            \node[ line width=0.6pt, dashed, draw opacity=0.5] (a) at (1.5,-0.2){$\scriptstyle h$};
            \node[ line width=0.6pt, dashed, draw opacity=0.5] (a) at (3,0.2){$\cdots$};
            \draw [fill = black] (4.5,0) circle (1.3pt);
            \draw [black, ->-=1, line width=0.6pt] (4.5,0.4) [partial ellipse=245:-66:0.3 and 0.4];
            \node[ line width=0.6pt, dashed, draw opacity=0.5] (a) at (4.5,-0.2){$\scriptstyle k$};
        \end{tikzpicture}
\end{equation*}
with all arrows being the identity morphisms. Note that the dual object $\bar{g} = g^{-1}$.

Because the map $G\to G, h\mapsto gh$ is a permutation, the corresponding matrix $N_g=(N_{g,h}^l)_{h,l\in G}$ is a permutation matrix, meaning that each row and each column contains exactly one entry of $1$, with all other entries being zero. The real eigenvalues of a permutation matrix can only be $1$ or $-1$,\,\footnote{This is shown as follows. Suppose $P$ is a permutation matrix and $\lambda$ is an eigenvalue with eigenvector $v$, i.e., $Pv=\lambda v$. Since $P$ only permutes the order of the components of $v$, the norm does not change. Hence $\|v\|=\|Pv\| = \|\lambda v\| = |\lambda|\cdot\|v\|$, which implies $|\lambda| = 1$. Clearly, $v:=(1,\cdots,1)$ satisfies $Pv=v$, so $P$ has $1$ as the largest real eigenvalue.} thus the dimension $d_g =1$ for all $g\in \underline{G}$, and the total dimension $d_{\underline{G}} = \sum_{g\in \underline{G}} d_g^2 = |G|$. 

Since  $\underline{G}$  is a  $\underline{G}|\underline{G}$-bimodule category via $(g\otimes h) \otimes k= ghk$, we can construct the corresponding domain wall tube algebra $\Tube({_{\underline{G}}}\underline{G}_{\underline{G}})$. But the labels in basis diagrams are not arbitrary as $\Hom(g,h) = \delta_{g,h}\id_g$. For example, the following trivalent graphs are valid only if $k=gh$:
\begin{equation}
    \begin{aligned}

    \end{aligned}\;\;,
\end{equation}
where the morphism over the vertex is the identity. Therefore, the loops move and parallel move are written as 
\begin{equation}
    \begin{aligned}
        %
    \end{aligned}. 
\end{equation}

Let us write down the structure morphisms of $\Tube({_{\underline{G}}}\underline{G}_{\underline{G}})$ in their explicit forms. A general element in this algebra is 
\begin{equation} \label{eq:group_basis}
    \begin{aligned}%
    \end{aligned}\;\;, \;\;a, b, c, d \in \underline{G}.
\end{equation}
Thus $\dim \Tube({_{\underline{G}}}\underline{G}_{\underline{G}}) = |G|^4$. Using the expansion \eqref{eq:tube_prod_2}, the multiplication is given by 
\begin{equation}
    \begin{aligned}%
    \end{aligned}\;\;.  
\end{equation}
The coalgebra structure is 
\begin{equation}
    \Delta\left(\begin{aligned}%
    \end{aligned}\right) = \delta_{c,d}\,. 
\end{equation}
The antipode and $*$-operation are 
\begin{equation}
    S\left(\begin{aligned}%
    \end{aligned}\;\;. 
\end{equation}
The Haar integral is provided by 
\begin{equation}
    \mathfrak{I} = \frac{1}{|G|^2}\sum_{k,i,s\in G}\;  
    \begin{aligned}%
    \end{aligned}\;\;. 
\end{equation}
Indeed, it is direct that $\varepsilon_R(\mathfrak{I}) = 1$. Denoting by $X$ the basis diagram in \eqref{eq:group_basis}, we have 
\begin{equation*}
    X\mathfrak{I}  = \frac{1}{|G|^2}\sum_{k,i,s}\delta_{d,kis} \delta_{c,kis} \;\begin{aligned}%
    \end{aligned} = \varepsilon_L(X)\mathfrak{I}.  
\end{equation*}
Thus $\mathfrak{I}$ is a left integral satisfying $\varepsilon_R(\mathfrak{I})=1$, implying the claim by \cite[Prop.~3.26]{BOHM1998weak}.

Here is another example from finite group. The category $\Vect$ is a $\underline{G}|\underline{G}$-bimodule category. The one-dimensional vector space $\one = \mathbb{C}$ is the unique simple object in $\Vect$. Thus the tube algebra $\Tube({_{\underline{G}}}\Vect_{\underline{G}})$ has elements 
\begin{equation}
    \begin{aligned}%
    \end{aligned}\;, \;\; g,h\in \underline{G}. 
\end{equation}
The product is clearly given by 
\begin{equation}
    \begin{aligned}%
    \end{aligned}\;\;. 
\end{equation}
Hence the tube algebra $\Tube({_{\underline{G}}}\Vect_{\underline{G}})$ is isomorphic to $\mathbb{C}[G]\otimes \mathbb{C}[G]$.

\subsubsection{Fibonacci tube algebra}

The Fibonacci category $\mathsf{Fib}$ is the simplest non-abelian anyon system. The set of simple objects consists of two elements $\Irr(\mathsf{Fib}) = \{1, \tau\}$, whose fusion rule is 
\begin{equation}
    1\otimes 1 = 1,\quad 1\otimes \tau = \tau,\quad \tau\otimes\tau = 1 \oplus \tau. 
\end{equation}
By the last fusion, one sees that % $N_\tau = %
\big)$, thus the dimension $d_\tau = \frac{1+\sqrt{5}}{2}$, and $\FPdim \mathsf{Fib} = \frac{5+\sqrt{5}}{2}$. The fusion rule also says that only the following trivalent graphs are valid
\begin{equation}
    \begin{aligned}
        %
    \end{aligned}
\end{equation}
(All arrows are omitted. Actually only the last two are trivalent, and the others are considered to be bivalent.) The loop moves are
\begin{equation}
    \begin{aligned}
        %
    \end{aligned}\;\;,
\end{equation}
and the parallel moves are
\begin{equation}
    \begin{aligned}
        %
    \end{aligned}\;\;. 
\end{equation}
The unique nontrivial F-symbols are 
% \begin{equation}
%     F_{\tau}^{\tau\tau\tau} = %
    \end{aligned}. 
\end{align}

We can write down all the basis diagrams in the Fibonacci tube algebra $\Tube({_{\mathsf{Fib}}}\mathsf{Fib}_{\mathsf{Fib}})$: 
\begin{align*}
    &\begin{aligned}%
    \end{aligned} 
\end{align*}
In these diagrams, the solid line is labeled by $\tau$, while the dotted line is labeled by $1$. For simplicity, let us denote the elements above by $a_1, \cdots, a_{16}$ from left to right. Then the unit element is $1=a_1+a_2+a_3+a_4$, and the counit takes values
% $\varepsilon(a_1) =\varepsilon(a_4) =\varepsilon(a_5) = 1,\; \varepsilon(a_8) = \varepsilon(a_9)=\varepsilon(a_{12})= \varepsilon(a_{13}) = \sqrt{d_\tau},\; \varepsilon(a_{16}) = d_\tau,\; \varepsilon(a_j) = 0$ (other $j$). 
\begin{equation}
    \varepsilon(a_j) = 
    \begin{cases}
        1 & j=1,4,5,9 \\
        \sqrt{d_\tau} & j = 8,12,13, \\
        d_\tau & j=16,\\
        0 & \text{else}. 
    \end{cases}
\end{equation}
It is direct but tedious to write down the multiplication table. We can look at some special cases. For example, 
\begin{align*}
    \begin{aligned}
    \end{aligned}\;. 
\end{align*}
As 
\begin{equation}
    \Delta(1) = (a_1+a_2)\otimes (a_1+a_3) + (a_3+a_4)\otimes (a_2+a_4), 
\end{equation}
the left and right counit maps are 
\begin{align}
    \varepsilon_L(a_j) &=  \begin{cases}
        a_1+a_3 & j=1, \\
        \varepsilon(a_j)(a_2+a_4) & j = 4,5,8,9,12,13,16, \\
        0 & \text{else}, 
    \end{cases} \\
     \varepsilon_R(a_j) &=  \begin{cases}
        a_1+a_2 & j=1,5,9, \\
        \varepsilon(a_j)(a_3+a_4) & j = 4,8,12,13,16, \\
        0 & \text{else}.  
    \end{cases}
\end{align}
The Haar integral is provided as 
\begin{equation}
    \mathfrak{I} = \frac{1}{4} \left(a_1+d_\tau^{-4}a_4+d_\tau^{-2}a_5+d_\tau^{-7/2}a_8+a_9+d_\tau^{-7/2}d_{12}+d_\tau^{-3/2}a_{13}+d_\tau^{-3}a_{16}\right).
\end{equation}

\subsubsection{Ising tube algebra}

The Ising category $\mathsf{Ising}$ has three simple objects $\Irr(\mathsf{Ising}) = \{1,\sigma,\psi\}$ with the fusion rule 
\begin{equation}
    \psi \otimes \psi = 1,\quad \psi\otimes \sigma = \sigma,\quad \sigma\otimes \sigma = 1 \oplus \psi. 
\end{equation}
It is easy to see that the dimension $d_\psi = 1, d_\sigma = \sqrt{2}$. 
Hence $\FPdim \mathsf{Ising} = 4$.
By the fusion rule, the possible cases of trivalent graph, to name a few, are 
\begin{equation}
    \begin{aligned}

    \end{aligned}
\end{equation}
The topological moves are presented in order. First, the loop moves are 
\begin{equation}
    \begin{aligned}
        %
    \end{aligned}\;\;.
\end{equation}
Second, the parallel moves are 
\begin{equation}
    \begin{aligned}
        %
    \end{aligned}. 
\end{equation}
The F-moves are much more complicated. Let us see some interesting cases. Three $\sigma$ can only fuse to $\sigma$, we have the following 
\begin{align}
    &\begin{aligned}
        %
    \end{aligned}. 
\end{align}

Now let us consider the Ising tube algebra $\Tube({_{\mathsf{Ising}}}\mathsf{Ising}_{\mathsf{Ising}})$. 
For the algebra structure, the multiplication is 
\begin{equation} 
    \begin{aligned}%
    \end{aligned}\;, 
\end{equation}
and the unit is 
\begin{equation}
    1 = \sum_{x,y\in \{1,\psi,\sigma\}}\begin{aligned}%
    \end{aligned}\;. 
\end{equation}
 For the coalgebra structure, the comultiplication is 
 \begin{align}
    \Delta\left(\begin{aligned}%
    \end{aligned} \right); 
\end{align}
besides some other cases containing $1$ in the labels, the non-trivial counit is 
\begin{align}
    \varepsilon\left(\begin{aligned}%
    \end{aligned}\right) = \sqrt{2}. 
\end{align}
For the antipode structure, using F-move we have 
\begin{align}
    S\left(\begin{aligned}%
    \end{aligned}\;. 
\end{align} 
Lastly, the Haar integral is given by 
\begin{align}
    \mathfrak{I} = & \; \frac{1}{9} \left(\begin{aligned}%
    \end{aligned} \right)\;.
\end{align}

\section{Domain wall tube algebra from multimodule category}
\label{sec:multi_tube}

\begin{figure}[t]
    \centering
    \includegraphics[width=1\linewidth]{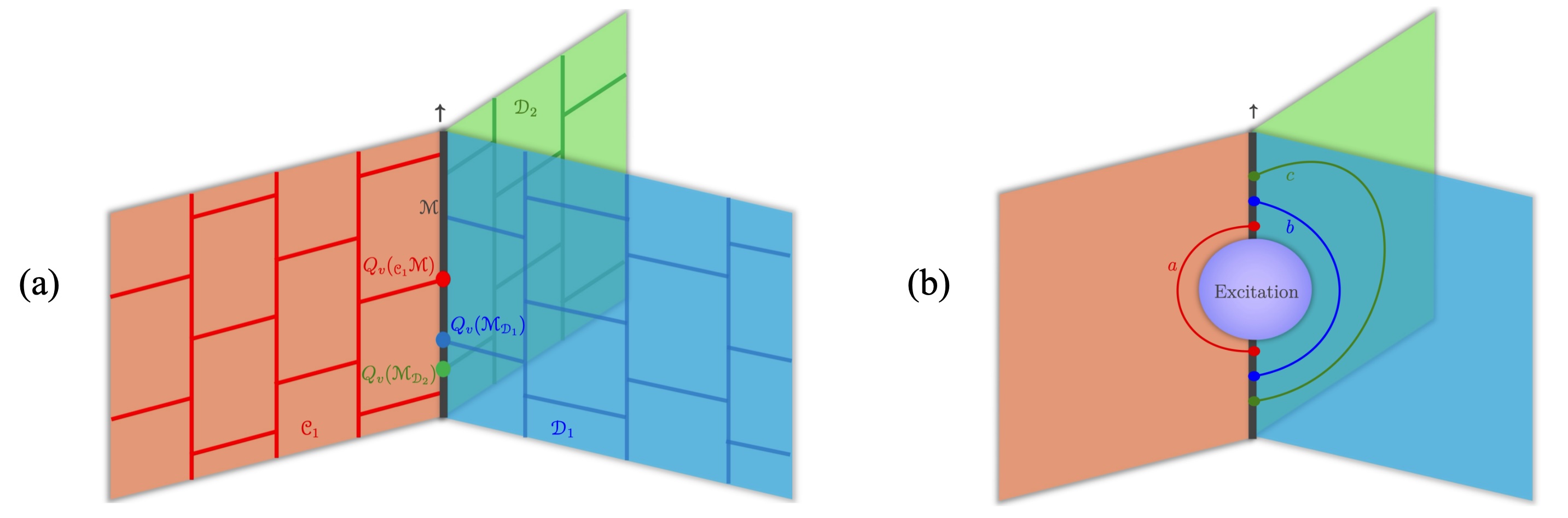}
\caption{Depiction of (a) string-net lattice model of a domain wall connecting multiple $2d$ phases; (b) domain wall excitation and the tube region surrounding the excitation.}
    \label{fig:Nwall}
\end{figure}

A domain wall between two topological phases can be naturally generalized to a domain wall among $N$ topological phases (See Figure \ref{fig:Nwall}). In this context, we introduce the concept of a multimodule category over a set of fusion categories $\{\EC_{\alpha}\}_{\alpha \in I}$. This serves as a categorification of the notion of a multimodule over multiple base rings. Multimodules were first described in Ref.~\cite[Chapter II, Section 1.14]{bourbaki2013algebra}. Since they can be reformulated as bimodules over the tensor product of base rings, they have received relatively limited attention in the literature \cite{kertesz1962multimodules,Bertozzini2022multimodule}. 
The multimodule has been previously used to understand the line defect of (2+1)D Reshetikhin-Turaev TQFT \cite{carqueville2018line}.

In the Levin-Wen string-net model which is the gapped phase corresponding to Turaev-Viro TQFT, it is natural to use the concept of multimodule categories to characterize topological domain walls and boundaries.
To our knowledge, there has been no systematic treatment of multimodule categories and its application in string-net model. Thus, in this section, we provide a brief discussion from a physical perspective and, more importantly, introduce the corresponding tube algebra.

\subsection{Multimodule category over $N$ base fusion categories}

Before discussing the construction of domain walls and the domain wall tube algebra from multimodule categories, we first introduce some basic definitions that will be used later.

\begin{definition}[Multimodule]
Let $M$ be an Abelian group, and let $\{R_{\alpha}\}_{\alpha \in I}$ and $\{S_{\beta}\}_{\beta \in J}$ be two families of rings. We say that $M$ is an $\{R_{\alpha}\}_{\alpha \in I}|\{S_{\beta}\}_{\beta \in J}$ multimodule if, for every pair $\alpha \in I$ and $\beta \in J$, $M$ is an $R_{\alpha}|S_{\beta}$ bimodule. Additionally, for any $\alpha, \alpha' \in I$, the left actions of $R_{\alpha}$ and $R_{\alpha'}$ commute, and for any $\beta, \beta' \in J$, the right actions of $S_{\beta}$ and $S_{\beta'}$ commute. When $\# I = \# J = 1$, $M$ is a bimodule. When $\# I = 1$ and $\# J = 0$, $M$ is a left module, and when $\# I = 0$ and $\# J = 1$, $M$ is a right module.\footnote{We use $\# I$ and $\# J$ to denote the cardinalities of $I$ and $J$, respectively.}
\end{definition}

The notion of a module category over a monoidal category generalizes the concept of a module over a ring \cite{etingof2016tensor}. We use $\EC^{\rm rev}$ to denote the \emph{reverse monoidal category}, with the tensor product defined as $X \otimes^{\rm rev} Y := Y \otimes X$, which categorifies the notion of the opposite ring $R^{\rm op}$.
The left action of $\EC$ on $\EM$ can be transformed into a right action of $\EC^{\rm rev}$ by defining $M \otimes^{\rm rev} X := X \otimes M$, where $X \in \EC$ and $M \in \EM$ \footnote{In this work, we will not distinguish between the module category action $X \otimes M$ for $X \in \EC$ and $M \in \EM$, and the monoidal structure $X \otimes Y$ for $X, Y \in \EC$. Additionally, we will use the same notation for the associator, F-symbols, and other related structures. However, the reader should be aware that these represent distinct mathematical objects residing in different categories.}.
The right module associator $\alpha_{M,X,Y}^{\rm rev}: (M\otimes^{\rm rev} X)\otimes^{\rm rev} Y \to M\otimes^{\rm rev} (X\otimes^{\rm rev} Y)$ is the same as the inverse of left module associator $\alpha^{-1}_{Y,X,M}: Y\otimes (X\otimes M) \to (Y\otimes X) \otimes M$.
Similarly, the right action of $\EM_{\beta}$ on $\EM$ can be transformed into a left action by $Y \otimes^{\rm rev} M := M \otimes Y$.
The notion of a multimodule can be categorified as follows:

\begin{definition}[Multimodule category]
    Consider two sets of monoidal categories, $\{\EC_{\alpha}\}_{\alpha\in I}$ and $\{\ED_{\beta}\}_{\beta\in J}$, along with an Abelian category $\EM$. We say that $\EM$ is a $\{\EC_{\alpha}\}_{\alpha\in I}|\{\ED_{\beta}\}_{\beta\in J}$ multimodule category if and only if it satisfies the following conditions:
    \begin{itemize}
        \item For each pair $\EC_{\alpha}$ and $\ED_{\beta}$, $\EM$ is a $\EC_{\alpha}|\ED_{\beta}$ bimodule category.
        \item For any pair $\EC_{\alpha}$ and $\EC_{\alpha'}$, $\EM$ is a $\EC_{\alpha}|\EC_{\alpha'}^{\rm rev}$ bimodule category.
        \item For any pair $\ED_{\beta}$ and $\ED_{\beta'}$, $\EM$ is a $\ED_{\beta}^{\rm rev}|\ED_{\beta'}$ bimodule category.
    \end{itemize}
   It is clear that left, right and bimodule categories are all special cases of multimodule categories.
\end{definition}

Hereinafter, we will assume that all $\{\EC_{\alpha}\}_{\alpha \in I}$ and $\{\ED_{\beta}\}_{\beta\in J}$ are fusion categories and that $\EM$ is a finitely semisimple Abelian category. 
For two fusion category $\EC$ and $\ED$, the Deligne tensor product is defined via the universal property \cite{deligne2007categories,etingof2016tensor,Lopez2013tensorproduct}
\begin{equation}
    \begin{tikzcd}
\EC \times \ED \arrow[rd, "F"'] \arrow[r, "\boxtimes"] & \EC \boxtimes \ED \arrow[d, dashed, "\exists ! \tilde{F}"] \\
& \EuScript{E}
\end{tikzcd}
\end{equation}
where $\boxtimes$ represents the Deligne tensor product, $F$ is a bilinear bifunctor which is right exact in both variables.
The monoidal structure is given by $(X_1\boxtimes Y_1) \otimes (X_2\boxtimes Y_2):= (X_1\otimes X_2)\boxtimes (Y_1\otimes Y_2)$. 

Notice that we have designed the notion of a multimodule category in a way that naturally aligns with our model of domain wall structures in topological phases. There are also equivalent ways to define them purely in terms of left or right actions. This can be seen from the following result:

\begin{proposition}\label{prop:foldingM}
A $\{\EC_{\alpha}\}_{\alpha \in I}|\{\ED_{\beta}\}_{\beta \in J}$ multimodule category structure over an Abelian category $\EM$ is equivalent to a $(\boxtimes_{\alpha \in I} \EC_{\alpha}) | (\boxtimes_{\beta\in J} \ED_{\beta})$ bimodule structure over $\EM$. Furthermore, it is also equivalent to the left $(\boxtimes_{\alpha \in I} \EC_{\alpha}) \boxtimes (\boxtimes_{\beta\in J} \ED_{\beta}^{\rm rev})$  module category structure and right $(\boxtimes_{\alpha \in I} \EC_{\alpha}^{\rm rev}) \boxtimes (\boxtimes_{\beta\in J} \ED_{\beta})$ module category structure over $\EM$. 
\end{proposition}

\begin{proof}
This follows directly from the definition~\cite{etingof2010fusion}. We emphasize that the action of the Deligne tensor product $\EC_{\alpha} \boxtimes \EC_{\alpha'}$ on $\EM$ characterizes the commutativity of the actions of $\EC_{\alpha}$ and $\EC_{\alpha'}$ on $\EM$.
\end{proof}

\begin{corollary}
The structure of a $\{\EC_{\alpha}\}_{\alpha \in I}|\{\ED_{\beta}\}_{\beta \in J}$ multimodule category over an Abelian category $\EM$ is in one-to-one correspondence with the monoidal functor
\begin{equation}
F: \left( \boxtimes_{\alpha \in I} \EC_{\alpha} \right) \boxtimes \left( \boxtimes_{\beta \in J} \ED_{\beta}^{\rm rev} \right) \to \mathsf{End}(\EM),
\end{equation}
or equivalently, the monoidal functor
\begin{equation}
G: \left( \boxtimes_{\alpha \in I} \EC_{\alpha}^{\rm rev} \right) \boxtimes \left( \boxtimes_{\beta \in J} \ED_{\beta} \right) \to \mathsf{End}(\EM),
\end{equation}
where $\mathsf{End}(\EM)$ is the category of endofunctors (with $\EM$ treated as an Abelian category), which is a monoidal category with tensor product given by composition of functors.
\end{corollary}

\begin{proof}
Using Proposition~\ref{prop:foldingM} above and Proposition 7.1.3 in Ref.~\cite{etingof2016tensor}, we obtain the desired result.
\end{proof}

In the physics literature, the above result is often referred to as the \emph{folding trick}. Using this technique, we can transform a domain wall into a boundary. 
By applying the folding trick, we can define the $\{\EC_{\alpha}\}_{\alpha \in I}|\{\ED_{\beta}\}_{\beta \in J}$ multimodule category functor as a left $(\boxtimes_{\alpha \in I} \EC_{\alpha}) \boxtimes (\boxtimes_{\beta \in J} \ED_{\beta}^{\rm rev})$ module category functor. This consists of a functor $F$ and a natural isomorphism $\gamma_{\bullet,\bullet}: F(\bullet \otimes \bullet) \to \bullet \otimes F(\bullet)$, which satisfies the pentagon relation. This relation is crucial for understanding domain wall excitations, and will be illustrated in detail later.
The category of all right exact multimodule functors will be denoted as $\Fun_{\{\EC_{\alpha}\}_{\alpha \in I}|\{\ED_{\beta}\}_{\beta \in J}} (\EM,\EN)$. 
Whenever necessary to avoid ambiguity, we will use $\Fun^L$ and $\Fun^R$ to denote the categories of all right exact left and right module functors, respectively. In cases where there is no ambiguity, we will simply write $\Fun_{\EC}(\EM,\EN)$ to denote the category of right exact left $\EC$-module functors from $\EM$ to $\EN$, as is standard in the literature.
The category of multimodule endofunctors will be denoted as $\mathsf{End}_{\{\EC_{\alpha}\}_{\alpha \in I}|\{\ED_{\beta}\}_{\beta \in J}} (\EM)$, to stress the left or right module structures, we will also use $\mathsf{End}^L,\mathsf{End}^R$.

% \begin{corollary}
% The following results can be proved directly via using folding trick:
% \begin{enumerate}
%     \item  We say a multimodule category $\EM$ is exact if for any projective $X\in \EC_{\alpha}$ or $Y\in \ED_{\beta}$, and any $M\in \EM$, objects $X\otimes M$ and $M\otimes Y$ are projective.
%     \item 
%     \item
%     \item
% \end{enumerate}
% \end{corollary}

% \begin{proof}
%     ddd
% \end{proof}

When constructing lattice models of domain walls, the unitary structure of the module category also plays a crucial role. Not every fusion category admits a unitary structure~\cite[Exercise 9.4.6]{etingof2016tensor}; however, when it exists, the unitary structure is unique~\cite{Reutter2023uniqueness}. For multimodule categories, this uniqueness also holds:

\begin{proposition}
If a multimodule category is unitarizable, meaning that it admits a compatible positive $\dagger$-structure, then its unitary structure is unique.  
More precisely, let $\EM$ and $\EN$ be unitary multimodule categories over unitary fusion categories $\{\EC_{\alpha}\}_{\alpha\in I}$ and $\{\ED_{\beta}\}_{\beta\in J}$. Any multimodule equivalence  
\begin{equation}
    F: {_{\{\EC_{\alpha}\}_{\alpha\in I}}} \EM_{\{\ED_{\beta}\}_{\beta\in J}} 
    \to {_{\{\EC_{\alpha}\}_{\alpha\in I}}} \EN_{\{\ED_{\beta}\}_{\beta\in J}}
\end{equation}
is naturally isomorphic, as a module functor, to a unitary module functor.
\end{proposition}

\begin{proof}
This can be proved by first applying the folding trick and then using Theorem~8 in Ref.~\cite{Reutter2023uniqueness}.  
The only remaining step is to show that the Deligne tensor product of unitary fusion categories remains unitary, which follows directly from the definition of the Deligne tensor product and $\Hom_{\EC\boxtimes \ED} (X_1\boxtimes Y_1, X_2\boxtimes Y_2)\cong \Hom_{\EC}(X_1,X_2)\otimes \Hom_{\ED}(Y_1,Y_2)$.
\end{proof}

Hereinafter, we assume that all module categories under consideration are unitarizable, and all functors between them are right exact.

\subsection{Domain wall string-net lattice model from multimodule categories}

We now turn to the construction of a domain wall model that connects multiple gapped phases of Turaev-Viro TQFT. This model follows a similar approach to the Kitaev-Kong construction for the bimodule category case~\cite{Kitaev2012boundary}, but requires an additional piece of data: the orientation of the domain wall and the compatibility of each $2d$ manifold's orientation with the one induced by the domain wall.

In the case of a bimodule category domain wall, where two $2d$ phases are placed on a plane, the wall naturally inherits an orientation. A key, though often implicit, assumption is that each $2d$ half-plane supporting a gapped phase has a well-defined orientation. One of these orientations aligns with that of the entire plane, following the right-hand rule, while the other is opposite. This perspective was emphasized in our previous work~\cite{jia2023boundary}.

For the multimodule category case, we first assign an orientation to the $1d$ domain wall connecting $N$ gapped phases. Each $2d$ manifold $\mathbb{M}_{\alpha}$ supporting a phase is also assigned an orientation $\nu_{\alpha}$\footnote{The orientation can be viewed as a function $\nu: I \to \mathbb{Z}_2$, where $I$ indexes all $2d$ manifolds supporting gapped phases, and $\nu_{\alpha} = 1$ indicates that the orientations are consistent.}, which may or may not align with that of the domain wall, denoted $\nu^{\rm DW}_{\alpha}$, according to the right-hand rule. If a $2d$ manifold has an orientation consistent with the domain wall ($\nu_{\alpha} = \nu^{\rm DW}_{\alpha}$), we assign a left $\EC_{\alpha}$ action on $\EM$; otherwise, we assign a right $\ED_{\beta}$ action on $\EM$.

To construct a lattice model from a multimodule category, assign a Levin-Wen string-net lattice for each $2d$ phase with input data $\EC_{\alpha}$ and $\ED_{\beta}$. The only modification needed is in the region near the domain wall, as shown in Figure~\ref{fig:Nwall}. The wall Hilbert space is spanned by all domain wall configurations labeled by simple objects in $\EM$, simple objects in $\EC_{\alpha}$ and $\ED_{\beta}$, with vertex labels coming from the multimodule Hom-space.
To illustrate this concept, let take the following configuration as an example:
\begin{equation}
\begin{aligned}
a_1\otimes m_1 \to m_2,\\
m_2\otimes b_1 \to m_3,\\
m_3\otimes b_2 \to m_4,\\
a_2\otimes m_4 \to m_5,\\
a_i \in \EC_i, b_j\in \ED_j, m_k\in \EM.
\end{aligned} \quad 
\begin{aligned}
    \scalebox{0.7}{ % Scale factor of 0.5 (adjust as needed)
    \begin{tikzpicture} 
    % left two face
    \fill[blue!50,opacity=0.5] (-3.7,-1.5) -- (0,0) -- (0,3) -- (-3.7,1.5) -- cycle;
    \fill[red!50,opacity=0.5] (-3.5,1) -- (0,0) -- (0,3) -- (-3.5,4) -- cycle;
    % right two face
    \fill[green!50,opacity=0.5] (3.7,-1.5) -- (0,0) -- (0,3) -- (3.7,1.5) -- cycle;
    \fill[yellow!50,opacity=0.5] (3.5,1) -- (0,0) -- (0,3) -- (3.5,4) -- cycle;
     % line in domain wall
    \draw[->-=0.6,thick] (0,0) -- (0,0.5);
    \draw[->-=0.6,thick] (0,0.5) -- (0,1);
    \draw[->-=0.6,thick] (0,1) -- (0,1.5);
    \draw[->-=0.6,thick] (0,1.5) -- (0,2);
    \draw[->-=0.6,thick] (0,2) -- (0,3);
    \draw[thick] (0,0) -- (0,3);
    \draw[->-=0.6,blue,thick] (-2,-0.2) -- (0,0.5);
    \draw[->-=0.6,green,thick] (2,0.3) -- (0,1);
    \draw[->-=0.6,brown,thick] (1.9,2) -- (0,1.5);
    \draw[->-=0.6,red,thick] (-1.9,2.7) -- (0,2);
    \node[ line width=0.6pt, dashed, draw opacity=0.5] (a) at (-3,-.6){$\nu_1=1$};
    \node[ line width=0.6pt, dashed, draw opacity=0.5] (a) at (3,-.6){$\nu_2=0$};
    \node[ line width=0.6pt, dashed, draw opacity=0.5] (a) at (2.7,2.6){$\nu_3=0$};
    \node[ line width=0.6pt, dashed, draw opacity=0.5] (a) at (-2.7,2.6){$\nu_4=1$};
    \node[ line width=0.6pt, dashed, draw opacity=0.5] (a) at (-0.8,-0.15){$a_1$};
    \node[ line width=0.6pt, dashed, draw opacity=0.5] (a) at (0.5,.2){$m_1$};
    \node[ line width=0.6pt, dashed, draw opacity=0.5] (a) at (-0.45,.65){$m_2$};
    \node[ line width=0.6pt, dashed, draw opacity=0.5] (a) at (1.2,.35){$b_1$};
    \node[ line width=0.6pt, dashed, draw opacity=0.5] (a) at (0.5,1.2){$m_3$};
    \node[ line width=0.6pt, dashed, draw opacity=0.5] (a) at (1.5,2.3){$b_2$};
    \node[ line width=0.6pt, dashed, draw opacity=0.5] (a) at (-0.45,1.85){$m_4$};
     \node[ line width=0.6pt, dashed, draw opacity=0.5] (a) at (0.5,2.7){$m_5$};
    \node[ line width=0.6pt, dashed, draw opacity=0.5] (a) at (-1.2,2.85){$a_2$};
    \end{tikzpicture}
    }
\end{aligned}
\end{equation}
Here, there are four $2d$ gapped phases labeled from 1 to 4. In this setup, $\mathbb{M}_1$ and $\mathbb{M}_4$ have orientations that are consistent with the orientation induced by the domain wall (upwards), while $\mathbb{M}_2$ and $\mathbb{M}_3$ have orientations that are inconsistent with the domain wall orientation.
All domain wall edges are labeled by simple objects in multimodule category $\EM$.

Each gapped phase is associated with its own input fusion category $\EC_{\alpha}$. When $\nu_{\alpha} = 1$, the corresponding left module category F-symbol is denoted as $[{_{\EC_{\alpha} \triangleright \EM}} F_{m}^{bcn}]_{e,\rho, \tau}^{s,\zeta,\gamma}$. When $\nu_{\beta} = 0$, we have the right module category F-symbol $[{_{\EM \triangleleft \ED_{\beta}}} F_{m}^{nbc}]_{e,\rho, \tau}^{s,\zeta,\gamma}$. The middle F-symbol for the $\EC_{\alpha} | \ED_{\beta}$ bimodule structure is denoted as $[{_{\EC_{\alpha} \triangleright \EM \triangleleft \ED_{\beta}}} F_{m}^{bcn}]_{e,\rho, \tau}^{s,\zeta,\gamma}$.
The multimodule category $\EM$ with $N$ base fusion categories will have $N + C(N,2)$ sets of F-symbols, where $C(N,2)$ represents the binomial coefficient. These F-symbols are compatible with each other through the pentagon relations. Hereinafter, to avoid clustering of equations, we will omit the left subscripts that indicate the F-symbol arises from different module actions. This should not cause significant ambiguity, as the information is clear from the labels of the objects appearing in the F-symbol.

For $\EuScript{C}_{\alpha}$ action over $\EuScript{M}$, we label the corresponding fusion vertex as morphisms in $_{\EuScript{C}_{\alpha}}\EuScript{M}$. The vertex stabilizer is defined as the projective operator
\begin{equation}
    Q_{v}({_{\EuScript{C}_{\alpha}}\EuScript{M}}) = \operatorname{Proj}_{\bigoplus_{a\in \EuScript{C}_{\alpha},\, m,n\in \EuScript{M}} \Hom(a\otimes m,n)}.
\end{equation}
The face operator ${B}_f^k({_{\EuScript{C}_{\alpha}}\EuScript{M}})$ can also be defined by inserting $k$ loops into the domain wall face $f$ and
\begin{equation}
    {B}_f({_{\EuScript{C}_{\alpha}}\EuScript{M}}) = \sum_{k\in \operatorname{Irr}(\EuScript{C}_{\alpha})} w_k {B}_f^k({_{\EuScript{C}_{\alpha}}\EuScript{M}}).
\end{equation}
The local stabilizers of $\EuScript{D}_{\beta}$ action can be defined similarly. 
We define the domain wall Hamiltonian as
\begin{equation}
\begin{aligned}
    H_{\rm wall}[{_{\{\EuScript{C}_{\alpha}\}_{\alpha\in I}}}\EM_{\{\EuScript{D}_{\beta}\}_{\beta\in J}}] 
    =& \sum_{\alpha \in I} \left[ -\sum_{v} Q_{v}({_{\EuScript{C}_{\alpha}}\EuScript{M}}) - \sum_f {B}_f({_{\EuScript{C}_{\alpha}}\EuScript{M}}) \right] \\
    &+ \sum_{\beta \in J} \left[ -\sum_{v} Q_{v}(\EuScript{M}_{\EuScript{D}_{\beta}}) - \sum_f {B}_f(\EuScript{M}_{\EuScript{D}_{\beta}}) \right].
\end{aligned}
\end{equation}
The multimodule category Levin-Wen Hamiltonian can then be expressed as
\begin{equation}
    H= \sum_{\alpha\in I }H_{\rm bulk}[\mathbb{M}_{\alpha},\EuScript{C}_{\alpha}]
    +\sum_{\beta\in J }H_{\rm bulk}[\mathbb{M}_{\beta},\EuScript{D}_{\beta}] 
    +  H_{\rm wall}[{_{\{\EuScript{C}_{\alpha}\}_{\alpha\in I}}}\EM_{\{\EuScript{D}_{\beta}\}_{\beta\in J}}],
\end{equation}
which is a local commutative projector (LCP) Hamiltonian, thus is an exactly solvable gapped model.

\begin{proposition}
    The domain wall topological excitation for the  $\{\EC_{\alpha}\}_{\alpha \in I}|\{\ED_{\beta}\}_{\beta \in J}$ multimodule category $\EM$ is characterized by the $\{\EC_{\alpha}\}_{\alpha \in I}|\{\ED_{\beta}\}_{\beta \in J}$ multimodule functors:
    \begin{equation}
        \mathsf{Phase}^{\rm wall}_{\EM}= \Fun_{\{\EC_{\alpha}\}_{\alpha \in I}|\{\ED_{\beta}\}_{\beta \in J}}(\EM,\EM).
    \end{equation}
    For two multimodule category $\EM$ and $\EN$ with the same base fusion categories, the topological defect between two domain walls are characterized by the functors between two multimodule categories:
   \begin{equation}
        \mathsf{Defect}^{\rm wall}_{\EM,\EN}= \Fun_{\{\EC_{\alpha}\}_{\alpha \in I}|\{\ED_{\beta}\}_{\beta \in J}}(\EM,\EN).
    \end{equation}
    Via folding trick (Proposition~\ref{prop:foldingM}), the domain wall $\EM$ can be transformed into a left $(\boxtimes_{\alpha \in I} \EC_{\alpha}) \boxtimes (\boxtimes_{\beta\in J} \ED_{\beta}^{\rm rev})$  module category boundary or a right $(\boxtimes_{\alpha \in I} \EC_{\alpha}^{\rm rev}) \boxtimes (\boxtimes_{\beta\in J} \ED_{\beta})$ module category boundary, the domain wall phase is equivalent to the boundary phase:
    \begin{equation}
    \begin{aligned}
        \mathsf{Phase}^{\rm wall}_{\EM} &\simeq  \mathsf{Phase}^{\rm leftbd}_{\EM} = \Fun^L_{(\boxtimes_{\alpha \in I} \EC_{\alpha}) \boxtimes (\boxtimes_{\beta\in J} \ED_{\beta}^{\rm rev})}(\EM,\EM)\\
        & \simeq   \mathsf{Phase}^{\rm rightbd}_{\EM} =\Fun^R_{(\boxtimes_{\alpha \in I} \EC_{\alpha}^{\rm rev}) \boxtimes (\boxtimes_{\beta\in J} \ED_{\beta})}(\EM,\EM).
    \end{aligned}
    \end{equation}
   Similarly, via the folding trick  (Proposition~\ref{prop:foldingM}), domain wall defects can also be transformed into boundary defects
       \begin{equation}
    \begin{aligned}
        \mathsf{Defect}^{\rm wall}_{\EM,\EN} &\simeq  \mathsf{Defect}^{\rm leftbd}_{\EM,\EN} = \Fun^L_{(\boxtimes_{\alpha \in I} \EC_{\alpha}) \boxtimes (\boxtimes_{\beta\in J} \ED_{\beta}^{\rm rev})}(\EM,\EN)\\
        & \simeq   \mathsf{Defect}^{\rm rightbd}_{\EM,\EN} =\Fun^R_{(\boxtimes_{\alpha \in I} \EC_{\alpha}^{\rm rev}) \boxtimes (\boxtimes_{\beta\in J} \ED_{\beta})}(\EM,\EN).
    \end{aligned}
    \end{equation}
    Notice that topological excitations are special case of topological defect.
\end{proposition}

\begin{proof}
    This follows immediately by applying the folding trick and then using the results in Refs.~\cite{Kitaev2012boundary,kong2012universal,jia2023boundary,Jia2023weak,jia2024weakTube}. 
\end{proof}

\begin{remark}
Multimodule domain walls can also be realized via a lattice Hamiltonian construction for the weak Hopf quantum double model \cite{Jia2023weak}. In this framework, each $2d$ bulk phase is assigned a weak Hopf gauge symmetry specified by a weak Hopf algebra $W_i$. A domain wall separating such bulk phases is described by a comultimodule algebra $\mathfrak{A}$ over the family $\{W_i\}_{i \in I}$. 
Excitations on the domain wall are then characterized by $\mathfrak{A}|\mathfrak{A}$ bimodules that are compatible with the coactions of $\{W_i\}_{i \in I}$ \cite{Jia2023weak}. The category of such bimodules is denoted by ${_{\mathfrak{A}}}\mathsf{Mod}_{\mathfrak{A}}^{\{W_i\}_{i \in I}}$. Similarly, domain wall defects between two domain walls labeled by $\mathfrak{A}$ and $\mathfrak{B}$ are described by the category ${_{\mathfrak{B}}}\mathsf{Mod}_{\mathfrak{A}}^{\{W_i\}_{i \in I}}$.
For a comparison between the weak Hopf algebra formulation and the fusion category description, see Table~\ref{tab:CorrspondenceQD_SN}.
\end{remark}

\begin{table}[t]
\centering \small 
\begin{tabular} {|l|c|c|} 
\hline
   &Weak Hopf quantum double model & (Multi)fusion string-net model   \\ \hline
 Bulk gauge symmetries  & Weak Hopf algebra $\{W_i\}_{i\in I}$ &  $\{\EC_i=\mathsf{Rep}(W_i) \}_{i\in I}$  \\ \hline
 Bulk phase  & 
 $\EP=\mathsf{Rep}(D(W_i))$ & $\mathsf{Fun}_{\EC_i |\EC_i}(\EC_i,\EC_i) $  \\ \hline
$\begin{aligned}
    &\text{Domain wall} \\
    &\text{gauge symmetry}
\end{aligned}$ & $\begin{aligned}
  &  \{W_i\}_{i\in I}-\text{comultimodule} \\
  &\text{algebra} \,\, \mathfrak{A}
\end{aligned} $ & $\begin{aligned}
   & \{\EC_i\}_{i\in I}-\text{multimodule category} \\
   & \EM={_{\mathfrak{A}}}\mathsf{Mod}
\end{aligned}$ \\\hline
 Domain wall phase & $\EB\simeq {_{\mathfrak{A}}}\mathsf{Mod}_{\mathfrak{A}}^{\{W_i\}_{i\in I}}$ & $\mathsf{Fun}_{\{\EC_i\}_{i\in I}}({_{\mathfrak{A}}}\EM,{_{\mathfrak{A}}}\EM)$ \\\hline
Domain wall defect & ${_{\mathfrak{B}}}\mathsf{Mod}_{\mathfrak{A}}^{\{W_i\}_{i\in I}}$ & $\mathsf{Fun}_{\{\EC_i\}_{i\in I}}({_{\mathfrak{A}}}\EM,{_{\mathfrak{B}}}\EM)$ \\
\hline
\end{tabular}
\caption{The correspondence between descriptions of non-chiral topologically ordered phases in the framework of the quantum double model and the multifusion string-net model can be elucidated through a dictionary.
\label{tab:CorrspondenceQD_SN}}
\end{table}

\subsection{Domain wall weak tube algebra from multimodule categories}

The key motivation for introducing the multimodule domain wall is that its tube algebra encodes richer information. In a sense, it generalizes Drinfeld's quantum double construction to the setting of multiple weak Hopf algebras, forming what can be referred to as a \emph{tube quantum $N$-tuple algebra}.

The domain wall tube algebra can be viewed as a quantum double of the left and right boundary tube algebras. For a multimodule category associated with three base fusion categories, the corresponding domain wall can be interpreted as a \emph{quantum triple algebra} of three boundary tube algebras (see Figure~\ref{fig:Nwall} (b)). More generally, for a multimodule category with $N$ base fusion categories, the associated domain wall tube algebra can be regarded as a quantum $N$-tuple algebra.

\begin{definition}
    Suppose $\#I=k$, and $\#J=l$. For a multimodule category domain wall determined by a  $\{\EC_{\alpha}\}_{\alpha \in I}|\{\ED_{\beta}\}_{\beta \in J}$ multimodule category $\EM$ ,  the domain wall tube algebra $\mathbf{Tube}({_{\{\EC_{\alpha}\}_{\alpha\in I}}} \EM_{\{\ED_{\beta}\}_{\beta\in J}} )$ is the vector space spanned by the following basis:
\begin{equation} \label{eq:mul_tube_basis}
 \left\{  
   \begin{aligned}

    \end{aligned}
   \;:\; 
   \begin{aligned}
     & a_i\in \Irr(\EC_i),b_j\in \Irr(\ED_j), \\
    & u_i,z_i,v_j,w_j \in \Irr(\EM), \\
     & \text{wall vertex label} \in \Hom_{\EM}.
   \end{aligned}
    \right\}.
\end{equation}
In the upper half, the wall labels are $u_1,\cdots,u_k,v_1,\cdots,v_l,v_{l+1}$ from bottom to top, while in the lower half, the wall labels are $z_1,\cdots,z_k,w_1,\cdots,w_l,w_{l+1}$ from top to bottom. The wall vertex labels are from the Hom-spaces listed as follows: 
\begin{gather*}
    \gamma_i\in \Hom_\EM(a_i\otimes u_i,u_{i+1}),\;\;\nu_i\in \Hom_\EM(z_{i+1},a_i\otimes z_i), \;\;1\leq i \leq k-1, \\
    \zeta_j\in \Hom_\EM(v_j\otimes b_j,v_{j+1}),\;\; \mu_j\in \Hom_\EM(w_{j+1},w_j\otimes b_j),\;\; 1\leq j\leq l, \\
    \gamma_k\in \Hom_\EM(a_k\otimes u_k,v_1),\;\; \nu_k\in \Hom_\EM(w_1,a_k\otimes z_k). 
\end{gather*}
One implicit condition is that these Hom-spaces are nonzero. 
\end{definition}

The following is the main result in this section. Note that the weak Hopf algebra structure is constructed for general case, but the $*$-structure is constructed based on the assumption that $d_m^L=d_m^R=d_m$ for all $m\in \EM$.

\begin{theorem}
    The multimodule category domain wall tube algebra $\mathbf{Tube}({_{\{\EC_{\alpha}\}_{\alpha\in I}}} \EM_{\{\ED_{\beta}\}_{\beta\in J}} )$ is a $C^*$ weak Hopf algebra by the following structure morphisms. 
    \begin{enumerate}
        \item The algebra structure: The unit element is defined by 
    \begin{equation} \label{eq:Ntube_unit}
        1=\sum_{x,z\in\Irr(\EM)} \;\begin{aligned}

        \end{aligned}\;\;, 
    \end{equation}
    where dotted lines are labeled by multiple unit objects in respective base fusion categories. 
    
    The multiplication map $\mu$ is defined as 
    \begin{equation} \label{eq:Ntube_prod_1}
    \begin{aligned}
     &     \mu\left( \begin{aligned}
        %
        \end{aligned}\;.
    \end{aligned}
    \end{equation}

        \item The coalgebra structure: The counit is given by  
            \begin{align} \label{eq:tube_counit}
                 \varepsilon\left(\begin{aligned}
        %
    \end{aligned} \\
    & =\delta_{v_{l+1},w_{l+1}}\prod_{i=1}^k \delta_{u_i,z_i}\delta_{\gamma_i,\nu_i}\prod_{j=1}^l\delta_{v_j,w_j}\delta_{\zeta_j,\mu_j} \nonumber \\
    & \quad \times \sqrt{\frac{d_{a_1}d_{u_1}^{L,1}}{d_{u_2}^{L,1}}}\sqrt{\frac{d_{a_2}d_{u_2}^{L,2}}{d_{u_3}^{L,2}}} \cdots \sqrt{\frac{d_{a_k}d_{u_k}^{L,k}}{d_{v_1}^{L,k}}} \nonumber \\
    & \quad \times \sqrt{\frac{d_{v_1}^{R,1}d_{b_1}}{d_{v_2}^{R,1}}} \sqrt{\frac{d_{v_2}^{R,2}d_{b_2}}{d_{v_3}^{R,2}}}\cdots \sqrt{\frac{d_{v_l}^{R,l}d_{b_l}}{d_{v_{l+1}}^{R,l}}} \nonumber \\
    & =\delta_{v_{l+1},w_{l+1}}\prod_{i=1}^k \delta_{u_i,z_i}\delta_{\gamma_i,\nu_i}\prod_{j=1}^l\delta_{v_j,w_j}\delta_{\zeta_j,\mu_j} \nonumber \\
    & \quad \times \sqrt{\frac{\prod_{i=1}^kd_{a_i}d_{u_i}^{L,i}}{\prod_{i=2}^kd_{u_i}^{L,i-1}d_{v_1}^{L,k}}} \sqrt{\frac{\prod_{j=1}^ld_{v_j}^{R,j}d_{b_j}}{\prod_{j=2}^{l+1}d_{v_j}^{R,j-1}}}\;. \nonumber 
    \end{align}
    Here, we use $d_x^{L,i}$ and $d_x^{R,j}$ to denote the quantum dimensions of $x$ in $\EM$ with respect to the fusion categories $\EC_i$ and $\ED_j$ respectively. 
    
    The comultiplication map $\Delta$ is defined as 
            \begin{equation} \label{eq:tube_coprod}
        \begin{aligned}
            \Delta\left(\begin{aligned}

    \end{aligned}\;,
        \end{aligned}
        \end{equation}
    where in the lower half of the first component the wall labels are $s_1,\cdots,s_k,t_1,\cdots,t_l,t_{l+1}$ from top to bottom, and in the upper half of the second component the wall labels are $s_1,\cdots,s_k,t_1,\cdots,t_l,t_{l+1}$ from bottom to top. 

    \item The antipode is given by 
    \begin{equation}
        \begin{aligned}
            S\left(\begin{aligned}
        %
    \end{aligned}
    \end{aligned}\;.
    \end{equation}
    The wall vertex labels on the right-hand side should be understood as the corresponding morphisms in isomorphic Hom-spaces. For example, the original label $\gamma_1$ is a morphism $a_1\otimes u_1 \to u_2$, then the label $\gamma_1$ on the right-hand side is the corresponding morphism $u_1\to \bar{a}_1\otimes u_2$ under the isomorphism $\Hom_\EM(a_1\otimes u_1,u_2)\simeq \Hom_\EM(u_1,\bar{a}_1\otimes u_2)$. 
    
    \item Assume that $d_x^L = d_x^R = d_x$ for all $x\in \Irr(\EM)$. The $*$-operation is defined as  
    \begin{equation}
        \begin{aligned}
            \left(\begin{aligned}

    \end{aligned}
    \end{aligned}\;.
    \end{equation}      
    \end{enumerate}
\end{theorem}

\begin{proof}
The proof is almost identical with that for bimodule case. The algebra structure and coalgebra structure are directly proved. Denote by $X$, $X'$, $X''$ the basis diagram in \eqref{eq:mul_tube_basis} label by $\{a_1,\cdots\}$, $\{a_1',\cdots\}$, and $\{a_1'',\cdots\}$ respectively. 

    1. Weak bialgebra structure. 
    Note that the comultiplication $\Delta(X)$ in Eq.~\eqref{eq:tube_coprod} is obtained by inserting a wall labeled by $\sum_{s_1} s_1$ and applying  parallel moves. 
    With this in mind, it is easy to see that $\Delta(XX') = \Delta(X)\Delta(X')$. 
    Also note that to compute the counit $\varepsilon(X)$, we actually connect the edges in $X$ labeled by $u_1,z_1$ and by $v_{l+1},w_{l+1}$, and then apply loop moves (there is a normalization factor $1/d_{v_{l+1}}^{R,l}$ to ensure the coalgebra conditions). 
    The same principle applies to $\varepsilon(XX'\cdots)$ by properties of F-symbols similar to Proposition~\ref{prop:F-symbols_property}. 
    More precisely, to compute $\varepsilon(XX'X'')$, we first compute the product, which introduces the factor $\delta_{u_1,v_{l+1}'}\delta_{z_1,w_{l+1}'}\delta_{u_1',v_{l+1}''}\delta_{z_1',w_{l+1}''}$. Then, computing the counit of the product diagram amounts to connecting the edges labeled by $u_1''$ and $z_1''$, and those labeled by $v_{l+1}$ and $w_{l+1}$, which introduces the factor $\delta_{u_1'',z_1''}\delta_{v_{l+1},w_{l+1}}/d_{v_{l+1}}^R$. Finally, we evaluate the resulting diagram using the loop move. Hence, one has 
    \begin{align*}
        \varepsilon(&XX'X'') \\ 
        &= \underbrace{\delta_{u_1,v_{l+1}'}\delta_{z_1,w_{l+1}'}\delta_{u_1',v_{l+1}''}\delta_{z_1',w_{l+1}''}}_{\text{factor~from~}XX'X''} \underbrace{\delta_{u_1'',z_1''}\delta_{v_{l+1},w_{l+1}}\frac{1}{d_{v_{l+1}}^R}}_{\text{factor~from~counit}}\\ 
        & \quad \times \prod_{i=1}^k \delta_{u_i'',z_i''}\delta_{\gamma_i'',\nu_i''}\delta_{u_i',z_i'}\delta_{\gamma_i',\nu_i'}\delta_{u_i,z_i}\delta_{\gamma_i,\nu_i}\\
        & \quad  \times \prod_{j=1}^l\delta_{v_j'',w_j''}\delta_{\zeta_j'',\mu_j''}\delta_{v_j',w_j'}\delta_{\zeta_j',\mu_j'}\delta_{v_j,w_j}\delta_{\zeta_j,\mu_j} \\
        & \quad \times \sqrt{\frac{\prod_{i=1}^kd_{a_i''}d_{u_i''}^{L,i}}{\prod_{i=2}^kd_{u_i''}^{L,i-1}d_{v_1''}^{L,k}}} \sqrt{\frac{\prod_{j=1}^ld_{v_j''}^{R,j}d_{b_j''}}{\prod_{j=2}^{l+1}d_{v_j''}^{R,j-1}}} \\
        & \quad \times \sqrt{\frac{\prod_{i=1}^kd_{a_i'}d_{u_i'}^{L,i}}{\prod_{i=2}^kd_{u_i'}^{L,i-1}d_{v_1'}^{L,k}}} \sqrt{\frac{\prod_{j=1}^ld_{v_j'}^{R,j}d_{b_j'}}{\prod_{j=2}^{l+1}d_{v_j'}^{R,j-1}}} \\
        & \quad \times \sqrt{\frac{\prod_{i=1}^kd_{a_i}d_{u_i}^{L,i}}{\prod_{i=2}^kd_{u_i}^{L,i-1}d_{v_1}^{L,k}}} \sqrt{\frac{\prod_{j=1}^ld_{v_j}^{R,j}d_{b_j}}{\prod_{j=2}^{l+1}d_{v_j}^{R,j-1}}} d_{v_{l+1}^R} \;, \\
        \sum_{(X')}&\varepsilon(XX'^{(1)})\varepsilon(X'^{(2)}X'') \\  
        & = \sum_{i=1}^k\sum_{j=1}^{l+1}\sum_{\sigma_i}\sum_{\rho_j} \sqrt{\frac{\prod_{i=1}^kd_{s_i}^{L,i-1}d_{t_1}^{L,k}}{\prod_{i=1}^kd_{a_i'}d_{s_i}^{L,i}}} \sqrt{\frac{\prod_{j=2}^{l+1}d_{t_j}^{R,j-1}}{\prod_{j=1}^ld_{t_j}^{R,j}d_{b_j'}}}\\ 
        & \quad \times \underbrace{\delta_{u_1,v_{l+1}'}\delta_{z_1,t_{l+1}}}_{\text{factor~from~}XX'^{(1)}}\underbrace{\delta_{u_1',s_1}\delta_{v_{l+1},w_{l+1}}\frac{1}{d_{v_{l+1}}^R}}_{\text{factor~from~counit}} \\
        & \quad \times \prod_{i=1}^k  \delta_{u_i',s_i}\delta_{\gamma_i',\sigma_i}\delta_{u_i,z_i}\delta_{\gamma_i,\nu_i} \prod_{j=1}^l \delta_{v_j',t_j}\delta_{\zeta_j',\rho_j}\delta_{v_j,w_j}\delta_{\zeta_j,\mu_j}\\
        & \quad \times \sqrt{\frac{\prod_{i=1}^kd_{a_i'}d_{s_i}^{L,i}}{\prod_{i=2}^kd_{s_i}^{L,i-1}d_{t_1}^{L,k}}} \sqrt{\frac{\prod_{j=1}^ld_{t_j}^{R,j}d_{b_j'}}{\prod_{j=2}^{l+1}d_{t_j}^{R,j-1}}} \\ 
        & \quad \times \sqrt{\frac{\prod_{i=1}^kd_{a_i}d_{u_i}^{L,i}}{\prod_{i=2}^kd_{u_i}^{L,i-1}d_{v_1}^{L,k}}} \sqrt{\frac{\prod_{j=1}^ld_{v_j}^{R,j}d_{b_j}}{\prod_{j=2}^{l+1}d_{v_j}^{R,j-1}}} d_{v_{l+1}}^R\\
        & \quad \times \underbrace{\delta_{s_1,v_{l+1}''}\delta_{z_1',w_{l+1}''}}_{\text{factor~from~}X'^{(2)}X''}\underbrace{\delta_{u_1'',z_1''}\delta_{t_{l+1},w_{l+1}'}\frac{1}{d_{w_{l+1}'}^R}}_{\text{factor~from~counit}} \\
        & \quad \times \prod_{i=1}^k  \delta_{u_i'',z_i''}\delta_{\gamma_i'',\nu_i''}\delta_{s_i,z_i'}\delta_{\sigma_i,\nu_i'} \prod_{j=1}^l \delta_{v_j'',w_j''}\delta_{\zeta_j'',\mu_j''}\delta_{t_j,w_j'}\delta_{\rho_j,\mu_j'}\\
        & \quad \times \sqrt{\frac{\prod_{i=1}^kd_{a_i''}d_{z_i''}^{L,i}}{\prod_{i=2}^kd_{z_i''}^{L,i-1}d_{w_1}^{L,k}}} \sqrt{\frac{\prod_{j=1}^ld_{w_j''}^{R,j}d_{b_j''}}{\prod_{j=2}^{l+1}d_{w_j''}^{R,j-1}}} \\ 
        & \quad \times \sqrt{\frac{\prod_{i=1}^kd_{a_i'}d_{z_i'}^{L,i}}{\prod_{i=2}^kd_{z_i'}^{L,i-1}d_{w_1'}^{L,k}}} \sqrt{\frac{\prod_{j=1}^ld_{w_j'}^{R,j}d_{b_j'}}{\prod_{j=2}^{l+1}d_{w_j'}^{R,j-1}}} d_{w_{l+1}'}^R\;,
    \end{align*}
    from which one shows $\varepsilon(XX'X'') =  \sum_{(X')}\varepsilon(XX'^{(1)})\varepsilon(X'^{(2)}X'')$. 
    Similarly, one can show the identity $\varepsilon(XX'X'') =  \sum_{(X')}\varepsilon(XX'^{(2)})\varepsilon(X'^{(1)}X'')$.
    The condition $(\Delta(1)\otimes 1)(1\otimes \Delta(1)) = \Delta^{[2]}(1) = (1\otimes \Delta(1))(\Delta(1)\otimes 1)$ is straightforward. 

    2. Antipode structure. First, we need to identify the left and right counits by $\varepsilon_L(x) = \sum_{(1_W)} \varepsilon(1_W^{(1)}x)1_W^{(2)}$ and $\varepsilon_R(x) = \sum_{(1_W)}1_W^{(1)}\varepsilon(x1_W^{(2)})$. It is easy to see that, for the tube algebra under consideration, they take the forms  
    \begin{align}
        \varepsilon_L(X) & = \varepsilon(X)\sum_{m\in\Irr(\EM)} \;\begin{aligned}
            \begin{tikzpicture}[scale=0.65]
                 \filldraw[black!60, fill=gray!15, dotted, even odd rule] (0,0) circle[radius=0.5] (0,0) circle[radius=2];
                 \draw[line width=.6pt,black] (0,0.5)--(0,2);
                 \draw[line width=.6pt,black] (0,-0.5)--(0,-2);
                 \draw[red, thick, dotted] (0,0.8) arc[start angle=90, end angle=270, radius=0.8];
                 \draw[red, thick, dotted] (0,1.2) arc[start angle=90, end angle=270, radius=1.2];
                 \draw[blue,thick,dotted] (0,1.4) arc[start angle=90, end angle=-90, radius=1.4];
                 \draw[blue,thick,dotted] (0,1.8) arc[start angle=90, end angle=-90, radius=1.8];
                \node[ line width=0.6pt, dashed, draw opacity=0.5] (a) at (-0.3,-1.6){$\scriptstyle m$};
                \node[ line width=0.6pt, dashed, draw opacity=0.5] (a) at (-0.5,1.6){$\scriptstyle w_{l+1}$};
            \end{tikzpicture}
        \end{aligned}\;\;, \\
        \varepsilon_R(X) & = \varepsilon(X)\sum_{m\in\Irr(\EM)} \;\begin{aligned}
            \begin{tikzpicture}[scale=0.65]
                 \filldraw[black!60, fill=gray!15, dotted, even odd rule] (0,0) circle[radius=0.5] (0,0) circle[radius=2];
                 \draw[line width=.6pt,black] (0,0.5)--(0,2);
                 \draw[line width=.6pt,black] (0,-0.5)--(0,-2);
                 \draw[red, thick, dotted] (0,0.8) arc[start angle=90, end angle=270, radius=0.8];
                 \draw[red, thick, dotted] (0,1.2) arc[start angle=90, end angle=270, radius=1.2];
                 \draw[blue,thick,dotted] (0,1.4) arc[start angle=90, end angle=-90, radius=1.4];
                 \draw[blue,thick,dotted] (0,1.8) arc[start angle=90, end angle=-90, radius=1.8];
                \node[ line width=0.6pt, dashed, draw opacity=0.5] (a) at (-0.3,-1.6){$\scriptstyle u_1$};
                \node[ line width=0.6pt, dashed, draw opacity=0.5] (a) at (-0.3,1.6){$\scriptstyle m$};
            \end{tikzpicture}
        \end{aligned}\;\;. 
    \end{align}
    Using the comultiplication $\Delta(X)$, one computes 
    \begin{align*}
     \sum_{(X)}X^{(1)}S(X^{(2)}) & = \; \sum_{i=1}^k\sum_{j=1}^{l+1}\sum_{\sigma_i}\sum_{\rho_j} \sqrt{\frac{\prod_{i=1}^kd_{s_i}^{L,i-1}d_{t_1}^{L,k}}{\prod_{i=1}^kd_{a_i}d_{s_i}^{L,i}}} \sqrt{\frac{\prod_{j=2}^{l+1}d_{t_j}^{R,j-1}}{\prod_{j=1}^ld_{t_j}^{R,j}d_{b_j}}}  \\
                & \quad \quad \quad \frac{\prod_{i=1}^kd_{s_i}^{L,i}\prod_{j=1}^ld_{t_j}^{R,j}}{\prod_{i=2}^kd_{s_i}^{L,i-1}d_{t_1}^{L,k}\prod_{j=2}^{l+1}d_{t_j}^{R,j-1}} \delta_{u_1,z_1}\; 
        \begin{aligned}

        \end{aligned} \;. 
    \end{align*}
    Rewrite the coefficients as
    \begin{align*}
     \sum_{i=1}^k\sum_{j=1}^{l}&\sum_{\sigma_i}\sum_{\rho_j}  \sqrt{\frac{\prod_{i=1}^kd_{s_i}^{L,i-1}d_{t_1}^{L,k}}{\prod_{i=1}^kd_{a_i}d_{s_i}^{L,i}}} \sqrt{\frac{\prod_{j=2}^{l+1}d_{t_j}^{R,j-1}}{\prod_{j=1}^ld_{t_j}^{R,j}d_{b_j}}} \frac{\prod_{i=1}^kd_{s_i}^{L,i}\prod_{j=1}^ld_{t_j}^{R,j}}{\prod_{i=2}^kd_{s_i}^{L,i-1}d_{t_1}^{L,k}\prod_{j=2}^{l+1}d_{t_j}^{R,j-1}} \\
    & =  \prod_{j=1}^l\left(\sum_{t_j,\rho_j} \sqrt{\frac{d_{t_j}^{R,j}}{d_{t_{j+1}}^{R,j}d_{b_j}}}\right)\sum_{s_k,\sigma_k}\sqrt{\frac{d_{s_k}^{L,k}}{d_{a_k}d_{t_1}^{L,k}}}\prod_{i=1}^{k-1}\left(\sum_{s_i,\sigma_i}\sqrt{\frac{d_{s_i}^{L,i}}{d_{a_i}d_{s_{i+1}}^{L,i}}}\right). 
\end{align*}
    Applying sequential parallel moves 
    \begin{equation*}
        \sum_{i,\sigma}\sqrt{\frac{d_i^L}{d_ad_j^L}}
	\begin{aligned}

	\end{aligned}
    \end{equation*}
    to the above diagram, the resulting diagram will have a lower half 
    \begin{equation*}
        \sum_{t_{l+1}} \; 
        \begin{aligned} 
            %
        \end{aligned}\;,
    \end{equation*}
    and an upper half  
    \begin{equation*}
        \delta_{u_1,z_1}\;\begin{aligned} 
            %
        \end{aligned}\;. 
    \end{equation*}
    Hence one sees that $ \sum_{(X)}X^{(1)}S(X^{(2)}) = \varepsilon_L(X)$. Similarly one can show $\sum_{(X)}S(X^{(1)})X^{(2)} = \varepsilon_R(X)$. Finally, one computes 
\begin{align*}
        & \quad \sum_{(X)}S(X^{(1)})X^{(2)}S(X^{(3)})  = \sum_{(X)}\varepsilon_R(X^{(1)})S(X^{(2)}) \\ 
        & = \; \sum_{i=1}^k\sum_{j=1}^{l+1}\sum_{\sigma_i}\sum_{\rho_j} \sqrt{\frac{\prod_{i=1}^kd_{s_i}^{L,i-1}d_{t_1}^{L,k}}{\prod_{i=1}^kd_{a_i}d_{s_i}^{L,i}}} \sqrt{\frac{\prod_{j=2}^{l+1}d_{t_j}^{R,j-1}}{\prod_{j=1}^ld_{t_j}^{R,j}d_{b_j}}}\frac{\prod_{i=1}^kd_{s_i}^{L,i}\prod_{j=1}^ld_{t_j}^{R,j}}{\prod_{i=2}^kd_{s_i}^{L,i-1}d_{t_1}^{L,k}\prod_{j=2}^{l+1}d_{t_j}^{R,j-1}}  \\
    & 
             \quad \quad \quad  \varepsilon_R\left( \begin{aligned}

    \end{aligned}  \\
    & = \; \sum_{i=1}^k\sum_{j=1}^{l+1}\sum_{\sigma_i}\sum_{\rho_j} \sqrt{\frac{\prod_{i=1}^kd_{s_i}^{L,i-1}d_{t_1}^{L,k}}{\prod_{i=1}^kd_{a_i}d_{s_i}^{L,i}}} \sqrt{\frac{\prod_{j=2}^{l+1}d_{t_j}^{R,j-1}}{\prod_{j=1}^ld_{t_j}^{R,j}d_{b_j}}}\frac{\prod_{i=1}^kd_{s_i}^{L,i}\prod_{j=1}^ld_{t_j}^{R,j}}{\prod_{i=2}^kd_{s_i}^{L,i-1}d_{t_1}^{L,k}\prod_{j=2}^{l+1}d_{t_j}^{R,j-1}}  \\
    & \quad  \times \prod_{i=1}^k\delta_{s_i,u_i}\delta_{\sigma_i,\gamma_i}\prod_{j=1}^{l}\delta_{t_j,v_j}\delta_{\rho_j,\zeta_j}\delta_{t_{l+1},v_{l+1}}\sqrt{\frac{\prod_{i=1}^kd_{a_i}d_{s_i}^{L,i}}{\prod_{i=1}^kd_{s_i}^{L,i-1}d_{t_1}^{L,k}}} \sqrt{\frac{\prod_{j=1}^ld_{t_j}^{R,j}d_{b_j}}{\prod_{j=2}^{l+1}d_{t_j}^{R,j-1}}} \\
    & \quad \quad \quad \begin{aligned}

    \end{aligned} = S(X).
    \end{align*}
    3. $C^*$-structure. First, $X^{**} = X$ is clear. 
    Second, one shows that 
    \begin{align*}
    (XX')^* & = \delta_{u,v'}\delta_{z,w'}\frac{d_{z}}{d_w} \frac{d_{z'}}{d_{w'}} 
        \begin{aligned}
            %
        \end{aligned} = X'^*X^*. 
\end{align*}
Finally, inserting a wall labeled by $\sum_i i$ inside $X^*$ to compute $\Delta(X^*)$ results in 
\begin{align*}
    \Delta(X^*) = &\; \sum_i\sum_{\alpha\in I}\sum_{i_\alpha,\sigma_\alpha}\sum_{\beta\in J}\sum_{j_\beta,\rho_\beta} \sqrt{\frac{d_{j_l}}{d_{\bar{a}_1}\cdots d_{\bar{a}_k}d_{i}d_{\bar{b}_1}\cdots d_{\bar{b}_l}}}  \frac{d_z}{d_w}\\
    & 
                \begin{aligned}
        %
    \end{aligned} = \Delta(X)^*, 
\end{align*} 
where $\Delta(X)^* = \sum_{(X)}(X^{(1)})^*\otimes (X^{(2)})^*$. The proof is completed. 
\end{proof}

The above multimodule weak Hopf tube algebra naturally contains many weak Hopf tube subalgebras, which can be obtained by setting certain bulk strings to the tensor unit (viz. set the corresponding string as dotted string). 
For instance, if we set the bulk strings in $\EC_{\alpha}$ and $\ED_{\beta}$ to be the tensor unit, we obtain a tube subalgebra that is isomorphic to the domain wall tube algebra
\[
\Tube({_{\EC_1 \boxtimes \cdots \boxtimes \hat{\EC}_{\alpha} \boxtimes \cdots \boxtimes \EC_{|I|}}} \EM_{\ED_1 \boxtimes \cdots \boxtimes \hat{\ED}_{\beta} \boxtimes \cdots \boxtimes \ED_{|J|}}),
\]
where the hat symbol $\hat{\bullet}$ indicates the removal of the corresponding fusion categories.
This observation implies that the multimodule tube algebra possesses a hierarchical structure. 
When $N = |I| + |J|$, it can be interpreted as a quantum $N$-tuple algebra. 
Every quantum $k$-tuple algebra constructed from the given multimodule setup corresponds to a weak Hopf subalgebra of this $N$-tuple algebra. 
This reveals an interesting and potentially novel quantum algebraic structure that, to our knowledge, has not been previously discussed in the literature. 
We refer to this structure as a tube quantum $N$-tuple algebra.
This will be discussed in detail in Section~\ref{sec:N-tupleAlgebra}.

\subsubsection{Folding trick}

A direct consequence of the domain wall tube algebra is that one can define a generalized boundary tube algebra by simply folding all domain wall tubes appearing in the defining data of the weak Hopf domain wall tube algebra into a single boundary tube
\begin{equation}
     \begin{aligned}

\end{aligned}\;.
\end{equation}

\begin{proposition}
The folded domain wall tube becomes a  boundary tube with two bulk strings
\begin{equation}
\Tube_{\rm bd}({_{(\boxtimes_{\alpha \in I} \EC_{\alpha}) \boxtimes (\boxtimes_{\beta\in J} \ED_{\beta}^{\rm rev})}} \EM) =\operatorname{span}_\mathbb{C}\left\{  
\begin{aligned}
 %
\end{aligned}
:\begin{aligned}
    &a_i\in \Irr(\EC_i),b_j\in \Irr(\ED_j), \\
    &\text{wall edge label} \in \Irr(\EM),\\
 &\text{wall vertex label} \in \Hom_{\EM}.
\end{aligned}
\right\}.
\label{eq:bdtubebasis}
\end{equation}
Notice that the vertices connecting the blue string and the domain wall (black) string now lie in the left module category ${_{ {\ED_{\beta}^{\mathrm{rev}}} }}\EM$. 
\end{proposition}
\begin{proof}
    This configuration clearly defines a $C^*$ weak Hopf algebra, as all domain wall tubes can be folded into a boundary tube, yielding the structure induced by the weak Hopf domain wall tube algebra $\Tube({_{\{\EC_{\alpha}\}_{\alpha \in I}}}\EM_{ \{\ED_{\beta}\}_{\beta \in J} })$.
\end{proof}

It is important to note that the unfolding process is generally not straightforward. Since the bulk strings originate from different fusion categories, certain topological local moves cannot be applied to these edges. For example, although $a \boxtimes b$ is a simple object in $\EC \boxtimes \ED^{\mathrm{rev}}$, the components $a$ and $b$ belong to distinct categories and therefore cannot fuse. However, when constructing a more general boundary tube algebra involving multiple bulk strings, we may choose not to impose this constraint.

\subsection{Morita theory of multimodule tube algebra}

Inspired by Ref. \cite{kong2012universal}, here we develop the Morita theory for domain wall tube algebra. 
The special case, Morita theory of bulk tube algebra, has be given in our previous work \cite{jia2024weakTube}.

Recall that for two algebras $A$ and $B$, a Morita context is a pair $(M,N)$ where $M$ is an $A|B$ bimodule and $N$ is a $B|A$ bimodule, such that $M\otimes_BN\simeq A$ as $A|A$ bimodules and $N\otimes_AM\simeq B$ as $B|B$ bimodules. By such data, the category ${_A}\Mod$ of left $A$-modules is equivalent to the category ${_B}\Mod$ of left $B$-modules; the correspondences are provided by ${_A}\Mod \ni R\mapsto N\otimes_AR\in {_B}\Mod \ni T\mapsto M\otimes_BT\in {_A}\Mod$. We say that $A$ and $B$ are Morita equivalent in this sense. 
 
For $\mathbf{s}=(s_1,\cdots,s_{k}),  \mathbf{r} = (r_1,\cdots,r_k) \in \mathbb{Z}^{|I|}_{\geq 0}$, and $\mathbf{m} = (m_1,\cdots,m_l), \mathbf{n} = (n_1,\cdots,n_l)\in \mathbb{Z}^{|J|}_{\geq 0}$, consider the vector space (called the tube space)
\begin{equation} \label{eq:tube_space}
    \mathbf{T}^{\mathbf{s},\mathbf{r};\mathbf{m},\mathbf{n}} = \operatorname{span}_\mathbb{C}
    \left\{
    \begin{aligned}

    \end{aligned}\; : \; \text{edge}\in \Irr, \text{vertex}\in \Hom 
    \right\}. 
\end{equation}
To be more precisely, for each $\alpha\in I$, there are $s_\alpha$ exterior edges and $r_\alpha$ interior edges along the red arc labeled by $\Irr(\EC_\alpha)$; and for each $\beta\in J$, there are $m_\beta$ exterior edges and $n_\beta$ interior edges along the blue arc labeled by $\Irr(\ED_\beta)$. 
Hence $\mathbf{T}^{\mathbf{0},\mathbf{0};\mathbf{0},\mathbf{0}}=\mathbf{Tube}({_{\{\EC_{\alpha}\}_{\alpha\in I}}} \EM_{\{\ED_{\beta}\}_{\beta\in J}})$ with $\mathbf{0} = (0,\cdots,0)$. 

When $\mathbf{s} = \mathbf{r}$, and $\mathbf{m} = \mathbf{n}$, the tube space $\mathbf{T}^{\mathbf{s},\mathbf{s};\mathbf{m},\mathbf{m}}$ can be equipped with the following algebra structure. The unit is the sum of all diagrams in \eqref{eq:tube_space} such that all edges along the arcs are labeled by unit objects; here the sum is taken over all irreducible objects labeling the extra edges and the vertical edges. The product $TT'$ of two such diagrams is defined by connecting the interior edges of $T$ with the exterior edges of $T'$, provided they have the same labels in the corresponding edges; otherwise the product is zero. The tube space $\mathbf{T}^{\mathbf{s},\mathbf{r};\mathbf{m},\mathbf{n}}$ has a left $\mathbf{T}^{\mathbf{s},\mathbf{s};\mathbf{m},\mathbf{m}}$-action: the action of $T\in \mathbf{T}^{\mathbf{s},\mathbf{s};\mathbf{m},\mathbf{m}}$ on $M\in \mathbf{T}^{\mathbf{s},\mathbf{r};\mathbf{m},\mathbf{n}}$ is given by connecting the interior edges of $T$ with the exterior edges of $M$, provided they have the same labels in the corresponding edges; otherwise the action is zero. Likewise, the tube space $\mathbf{T}^{\mathbf{s},\mathbf{r};\mathbf{m},\mathbf{n}}$ has a right $\mathbf{T}^{\mathbf{r},\mathbf{r};\mathbf{n},\mathbf{n}}$-action by connecting corresponding edges with the same labels. These two actions clearly commute, hence endow $\mathbf{T}^{\mathbf{s},\mathbf{r};\mathbf{m},\mathbf{n}}$ a bimodule structure. All the constructions are similar to the bimodule case in Ref.~\cite{jia2024weakTube}. 

\begin{proposition}
    For any $\mathbf{s},\mathbf{r}$ and $\mathbf{m},\mathbf{n}$, the tube space $\mathbf{T}^{\mathbf{s},\mathbf{s};\mathbf{m},\mathbf{m}}$ is an algebra, and the tube space $\mathbf{T}^{\mathbf{s},\mathbf{r};\mathbf{m},\mathbf{n}}$ is a $\mathbf{T}^{\mathbf{s},\mathbf{s};\mathbf{m},\mathbf{m}}|\mathbf{T}^{\mathbf{r},\mathbf{r};\mathbf{n},\mathbf{n}}$ bimodule. The pair $(\mathbf{T}^{\mathbf{s},\mathbf{r};\mathbf{m},\mathbf{n}},\mathbf{T}^{\mathbf{r},\mathbf{s};\mathbf{n},\mathbf{m}})$ forms a Morita context in the sense that 
    \begin{equation}
        \mathbf{T}^{\mathbf{s},\mathbf{r};\mathbf{m},\mathbf{n}}\otimes_{\mathbf{T}^{\mathbf{r},\mathbf{r};\mathbf{n},\mathbf{n}}}\mathbf{T}^{\mathbf{r},\mathbf{s};\mathbf{n},\mathbf{m}} \simeq \mathbf{T}^{\mathbf{s},\mathbf{s};\mathbf{m},\mathbf{m}},\quad \mathbf{T}^{\mathbf{r},\mathbf{s};\mathbf{n},\mathbf{m}}\otimes_{\mathbf{T}^{\mathbf{s},\mathbf{s};\mathbf{m},\mathbf{m}}}\mathbf{T}^{\mathbf{s},\mathbf{r};\mathbf{m},\mathbf{n}} \simeq \mathbf{T}^{\mathbf{r},\mathbf{r};\mathbf{n},\mathbf{n}}. 
    \end{equation}
    Therefore, the algebras $\mathbf{T}^{\mathbf{s},\mathbf{s};\mathbf{m},\mathbf{m}}$ are Morita equivalent for all $\mathbf{s}\in \mathbb{Z}^{|I|}_{\geq 0},\mathbf{m}\in \mathbb{Z}^{|J|}_{\geq 0}$. 
\end{proposition}

\begin{proof}
    The first statement is already clear. The second statement is also clear by the actions defined above, which can be similarly proved as that in the bimodule case in Ref.~\cite{jia2024weakTube}. 
\end{proof}
\section{Domain wall tube algebra as generalization of quantum double algebra}
\label{sec:tubeDouble}

Our goal in this section and the next is to establish that the domain wall tube algebra can be understood as a Drinfeld quantum double of the boundary tube algebras. Motivated by this perspective, we further propose the notion of a \emph{quantum $N$-tuple algebra}.

Recall that the Drinfeld quantum double is constructed from two weak Hopf algebras, both of which embed into the resulting double. In the context of multimodule domain wall tube algebras, when the construction involves a multimodule category over $N$ base fusion categories, the resulting tube algebra can be interpreted as a \emph{quantum $N$-tuple algebra}—a generalization of the quantum double built from $N$ weak Hopf algebras, each of which embeds into the $N$-tuple algebra.
For instance, a bimodule category yields a quantum double, a trimodule category gives rise to a \emph{quantum triple algebra}, and so forth. Fixing $N \in \mathbb{Z}_+$, we observe that for any $M < N$, there exist numerous quantum $M$-tuple subalgebras contained within a given quantum $N$-tuple algebra. This hierarchical structure provides a natural generalization of the Drinfeld quantum double within the framework of domain wall tube algebras.

\subsection{Tube quantum double algebra}

There are two canonical subalgebras of $\mathbf{Tube}({_{\EC}}\EM_{\ED})$ which we will denote them as $\mathbf{Tube}_L({_{\EC}}\EM)$ and $\mathbf{Tube}_R(\EM_{\ED})$: 
\begin{align}
    \mathbf{Tube}_L({_{\EC}}\EM)& = \mathrm{span}_\mathbb{C}\left\{  
   \begin{aligned}

    \end{aligned}
   \;:\; 
   \begin{aligned}
     & b\in \Irr(\ED), x,y,v,w\in \Irr(\EM), \\
     & \mu\in\Hom_{\EM}(x,y\otimes b)\neq 0,\\
     &\gamma\in \Hom_{\EM}(v\otimes b,w)\neq 0  
   \end{aligned}
    \right\},
\end{align}
which means we identify a half-tube (or boundary tube) with a tube that is a trivial (tensor unit) bulk string.
They can be regarded as the embeddings of boundary tube algebras $\mathbf{Tube}_{\rm bd}(_{\EC}\EM) $ and  $\mathbf{Tube}_{\rm bd}(\EM_{\ED}) $ into domain wall tube algebra $\Tube(_{\EC}\EM_{\ED})$ via the following one-to-one map
\begin{equation}
    \begin{aligned}

    \end{aligned}.
\end{equation}
This embedding will equip the boundary tube algebra with a $C^*$ weak Hopf algebra structure induced by the domain wall tube algebra.

\begin{proposition}
        The boundary tube algebras $\mathbf{Tube}_{\rm bd}({_{\EC}}\EM)$ and $\mathbf{Tube}_{\rm bd}(\EM_{\ED})$ are $C^*$ weak Hopf algebras, which embed into $\mathbf{Tube}({_{\EC}}\EM_{\ED})$ as $C^*$ weak Hopf subalgebras, and they are isomorphic to  $\mathbf{Tube}_{L}({_{\EC}}\EM)$ and $\mathbf{Tube}_{R}(\EM_{\ED})$, respectively. 
     The structure morphisms of $\mathbf{Tube}_{\rm bd}({_{\EC}}\EM)$ are as follows: 
\begin{gather}
    1=\sum_{x,z\in\Irr(\EM)} \;\begin{aligned}

    \end{aligned}\;\;. 
\end{gather}
    The structure morphisms of $\mathbf{Tube}_{\rm bd}(\EM_{\ED})$ are given similarly. 
\end{proposition}

\begin{proof}
    Via the embedding, the structure morphisms are inherited from $\mathbf{Tube}_L({_{\EC}}\EM) \hookrightarrow \mathbf{Tube}({_{\EC}}\EM_{\ED})$.
    Alternatively, they can be derived directly in the same manner as for the domain wall tube algebra $\mathbf{Tube}({_{\EC}}\EM_{\ED})$, as we now proceed to show. For convenience, we use $X, X', X''$ to denote basis diagrams labeled by $\{a,\cdots,\nu\}$, $\{a',\cdots,\nu'\}$, and $\{a'',\cdots,\nu''\}$, respectively. 

    The algebra and coalgebra structures are evident from the construction. Let us prove the weak bialgebra axioms. First, using F-move, one rewrites the multiplication as 
    \begin{align}
   XX'&=\begin{aligned}

    \end{aligned}.  \label{eq:mul_XX'}
\end{align}
Therefore, the comultiplication of a product is 
\begin{align*}
    \Delta(XX') & = \delta_{u,v'}\delta_{z,y'}\sum_{t,\alpha,\beta}[F_{aa'u'}^{v}]_{u\zeta\zeta'}^{t\alpha\beta}\sum_{\theta}[F_y^{aa'z'}]_{z\nu\nu'}^{t\theta\beta}\sqrt{\frac{d_{a}d_{a'}}{d_t}} \\
    & \quad \quad \sum_{i,k,\sigma}\sqrt{\frac{d_k^L}{d_td_i^L}}\;\begin{aligned}
        %
    \end{aligned}. 
\end{align*}
Then the multiplicative property of $\Delta$ is shown by the computation: 
\begin{align*}
    \Delta(X)\Delta(X') & = \sum_{i,k,\sigma}\sqrt{\frac{d_k^L}{d_ad_i^L}}\sum_{i',k',\sigma'}\sqrt{\frac{d_{k'}^L}{d_{a'}d_{i'}^L}}\;\begin{aligned}
        %
    \end{aligned} \\
    & = \sum_{i,k,\sigma}\sqrt{\frac{d_k^L}{d_ad_i^L}}\sum_{i',k',\sigma'}\sqrt{\frac{d_{k'}^L}{d_{a'}d_{i'}^L}}\delta_{u,v'}\delta_{z,y'}\delta_{i,k'} \sum_{t,\alpha,\beta}[F_{aa'u'}^{v}]_{u\zeta\zeta'}^{t\alpha\beta}\sum_{\theta}[F_k^{aa'i'}]_{i\sigma\sigma'}^{t\theta\beta}\sqrt{\frac{d_{a}d_{a'}}{d_t}}  \\
    & \quad
     \sum_{t',\alpha',\beta'}[F_{aa'i'}^{k}]_{i\sigma\sigma'}^{t'\alpha'\beta'}\sum_{\theta'}[F_y^{aa'z'}]_{z\nu\nu'}^{t'\theta'\beta'}\sqrt{\frac{d_{a}d_{a'}}{d_{t'}}} \;\begin{aligned}

    \end{aligned}\\
    & = \delta_{u,v'}\delta_{z,y'}\sum_{k}\sum_{i'}\sum_{t,\alpha,\beta}\sum_{t',\alpha',\beta'}\sum_{\theta}\sum_{\theta'}\underbrace{\sum_{i,\sigma,\sigma'}[F_k^{aa'i'}]_{i\sigma\sigma'}^{t\theta\beta}[F_{aa'i'}^{k}]_{i\sigma\sigma'}^{t'\alpha'\beta'}}_{\delta_{t,t'}\delta_{\theta,\alpha'}\delta_{\beta,\beta'}}  \\
    & \quad\;
     [F_{aa'u'}^{v}]_{u\zeta\zeta'}^{t\alpha\beta}[F_y^{aa'z'}]_{z\nu\nu'}^{t'\theta'\beta'}\sqrt{\frac{d_{a}d_{a'}}{d_t}}\sqrt{\frac{d_{a}d_{a'}}{d_{t'}}}\sqrt{\frac{d_k^L}{d_ad_i^L}}\sqrt{\frac{d_i^L}{d_{a'}d_{i'}^L}} \;\begin{aligned}

    \end{aligned} = \Delta(XX'). 
\end{align*}
The weak comultiplicative property of the unit $(\Delta\otimes\id)\comp \Delta(1) = (\Delta(1)\otimes 1)(1\otimes \Delta(1)) = (1\otimes \Delta(1))(\Delta(1)\otimes 1)$ is immediate. To show the weak multiplicative property of the counit, we first compute the counit of a product by Eq.~\eqref{eq:mul_XX'}: 
\begin{align}
    \varepsilon(XX') & =\delta_{u,v'}\delta_{z,y'}\sum_{t,\alpha,\beta}[F_{aa'u'}^{v}]_{u\zeta\zeta'}^{t\alpha\beta}\sum_{\theta}[F_y^{aa'z'}]_{z\nu\nu'}^{t\theta\beta}\sqrt{\frac{d_{a}d_{a'}}{d_t}} \delta_{u',z'}\delta_{\theta,\alpha}\delta_{v,y} \sqrt{\frac{d_td_{u'}^L}{d_y^L}} \nonumber \\
    & =\delta_{u,v'}\delta_{z,y'}\delta_{u',z'}\delta_{v,y}\underbrace{\sum_{t,\alpha,\beta}[F_{aa'u'}^{v}]_{u\zeta\zeta'}^{t\alpha\beta}[F_y^{aa'z'}]_{z\nu\nu'}^{t\alpha\beta}}_{\delta_{u,z}\delta_{\zeta,\nu}\delta_{\zeta',\nu'}}\sqrt{\frac{d_{a}d_{a'}d_{u'}^L}{d_y^L}}   \nonumber \\
    & =\delta_{u,v'}\delta_{z,y'}\delta_{u',z'}\delta_{v,y}\delta_{u,z}\delta_{\zeta,\nu}\delta_{\zeta',\nu'}\sqrt{\frac{d_{a}d_{a'}d_{u'}^L}{d_y^L}} . 
\end{align}
Thus, one computes 
\begin{align}
    \sum_{(X')}\varepsilon(XX'^{(1)})\varepsilon(X'^{(2)}X'') & = \sum_{i,k,\sigma}\sqrt{\frac{d_k^L}{d_{a'}d_i^L}}\delta_{u,v'}\delta_{z,k}\delta_{u',i}\delta_{v,y}\delta_{u,z}\delta_{\zeta,\nu}\delta_{\zeta',\sigma}\sqrt{\frac{d_ad_{a'}d_{i}^L}{d_y^L}} \nonumber \\
    &\quad \quad \times  \delta_{i,v''}\delta_{z',y''}\delta_{u'',z''}\delta_{k,y'}\delta_{i,z'}\delta_{\sigma,\nu'}\delta_{\zeta'',\nu''}\sqrt{\frac{d_{a'}d_{a''}d_{u''}^L}{d_{y'}^L}}, \label{eq:count_L_1}\\
    \sum_{(X')}\varepsilon(XX'^{(2)})\varepsilon(X'^{(1)}X'') & = \sum_{i,k,\sigma}\sqrt{\frac{d_k^L}{d_{a'}d_i^L}}\delta_{u,k}\delta_{z,y'}\delta_{i,z'}\delta_{v,y}\delta_{u,z}\delta_{\zeta,\nu}\delta_{\sigma,\nu'}\sqrt{\frac{d_ad_{a'}d_{i}^L}{d_y^L}} \nonumber \\
    & \quad \quad \times \delta_{u',v''}\delta_{i,y''}\delta_{u'',z''}\delta_{v',k}\delta_{u',i}\delta_{\zeta',\sigma}\delta_{\zeta'',\nu''}\sqrt{\frac{d_{a'}d_{a''}d_{u''}^L}{d_k^L}}. \label{eq:count_L_2}
\end{align}
Now we compute the counit of $XX'X''$. It follows from Eq.~\eqref{eq:mul_XX'} that
\begin{align*}
    XX'X'' & = \delta_{u,v'}\delta_{z,y'}\sum_{t,\alpha,\beta}[F_{aa'u'}^{v}]_{u\zeta\zeta'}^{t\alpha\beta}\sum_{\theta}[F_y^{aa'z'}]_{z\nu\nu'}^{t\theta\beta}\sqrt{\frac{d_{a}d_{a'}}{d_t}} \\
    & \quad \delta_{u',v''}\delta_{z',y''}\sum_{t',\alpha',\beta'}[F_{ta''u''}^{v}]_{u'\alpha\zeta''}^{t'\alpha'\beta'}\sum_{\theta'}[F_y^{ta''z''}]_{z'\theta\nu''}^{t'\theta'\beta'}\sqrt{\frac{d_{t}d_{a''}}{d_{t'}}} \;\begin{aligned}
        \begin{tikzpicture}[scale=0.65]
        \begin{scope}
            \fill[gray!15]
                (0,1.5) arc[start angle=90, end angle=270, radius=1.5] -- 
                (0,-0.5) arc[start angle=270, end angle=90, radius=0.5] -- cycle;
        \end{scope}
        \draw[dotted] (0,1.5) arc[start angle=90, end angle=270, radius=1.5]; 
        \draw[dotted] (0,0.5) arc[start angle=90, end angle=270, radius=0.5];
             \draw[line width=.6pt,black] (0,0.5)--(0,1.5);
             \draw[line width=.6pt,black] (0,-0.5)--(0,-1.5);
             \draw[red] (0,0.8) arc[start angle=90, end angle=270, radius=0.8];
            \node[ line width=0.6pt, dashed, draw opacity=0.5] (a) at (-1,0){$\scriptstyle t'$};
            \node[ line width=0.6pt, dashed, draw opacity=0.5] (a) at (-0.2,-1.1){$\scriptstyle y$};
            \node[ line width=0.6pt, dashed, draw opacity=0.5] (a) at (0.3,-0.8){$\scriptstyle \theta'$};
            \node[ line width=0.6pt, dashed, draw opacity=0.5] (a) at (0,-0.3){$\scriptstyle z''$};
            \node[ line width=0.6pt, dashed, draw opacity=0.5] (a) at (0,0.3){$\scriptstyle u''$};
            \node[ line width=0.6pt, dashed, draw opacity=0.5] (a) at (-0.2,1.1){$\scriptstyle v$};
            \node[ line width=0.6pt, dashed, draw opacity=0.5] (a) at (0.3,0.8){$\scriptstyle \alpha'$};
        \end{tikzpicture}
    \end{aligned};
\end{align*}
so then we compute its counit
\begin{align*}
    \varepsilon(XX'X'') & = \delta_{u,v'}\delta_{z,y'}\sum_{t,\alpha,\beta}[F_{aa'u'}^{v}]_{u\zeta\zeta'}^{t\alpha\beta}\sum_{\theta}[F_y^{aa'z'}]_{z\nu\nu'}^{t\theta\beta}\sqrt{\frac{d_{a}d_{a'}}{d_t}} \\
    & \quad \delta_{u',v''}\delta_{z',y''}\sum_{t',\alpha',\beta'}[F_{ta''u''}^{v}]_{u'\alpha\zeta''}^{t'\alpha'\beta'}\sum_{\theta'}[F_y^{ta''z''}]_{z'\theta\nu''}^{t'\theta'\beta'}\sqrt{\frac{d_{t}d_{a''}}{d_{t'}}} \delta_{z'',u''}\delta_{v,y}\delta_{\theta',\alpha'}\sqrt{\frac{d_{t'}d_{u''}^L}{d_y^L}} \\
    & = \delta_{u,v'}\delta_{z,y'}\delta_{z'',u''}\delta_{v,y}\delta_{u',v''}\delta_{z',y''}\sum_{t,\alpha,\beta}[F_{aa'u'}^{v}]_{u\zeta\zeta'}^{t\alpha\beta}\sum_{\theta}[F_y^{aa'z'}]_{z\nu\nu'}^{t\theta\beta} \\
    & \quad \underbrace{\sum_{t',\alpha',\beta'}[F_{ta''u''}^{v}]_{u'\alpha\zeta''}^{t'\alpha'\beta'}[F_y^{ta''z''}]_{z'\theta\nu''}^{t'\alpha'\beta'}}_{\delta_{u',z'}\delta_{\alpha,\theta}\delta_{\zeta'',\nu''}} \sqrt{\frac{d_ad_{a'}d_{a''}d_{u''}^L}{d_y^L}} \\
    & = \delta_{u,v'}\delta_{z,y'}\delta_{z'',u''}\delta_{v,y}\delta_{u',v''}\delta_{z',y''} \delta_{u',z'}\delta_{\zeta'',\nu''}  \underbrace{\sum_{t,\alpha,\beta}[F_{aa'u'}^{v}]_{u\zeta\zeta'}^{t\alpha\beta}[F_y^{aa'z'}]_{z\nu\nu'}^{t\alpha\beta}}_{\delta_{u,z}\delta_{\zeta,\nu}\delta_{\zeta',\nu'}}\sqrt{\frac{d_ad_{a'}d_{a''}d_{u''}^L}{d_y^L}} \\
    & = \delta_{u,v'}\delta_{z,y'}\delta_{z'',u''}\delta_{v,y}\delta_{u',v''}\delta_{z',y''} \delta_{u',z'}\delta_{\zeta'',\nu''}\delta_{u,z}\delta_{\zeta,\nu}\delta_{\zeta',\nu'}\sqrt{\frac{d_ad_{a'}d_{a''}d_{u''}^L}{d_y^L}}\;. 
\end{align*}
which together with Eqs.~\eqref{eq:count_L_1} and \eqref{eq:count_L_2} implies 
\begin{equation}
    \varepsilon(XX'X'') = \sum_{(X')}\varepsilon(XX'^{(1)})\varepsilon(X'^{(2)}X'') = \sum_{(X')}\varepsilon(XX'^{(2)})\varepsilon(X'^{(1)}X''). 
\end{equation}

Then let us show the antipode axioms: 
\begin{align*}
    \sum_{(X)}X^{(1)}S(X^{(2)}) & = \sum_{i,k,\sigma}\sqrt{\frac{d_k^L}{d_ad_i^L}}\frac{d_i^L}{d_k^L}\delta_{u,z} \; 
    \begin{aligned}

    \end{aligned} = S(X).
\end{align*}

Finally, we verify the $*$-structure axioms. The identities $X^{**} = X$ and $\Delta(X^*) = \Delta(X)^*$ are evident by definition. It suffices to verify the identity $(XX')^* = X'^*X^*$. Using Eq.~\eqref{eq:mul_XX'}, we have 
\begin{align*}
    (XX')^* & = \delta_{u,v'}\delta_{z,y'}\sum_{t,\alpha,\beta}\overline{[F_{aa'u'}^{v}]_{u\zeta\zeta'}^{t\alpha\beta}}\sum_{\theta} \,\overline{[F_y^{aa'z'}]_{z\nu\nu'}^{t\theta\beta}}\sqrt{\frac{d_{a}d_{a'}}{d_t}} \;\left(\begin{aligned}
        %
    \end{aligned}\;. 
\end{align*}
The second equality holds because $(F_{aa'u'}^v)^\dagger = (F_{aa'u'}^v)^{-1}$ and $(F^{aa'z'}_y)^\dagger = (F^{aa'z'}_y)^{-1}$ by unitarity. Note that we have the following properties for F-symbols:
\begin{equation}
    [(F_{aa'u'}^{v})^{-1}]^{u\zeta\zeta'}_{t\alpha\beta} = [F^{aa'u'}_{v}]_{u\zeta\zeta'}^{t\alpha\beta},\quad [(F_y^{aa'z'})^{-1}]^{z\nu\nu'}_{t\theta\beta}= [F^y_{aa'z'}]_{z\nu\nu'}^{t\theta\beta}. 
\end{equation}
These identities come from evaluating the following diagrams using different F-moves: 
\begin{equation*}
    \begin{aligned}

    \end{aligned}. 
\end{equation*}
By virtue of these properties, one computes that 
\begin{align*}
    X'^* X^* & = \frac{d_{z'}^Ld_z^L}{d_{y'}^Ld_y^L}\delta_{v',u}\delta_{y',z}\; \begin{aligned}
        %
    \end{aligned} = (XX')^*, 
\end{align*}
which finishes the proof. 
\end{proof}

\begin{corollary}
    The Haar integral of  $\mathbf{Tube}_{\rm bd}({_{\EC}}\EM)$ is of the form 
    \begin{equation} \label{eq:bdHaarL}
   \mathfrak{I}_{\rm bd}^L =\frac{1}{\operatorname{rank} \EM} \sum_{a,x,z,\nu} \frac{1}{d_z^L}\sqrt{\frac{d_a}{d_x^Ld_z^L}} \;\;
    \begin{aligned}%
    \end{aligned}\;.
\end{equation}
The canonical group-like element and its inverse are 
\begin{equation} 
    \xi_{\rm bd}^L =\sum_{x,z\in\Irr(\EM)} \frac{d_x}{d_z} \;\begin{aligned}
        %
    \end{aligned}\;\;.
\end{equation}
Similar results hold for $\mathbf{Tube}_{\rm bd}(\EM_{\ED})$. 
\end{corollary}

\begin{proof}
   First we reformulate the expression \eqref{eq:bdHaarL} as follows:  \begin{equation}
    \mathfrak{I}_{\rm bd}^L  =\frac{1}{\operatorname{rank} \EM} \sum_{x,z} \frac{1}{d_z^L}\sum_{a,\nu}\sqrt{\frac{d_a}{d_x^Ld_z^L}} \;\;
    \begin{aligned}%
\end{aligned}\;.
\end{equation} 
See Remark~\ref{rmk:module_dual_obj} for the validity of the last equality. 
For any $X = \begin{aligned}%
\end{aligned}\; = \varepsilon_L(X)\mathfrak{I}_{\rm bd}^L. 
\end{align*}
Thus, $\mathfrak{I}_{\rm bd}^L$ is a left integral. Moreover, it is right-normalized in the following sense: 
\begin{align*}
    \varepsilon_R(\mathfrak{I}_{\rm bd}^L) & =\frac{1}{\operatorname{rank} \EM}\sum_{a,x,z} \frac{1}{d_z^L}\sqrt{\frac{d_a}{d_x^Ld_z^L}}N_{az}^x \sum_{x'} \sqrt{\frac{d_ad_z^L}{d_x^L}}\;\;
    \begin{aligned}%
    \end{aligned} \; = 1_{\mathbf{Tube}_{\rm bd}({_{\EC}}\EM)},
\end{align*}
which together with the previous assertion implies that $\mathfrak{I}_{\rm bd}^L$ is the Haar integral of the left boundary tube algebra. The statement for the canonical group-like element is derived similar to that for \(\mathbf{Tube}({_{\EC}}\EM_{\ED})\). 
\end{proof}

The domain wall tube algebra \(\mathbf{Tube}({_{\EC}}\EM_{\ED})\) can be regarded as the generalized Drinfeld (internal\footnote{By `internal' we mean that there is a big algebra, and we construct the double in this bigger algebra.}) quantum double of  boundary tube algebras $\mathbf{Tube}_{\rm bd}(_{\EC}\EM) $ and  $\mathbf{Tube}_{\rm bd}(\EM_{\ED}) $. We first embed the boundary tube algebra $\mathbf{Tube}_L({_{\EC}}\EM)$ and $\mathbf{Tube}_R(\EM_{\ED})$.  
Then we define a map from $ \mathbf{Tube}_R(\EM_{\ED}) \otimes  \mathbf{Tube}_L({_{\EC}}\EM) $ to $\mathbf{Tube}({_{\EC}}\EM_{\ED})$ as 
 \begin{equation}\label{eq:gluing}
        \Phi: 
         \begin{aligned}

    \end{aligned}\;.
    \end{equation}
It is clear that this is a surjective map and the boundary tube algebras $\mathbf{Tube}_{\rm bd}(_{\EC}\EM) $ and  $\mathbf{Tube}_{\rm bd}(\EM_{\ED})$ are embedded into the domain wall tube algebra \(\mathbf{Tube}({_{\EC}}\EM_{\ED})\) as weak Hopf algebras. We will denote the resultant weak Hopf algebra as $\mathbf{Tube}_{\rm bd}(\EM_{\ED}) \bicrosslr  \mathbf{Tube}_{\rm bd}({_{\EC}}\EM)$.

From a physical perspective, for a non-chiral topological phase, the boundary tube algebra can be interpreted as a gauge symmetry, while the bulk tube algebra corresponds to a charge symmetry~\cite{jia2024weakTube}. Since gauge and charge symmetries are typically related through the quantum double construction, this observation motivates the following conjecture: the domain wall tube algebra is isomorphic to the quantum double of the boundary tube algebras. This conjecture will be established in the next section.

\subsection{Quantum triple and $N$-tuple algebras from tube algebra}
\label{sec:N-tupleAlgebra}

The preceding discussion motivates a natural generalization of the notion of the `quantum double algebra' which is constructed from two algebras. 
In particular, the domain wall tube algebra $\TubeM$ can be interpreted as `double' of the boundary tube algebras $\mathbf{Tube}_{\rm bd}({_{\EC}}\EM)$ and $\mathbf{Tube}_{\rm bd}(\EM_{\ED})$ (See Theorem~\ref{thm:QD_tube} in the next section).
In Section~\ref{sec:multi_tube},  we develop the theory of tube algebras associated with multimodule categories. In this setting, there are $N$ boundary tube algebras of the form $\mathbf{Tube}_{\rm bd}({_{\EC_{\alpha}}}\EM)$ and $\mathbf{Tube}_{\rm bd}(\EM_{\ED_{\beta}})$. 
The corresponding domain wall tube algebra can then be regarded as an $N$-tuple generalization of the quantum double construction, in the sense that the tube double algebra is constructed from two boundary tube algebras, while the tube $N$-tuple algebra is constructed from $N$ boundary tube algebras.

Given a multimodule category ${_{\{\EC_{\alpha}\}_{\alpha\in I}}} \EM_{\{\ED_{\beta}\}_{\beta\in J}}$, we can construct embeddings of the boundary tube algebras into the domain wall tube algebra as follows. For each $\alpha \in I$, by fixing all bulk edges to be the trivial object $\mathbb{1}_{\alpha'}$ for $\alpha' \neq \alpha$, the boundary tube algebra $\Tube_{\rm bd}({_{\EC_{\alpha}}}\EM)$ can be embedded into the domain wall tube algebra, yielding a weak Hopf subalgebra $\Tube_L({_{\EC_{\alpha}}}\EM)$:
\begin{equation}
   \Tube_{\rm bd}({_{\EC_{\alpha}}}\EM) \simeq \Tube_L({_{\EC_{\alpha}}}\EM) \hookrightarrow \Tube ({_{\{\EC_{\alpha}\}_{\alpha\in I}}} \EM_{\{\ED_{\beta}\}_{\beta\in J}}).
\end{equation}
Similarly, for each $\beta \in J$, we obtain an embedding 
\begin{equation}
    \Tube_{\rm bd}(\EM_{\ED_{\beta}}) \simeq \Tube_R(\EM_{\ED_{\beta}}) \hookrightarrow \Tube ({_{\{\EC_{\alpha}\}_{\alpha\in I}}} \EM_{\{\ED_{\beta}\}_{\beta\in J}}).
\end{equation}
These embeddings allow us to define an $N$-tuple generalization of the gluing map, analogous to Eq.~\eqref{eq:gluing}, by taking products of the embedded basis elements. The resulting algebra is referred to as the \emph{tube $N$-tuple algebra}, and is given by:
\begin{equation}
    \left( \bicrosslr_{\alpha\in I} \Tube_L({_{\EC_{\alpha}}}\EM) \right)  \bicrosslr  \left( \bicrosslr_{\beta\in J} \Tube_R(\EM_{\ED_{\beta}}) \right).
\end{equation}
We adopt the notation `$\bicrosslr$' to highlight its resemblance to the bicross product structure\footnote{To distinguish it from the standard bicross product, we reserve the notation `$\Join$' to denote the Drinfeld bicross product.}.

For a multimodule category with three base fusion categories, we refer to the resulting algebra as the \emph{tube triple algebra}. Similarly, we have the \emph{tube quadruple algebra}, \emph{quantum quintuple algebra}, and so on.

\begin{proposition}
    Let ${_{\{\EC_{\alpha}\}_{\alpha\in I}}} \EM_{\{\ED_{\beta}\}_{\beta\in J}}$ be a multimodule category, and consider the associated tube $N$-tuple algebra
    $\left( \bicrosslr_{\alpha\in I} \Tube_L({_{\EC_{\alpha}}}\EM) \right) \bicrosslr \left( \bicrosslr_{\beta\in J} \Tube_R(\EM_{\ED_{\beta}}) \right)$.
    Then, for any tube $k$-tuple algebra with $1 \leq k \leq N$ constructed from the sub-multimodule category ${_{\{\EC_{\alpha}\}_{\alpha\in I'}}} \EM_{\{\ED_{\beta}\}_{\beta\in J'}}$, where $I' \subset I$, $J' \subset J$, and $|I'| + |J'| = k$, there exists a weak Hopf algebra embedding into the above tube $N$-tuple algebra. 
    In particular, the tube $N$-tuple algebra admits a hierarchical structure with respect to these subalgebras.
\end{proposition}

\begin{proof}
    This follows directly from the construction: the embeddings of lower-order tube algebras are obtained by restricting to a subset of bulk labels while setting the remaining ones to tensor unit objects. The algebraic operations are preserved under this restriction, and the image forms a weak Hopf subalgebra.
\end{proof}

\section{Skew-pairing of weak Hopf tube algebras and Drinfeld quantum double}
\label{sec:QDTube}

In this section, we establish the connection between the boundary and domain wall tube algebras by demonstrating that the domain wall tube algebra can be understood as the Drinfeld quantum double of the boundary tube algebras. The key technical tool in this construction is the notion of a skew-pairing between weak Hopf algebras. Throughout this section, we assume that the underlying bimodule category is pseudo-unitary.

The following definition (see Refs.~\cite{schauenburg2000duals,Schauenburg2003weakHoopf}) is a natural generalization of the notion of a skew-pairing for Hopf algebras. For a comprehensive discussion on pairings and its application in constructing quantum double, see, e.g., Ref.~\cite{majid2000foundations}.

\begin{definition}[Skew-pairing] \label{def:skew-pairing}
    Let $U, V$ be two weak Hopf algebras with invertible antipodes $S_U,S_V$. A skew-pairing between them is a bilinear map 
    $p:U\times V\to \mathbb{C}, (u,v)\mapsto p(u,v)=:\langle u,v \rangle_p$, satisfying\,
    % \footnote{In this section, we adopt a convention analogous to the Einstein summation convention: all summation symbols for comultiplication are omitted. The appearance of superscripts such as $\cone$, $\ctwo$, etc., implicitly indicates that comultiplication is being applied.}
    \begin{gather}
        p(1_{U},v) =\varepsilon_{V}(v),\quad p(u,1_{V}) =\varepsilon_{U}(u),  \label{eq:pairing-unit} \\
        p(uu',v) =  \sum_{(v)}p(u,v^{(1)})p(u',v^{(2)}),\label{eq:pairing-left}\\
     p(u,vv') =  \sum_{(u)}p(u^{(2)},v)  p(u^{(1)},v'), \label{eq:pairing-right}\\
        p(S_U(u),v) = p(u,S_V^{-1}(v)). \label{eq:pairing-S}
    \end{gather}
\end{definition}

Note that Eq.~\eqref{eq:pairing-S} can be derived from Eqs.~\eqref{eq:pairing-unit}–\eqref{eq:pairing-right}, together with the fact that the antipodes of both $U$ and $V$ are invertible. 
This follows from our assumption that $U$ and $V$ are finite-dimensional, in which case the invertibility of the antipode is guaranteed by~\cite[Theorem 2.10]{BOHM1998weak}. 
For convenience in later discussions, we include Eq.~\eqref{eq:pairing-S} as part of the definition. 
Also note that Eq.~\eqref{eq:pairing-S} is equivalent to
\begin{equation}
    p(S^{-1}_U(u),v) = p(u,S_V(v)). 
\end{equation}
Indeed, as $S$ is of finite order, $S^N = \id$ for some positive integer $N$, which together with Eq.~\eqref{eq:pairing-S} implies 
\begin{equation*}
    p(S^{-1}_U(u),v) = p(S^{-2}_U(u),S^{-1}_V(v)) = \cdots = p(S^{-N}_U(u),S^{-(N-1)}_V(v))=p(u,S_V(v)). 
\end{equation*} 
The converse is verifed similarly.

\subsection{Weak Hopf Drinfeld quantum double from skew-pairing}

Similar to the discussion in Refs.~\cite{nikshych2003invariants,Jia2023weak,majid2000foundations}, we can obtain a generalized quantum double for a given skew-pairing. In this section, we will show that the domain wall tube algebra has a quantum double structure constructed from a pairing.  Before getting into the details, let us first discuss the general construction. Although the discussion has been well known in the Hopf algebra case, it is still worth reviewing the construction for the weak Hopf algebra case. 

\begin{lemma}
    Let $p:U\times V\to \mathbb{C}$ be a skew-pairing. Then the map $U\to \hat{V}^{\rm cop}$, $u\mapsto \hat{u} := p(u,-)$ is a weak Hopf algebra morphism. Hence, if $p$ is non-degenerate, we have an embedding $U\hookrightarrow \hat{V}^{\rm cop}$. 
\end{lemma}

\begin{proof}
    This follows from the Eqs.~\eqref{eq:pairing-unit}-\eqref{eq:pairing-S}. Note that the structure morphisms of $\hat{V}^{\rm cop}$ are $(\mu_{\hat{V}},1_{\hat{V}},\Delta^{\rm op}_{\hat{V}} = \tau\comp \Delta_{\hat{V}}, \varepsilon_{\hat{V}}, S^{-1}_{\hat{V}})$, where $\tau(\sigma\otimes \sigma') = \sigma'\otimes \sigma$ is the swapping map. 
    Denote by $\Phi(u) = \hat{u}$ the map in the statement. 
    The first identity in \eqref{eq:pairing-unit} and \eqref{eq:pairing-left} imply that $\Phi(1_U) = 1_{\hat{V}}$ and $\Phi(uu') = \Phi(u)\Phi(u')$, and the second identity in \eqref{eq:pairing-unit} and \eqref{eq:pairing-right} imply that $\varepsilon_U = \varepsilon_{\hat{V}}\comp \Phi$ and $(\Phi\otimes \Phi)\comp\Delta_U= \Delta^{\rm op}_{\hat{V}}\comp\Phi$. 
    The identity \eqref{eq:pairing-S} says that $\Phi\comp S_U = S^{-1}_{\hat{V}}\comp\Phi$. Thus, $\Phi$ is a homomorphism of weak Hopf algebras. 
\end{proof}

Given a skew-pairing $p:U\times V\to \mathbb{C}$, it induces a canonical pairing $\bar{p}:U^{\rm cop}\times V \to \mathbb{C}$, $(u,v)\mapsto p(u,v)$. The pairing $\bar{p}$ introduces a generalized quantum double in the sense of Refs.~\cite{nikshych2003invariants,Jia2023weak,majid2000foundations}, which is built on the vector space $(U^{\rm cop})^{\rm cop}\otimes V$ modulo some ideal. This motivates the following construction of a generalized quantum double model.

First, $U^{\rm cop}$ has the following left and right $V$-module structures 
\begin{equation}
    v \rightharpoonup u := \sum_{(u)}u^{(1)}p(u^{(2)},v),\quad u \leftharpoonup v := \sum_{(u)}p(u^{(1)},v)u^{(2)},\quad u\in U^{\rm cop}, v\in V. 
\end{equation}
Here the comultiplication $\Delta(u) = \sum_{(u)}u^{(1)}\otimes u^{(2)}$ is taken in $U^{\rm cop}$. In what follows, we will adhere to this convention unless otherwise stated. This is a generalization of the Sweedler's arrow notation.

Then, consider the set $J$ of $U^{\rm cop}\otimes V$ generated by the elements 
\begin{equation}
        u\otimes xv - u(x\rightharpoonup 1_U)\otimes v,\quad u\otimes yv - u(1_U\leftharpoonup y)\otimes v,\quad \forall~ x\in V_L,\; \forall~y\in V_R. 
\end{equation}
Endow $U^{\rm cop}\otimes V$ with the multiplication
\begin{equation}
    (u\otimes v)(u'\otimes v') = \sum_{(v),(u')}(uu'^{(2)}\otimes v^{(2)}v')p(u'^{(1)},S^{-1}(v^{(3)}))p(u'^{(3)},v^{(1)}),
\end{equation}
where the comultiplication $\Delta(u')$ is taken in $U^{\rm cop}$. Then $J$ is a two-sided ideal. We note that the proof relies on the non-degeneracy of $p$.

\begin{theorem}\label{thm:QD-algebra}
        Given a non-generate skew-pairing $p:U\times V\to \mathbb{C}$ between weak Hopf algebras, define the following structure morphisms on $(U^{\rm cop}\otimes V)/J$:
    \begin{align}
        &1 = [1_U\otimes 1_V], \\
        &[u\otimes v][u'\otimes v'] = \sum_{(v),(u')}[uu'^{(2)}\otimes v^{(2)}v']p(u'^{(3)},v^{(1)})p(u'^{(1)},S^{-1}(v^{(3)})),  \\
        & \Delta([u\otimes v])  = \sum_{(u),(v)}[u^{(2)}\otimes v^{(1)}] \otimes [u^{(1)}\otimes v^{(2)}], \\
        & \varepsilon([u\otimes v]) = p(u,\varepsilon_R(S^{-1}(v))), \\
        & S([u\otimes v]) = \sum_{(u),(v)}[S(u^{(2)})\otimes S(v^{(2)})] p(u^{(1)},v^{(3)})p(u^{(3)},S^{-1}(v^{(1)})), 
    \end{align}
    where the comultiplications $\Delta(u)$ and $\Delta(u')$ are taken in $U^{\rm cop}$. Then $(U\otimes V)/J$ is equipped with a weak Hopf algebra structure. Here by $[u\otimes v]$ we denote the equivalence class of $u\otimes v\in U^{\rm cop}\otimes V$ in the quotient space.
\end{theorem}

\begin{proof}
    Since the skew-pairing $p:U\times V\to \mathbb{C}$ induces a canonical pairing $\bar{p}:U^{\rm cop}\times V\to \mathbb{C}$, the proof is exactly the same as that for the usual quantum double $D_{\bar{p}}(U^{\rm cop},V)$. For completeness, we include a proof in the appendix. Readers interested in the details may refer to Appendix~\ref{sec:app_QD} or Ref.~\cite{Jia2023weak}. 
\end{proof}

\begin{definition}[Generalized quantum double] \label{def:QD-weak}
    The weak Hopf algebra given by the theorem is called the generalized quantum double of $p:U\otimes V\to \mathbb{C}$, denoted by $D_p(U,V)=U\Join_p V$.
\end{definition}

\begin{remark} 
    (1) Let $W$ be a weak Hopf algebra. Then its dual $\hat{W}$ is also a weak Hopf algebra. Consider the pairing $p:\hat{W}\times W\to \mathbb{C}$, $(\varphi,w)\mapsto \varphi(w)$. Then the associated generalized quantum double $D_p(\hat{W},W)$ is just the quantum double $D(W)$ \cite{nikshych2003invariants,Jia2023weak}. 

    (2) Two factors $U^{\rm cop}, V$ can be embedded into $D_p(U,V)$ as weak Hopf subalgebras by 
    \begin{equation}
        i_U:U\hookrightarrow D_p(U,V), \quad i_V:V\hookrightarrow D_p(U,V),
    \end{equation}
    where $i_U(u) = [u\otimes 1_V]$, $i_V(v) = [1_U\otimes v]$. This is because it holds that 
    \begin{equation}
        [u\otimes 1_V][u'\otimes 1_V'] = [uu'\otimes 1_V], \quad [1_U\otimes v][1_U'\otimes v'] = [1_U\otimes vv'].  
    \end{equation}
\end{remark}

\begin{definition}
    A quasitriangular structure on a weak Hopf algebra $W$ is an element $R=\sum_jx_j\otimes y_j\in \Delta^{\mathrm{op}}(1)(W \otimes W)\Delta(1)$ such that the following hold: 
    \begin{itemize}
        \item[(1)] $(\id\otimes \Delta)(R) = R_{13}R_{12}$, $(\Delta\otimes \id)(R)= R_{13}R_{23}$.
        \item[(2)] $\Delta^{\mathrm{op}}(w)R = R\Delta(w)$, $\forall~w\in W$. 
        \item[(3)] There exists $\tilde{R}\in \Delta(1)(W\otimes W)\Delta^{\mathrm{op}}(1)$ such that $\tilde{R}R = \Delta(1)$, and $R\tilde{R} = \Delta^{\mathrm{op}}(1)$.  
    \end{itemize}
    Here, $R_{12} = \sum_j x_j\otimes y_j\otimes 1$, $R_{13} = \sum_jx_j\otimes 1\otimes y_j$ and $R_{23} = \sum_j 1\otimes x_j\otimes y_j$. 
\end{definition}

\begin{proposition} \label{prop:R-matrix}
    Suppose $p:U\times V\to \mathbb{C}$ is a non-degenerate skew-pairing. Then the associated generalized quantum double $D_p(U,V)$ has a quasitriangular structure 
    \begin{equation}
        R = \sum_i[1_U\otimes v_i]\otimes [u_i\otimes 1_V],\quad \tilde{R} = \sum_j [1_U\otimes S(v_j)]\otimes [u_j\otimes 1_V],
    \end{equation}
    where $\{u_i\}$ and $\{v_j\}$ are basis of $U$ and $V$ respectively, such that $p(u_i,v_j)=\delta_{i,j}$. 
\end{proposition}

\begin{proof}
    The proof is the same as that for the quantum double $D_{\bar{p}}(U^{\rm cop},V)$ as in Ref.~\cite{Jia2023weak}. 
\end{proof}

\subsection{Example of pairing for weak Hopf boundary tube algebras}

The main goal of this section is to interpret the weak Hopf tube algebra for domain wall as a generalized quantum double in the above sense. 
Throughout this section, we assume that $d_m^L=d_m^R=d_m$ for all $m\in \EM$.

Consider the bilinear map $p: \mathbf{Tube}_{\rm bd}(\EM_{\ED}) \times \mathbf{Tube}_{\rm bd}({_{\EC}}\EM)    \to \mathbb{C}$ defined by 
\begin{equation} \label{eq:pairing_def}
\begin{aligned}
    p\left(\begin{aligned}

    \end{aligned}\;. 
\end{aligned}
\end{equation}

\begin{proposition}
    The map $p$ defined above is a skew-pairing of weak Hopf algebras. 
\end{proposition}

\begin{proof}
    The identities in Eq.~\eqref{eq:pairing-unit} are easy to verify as computing as follows:  
    \begin{align*}
\sum_{s,v}\, 
    p\left(\begin{aligned}
    %
\end{aligned}\right). 
\end{align*}

Let us verify the Eq.~\eqref{eq:pairing-right}. In fact, on the one hand, applying F-move and loop move results in 
\begin{align*}
    &\quad p\left(\begin{aligned}
%
    \end{aligned} \\
    & = \delta_{z,w'}\delta_{y,x'} \delta_{s,w}\delta_{t,x}\delta_{u,y'}\delta_{v,z'} \frac{1}{d_s} \sum_{k,\sigma} \sqrt{\frac{d_k}{d_yd_b}} \sum_{c,\alpha,\beta} [F^{a'ub}_k]_{y\sigma\nu'}^{c\alpha\beta} \sum_{e,\tau,\lambda}[F_s^{ayb}]_{t\mu\nu}^{e\tau\lambda} \\
    & \quad \times \delta_{k,e}\delta_{\sigma,\lambda}\sqrt{\frac{d_yd_b}{d_k}}\delta_{c,v}\delta_{\beta,\gamma}\sqrt{\frac{d_ud_b}{d_v}}\delta_{k,z}\delta_{\alpha,\zeta'}\sqrt{\frac{d_{a'}d_v}{d_z}}\delta_{\tau,\zeta} \sqrt{\frac{d_ad_z}{d_s}}d_s \\ 
    & = \delta_{z,w'}\delta_{y,x'} \delta_{s,w}\delta_{t,x}\delta_{u,y'}\delta_{v,z'}  \sum_{\sigma}  [F^{a'ub}_z]_{y\sigma\nu'}^{v\zeta'\gamma} [F_s^{ayb}]_{t\mu\nu}^{z\zeta\sigma} \sqrt{\frac{d_bd_ud_{a'}d_a}{d_s}}\;. 
\end{align*}
On the other hand, using the comultiplication and the definition \eqref{eq:pairing_def} one gets 
\begin{align*}
&\quad \sum_{i,k,\sigma}\sqrt{\frac{d_k}{d_id_b}}\;
p\left(\begin{aligned}

    \end{aligned}\\
    & \quad \times \delta_{z,w'}\delta_{y,x'}\delta_{u,y'}\delta_{v,z'}\delta_{s,w}\delta_{t,x}\sqrt{\frac{d_z}{d_yd_b}}\frac{1}{d_zd_s} \\
    & = \delta_{z,w'}\delta_{y,x'}\delta_{u,y'}\delta_{v,z'}\delta_{s,w}\delta_{t,x}\sum_{\sigma} \sum_{c,\alpha,\beta} [F_z^{a'ub}]_{y\sigma\nu'}^{c\alpha\beta} \delta_{\beta,\gamma} \delta_{c,v}\sqrt{\frac{d_ud_b}{d_v}}\delta_{\alpha,\zeta'}\sqrt{\frac{d_{a'}d_v}{d_z}}d_z  \\ 
    & \quad \times \sum_{e,\tau,\lambda}[F_s^{ayb}]_{t\mu\nu}^{e\tau\lambda} \delta_{\lambda,\sigma} \delta_{e,z}\sqrt{\frac{d_yd_b}{d_z}}\delta_{\tau,\zeta}\sqrt{\frac{d_ad_z}{d_s}}d_s \sqrt{\frac{d_z}{d_yd_b}}\frac{1}{d_zd_s} \\
    & = \delta_{z,w'}\delta_{y,x'}\delta_{u,y'}\delta_{v,z'}\delta_{s,w}\delta_{t,x}\sum_{\sigma}[F_z^{a'ub}]_{y\sigma\nu'}^{v\zeta'\gamma}[F_s^{ayb}]_{t\mu\nu}^{z\zeta\sigma}\sqrt{\frac{d_ud_bd_{a'}d_a}{d_s}}\;,
    \end{align*}
    which is equal to the expression above, as expected. One uses a similar technique to show the Eq.~\eqref{eq:pairing-left}. 

    In remains to show the antipode identity \eqref{eq:pairing-S}. First, the left-hand side is equal to 
\begin{align*}
\begin{aligned}
    & p\left(S\left(\begin{aligned}

    \end{aligned} \\
    & = \frac{\delta_{u,w}\delta_{v,x}\delta_{s,y}\delta_{t,z}}{d_v} \sum_{t,\mu,\zeta}[F_s^{\bar{a}ub}]_{t\mu\zeta}^{t'\mu'\zeta'} \delta_{t',v}\delta_{\mu',\nu}\sqrt{\frac{d_ad_s}{d_v}}\delta_{\zeta',\gamma}\sqrt{\frac{d_vd_b}{d_u}}d_u \\ 
    & = \delta_{u,w}\delta_{v,x}\delta_{s,y}\delta_{t,z}[F_s^{\bar{a}ub}]_{t\mu\zeta}^{v\nu\gamma}  \frac{1}{d_v}\sqrt{d_ad_sd_bd_u}\;.
\end{aligned}
\end{align*}
Second, the right-hand side is equal to 
\begin{align*}
\begin{aligned}
    &p\left(\begin{aligned}
    %
    \end{aligned} \\
    & =  \frac{\delta_{s,y}\delta_{t,z}\delta_{u,w}\delta_{v,x}}{d_x} \sum_{t',\mu',\zeta'} [F_s^{\bar{a}ub}]_{t\mu\zeta}^{t'\mu'\zeta'} \delta_{t',v}\delta_{\zeta',\gamma}\sqrt{\frac{d_ud_b}{d_v}}\delta_{\mu',\nu}\sqrt{\frac{d_ad_v}{d_s}}d_s \\
    & = \delta_{s,y}\delta_{t,z}\delta_{u,w}\delta_{v,x} [F_s^{\bar{a}ub}]_{t\mu\zeta}^{v\nu\gamma} \frac{1}{d_x}\sqrt{d_ud_bd_ad_s}\;.
\end{aligned}
\end{align*}
Both expressions are identical. Hence the proof is completed. 
\end{proof}

\begin{remark} \label{rmk:general_pairing}
    One can also show that the map $p$ defined above is still a skew-pairing for general case where $d^L_m$ is not necessarily the same as $d^R_m$ for $m\in \EM$. See Appendix~\ref{app:general_pairing}. 
\end{remark}

Note that the skew-pairing $p$ defined by \eqref{eq:pairing_def} is non-degenerate, and hence introduces a generalized quantum double $\mathbf{Tube}_{\rm bd}(\EM_{\ED}) \Join_p \mathbf{Tube}_{\rm bd}({_{\EC}}\EM)$ by Theorem~\ref{thm:QD-algebra}. 
The following is the main result of this section, which interprets the domain wall tube algebra $\mathbf{Tube}({_\EC}\EM_\ED)$ as a Drinfeld quantum double algebra. 

\begin{theorem} \label{thm:QD_tube}
    The domain wall tube algebra $\mathbf{Tube}({_\EC}\EM_\ED)$ has a quantum double structure. 
    More precisely, we have the following isomorphism 
    \begin{equation}
        \mathbf{Tube}({_\EC}\EM_\ED) \simeq \mathbf{Tube}_{\rm bd}(\EM_{\ED}) \Join_p \mathbf{Tube}_{\rm bd}({_{\EC}}\EM)
    \end{equation}
    as weak Hopf algebras. 
\end{theorem}

Before delving into the proof, let us clarify some concepts involved. Denote $U:=\mathbf{Tube}_{\rm bd}(\EM_{\ED})$ and $V:=\mathbf{Tube}_{\rm bd}({_{\EC}}\EM)$ for simplicity. It is immediate from the definition that the left and right counital algebras of $V$ are the following 
\begin{align}
    V_L & = \operatorname{span}_{\mathbb{C}} \left\{X:=\sum_{m\in \Irr(\EM)} \begin{aligned}

\end{aligned} \; :\; y'\in\Irr(\EM) \right\}.  \label{eq:right-counit}
\end{align}
Therefore, the actions reads 
\begin{equation}
        X\rightharpoonup 1_U  = \sum_s\;\begin{aligned}
    %
\end{aligned}$. Then one has 
\begin{align}
    Z\otimes XW - Z(X\rightharpoonup 1_U)\otimes W & = \delta_{w,x'}Z\otimes W - \delta_{u,x'}Z\otimes W, \label{eq:ideal-1} \\
    Z\otimes YW - Z(1_U\leftharpoonup Y)\otimes W & = \delta_{t,y'}Z\otimes W - \delta_{x,y'}Z\otimes W. \label{eq:ideal-2}
\end{align}

\begin{proof}[Proof of Theorem~\ref{thm:QD_tube}]
    Define a map $\Phi:\mathbf{Tube}_{\rm bd}(\EM_{\ED}) \otimes \mathbf{Tube}_{\rm bd}({_{\EC}}\EM) \to \mathbf{Tube}({_\EC}\EM_\ED)$ by 
    \begin{equation}
        \Phi\left(\begin{aligned}
    %
    \end{aligned}\;. 
    \end{equation}
    By Eqs.~\eqref{eq:ideal-1} and \eqref{eq:ideal-2}, it follows that 
    \begin{align*}
        \Phi(Z\otimes XW - Z(X\rightharpoonup 1_U)\otimes W) & = (\delta_{w,x'} - \delta_{u,x'})\delta_{u,w}\delta_{t,x}\; \begin{aligned}%
    \end{aligned}=0\;. 
    \end{align*}
    Thus, $\Phi$ passes to the generalized quantum double 
    \begin{equation}
        \Phi:\mathbf{Tube}_{\rm bd}(\EM_{\ED}) \Join_p \mathbf{Tube}_{\rm bd}({_{\EC}}\EM)\to \mathbf{Tube}({_\EC}\EM_\ED). 
    \end{equation}

    Now we will show that $\Phi$ is an isomorphism of weak Hopf algebras. To this end, let us compute the structure morphisms of the quantum double. First, for the algebra structure, we have 
\begin{align*}
    [Z\otimes& W][Z'\otimes W'] \\
    & = \sum_{(Z'),(W)}[ZZ'^{(2)}\otimes W^{(2)}W'] p(Z'^{(3)},W^{(1)}) p(Z'^{(1)},S^{-1}(W^{(3)})) \\
    & = \sum_{i,k,\sigma,j,l,\tau}\sum_{i',k',\sigma',j',l',\tau'}  \sqrt{\frac{d_k}{d_ad_i}}\sqrt{\frac{d_l}{d_ad_j}} \sqrt{\frac{d_{k'}}{d_{i'}d_{b'}}}\sqrt{\frac{d_{l'}}{d_{j'}d_{b'}}} \\
    & \quad  \quad \frac{\delta_{v',z}\delta_{u',i}\delta_{i',k}\delta_{k',w}}{d_{k'}}\;\begin{aligned}

    \end{aligned} .
\end{align*}
    We emphasize that the comultiplication $\Delta(Z')$ is taken in $\mathbf{Tube}_{\rm bd}(\EM_{\ED})^{\rm cop}$. 
    Using F-move, one computes 
    \begin{align*}
    \begin{aligned}
        %
    \end{aligned} = [(F_{ajb'}^x)^{-1}]_{l\tau\tau'}^{s'\mu'\nu}\sqrt{d_ad_xd_jd_{b'}}\;. 
\end{align*}
    It follows that 
    \begin{align*}
    \Phi([Z\otimes & W][Z'\otimes W']) \\
    & = \sum_{i,k,\sigma,j,l,\tau}\sum_{i',k',\sigma',j',l',\tau'}  \sqrt{\frac{d_k}{d_ad_i}}\sqrt{\frac{d_l}{d_ad_j}} \sqrt{\frac{d_{k'}}{d_{i'}d_{b'}}}\sqrt{\frac{d_{l'}}{d_{j'}d_{b'}}} \frac{d_y}{d_x}\\
    & \quad  \quad \frac{\delta_{v',z}\delta_{u',i}\delta_{i',k}\delta_{k',w}}{d_{k'}}[F_{aib'}^{k'}]_{z\gamma'\zeta}^{k\sigma\sigma'} \sqrt{d_ad_id_{b'}d_{k'}}\\
    & \quad \quad \frac{\delta_{l',x}\delta_{j',l}\delta_{t',j}\delta_{s',y}}{d_{s'}}[(F_{ajb'}^x)^{-1}]_{l\tau\tau'}^{s'\mu'\nu}\sqrt{d_ad_xd_jd_{b'}} \\
    & \quad \quad \delta_{k',u}\delta_{l',t} \delta_{i,w'}\delta_{j,x'}\delta_{i',k}\delta_{j',l} \; \begin{aligned}

    \end{aligned} \\
    & = \sum_{i,k,\sigma,j,l,\tau}\sum_{i',k',\sigma',j',l',\tau'}  \delta_{k',u}\delta_{l',t} \delta_{i,w'}\delta_{j,x'}\delta_{i',k}\delta_{j',l}  \\ 
    & \quad \quad  \delta_{v',z}\delta_{u',i}\delta_{i',k}\delta_{k',w}[F_{aib'}^{k'}]_{z\gamma'\zeta}^{k\sigma\sigma'} \delta_{l',x}\delta_{j',l}\delta_{t',j}\delta_{s',y}[(F_{ajb'}^x)^{-1}]_{l\tau\tau'}^{s'\mu'\nu} \\
    & \quad \quad \sum_{k'',\kappa,\sigma''} [(F_{aw'b'}^u)^{-1}]_{k\sigma\sigma'}^{k''\kappa\sigma''} \sum_{l'',\theta,\tau''} [(F_t^{ax'b'})^{-1}]_{l\tau\tau'}^{l''\theta\tau''} \;  \begin{aligned}

    \end{aligned}\\
    & =   \sum_{k'',\kappa,\sigma''}  \underbrace{\sum_{k,\sigma,\sigma'}[(F_{aw'b'}^u)^{-1}]_{k\sigma\sigma'}^{k''\kappa\sigma''} [F_{aw'b'}^{u}]_{z\gamma'\zeta}^{k\sigma\sigma'}}_{\delta_{k'',z}\delta_{\kappa,\gamma'}\delta_{\sigma'',\zeta}}\sum_{l'',\theta,\tau''} \underbrace{\sum_{l,\tau,\tau'}[(F_x^{ax'b'})^{-1}]_{l\tau\tau'}^{l''\theta\tau''}[(F_{ax'b'}^x)^{-1}]_{l\tau\tau'}^{s'\mu'\nu}}_{\delta_{l'',s'}\delta_{\theta,\mu'}\delta_{\tau'',\nu}}   \\
    & \quad \quad \delta_{v',z}\delta_{s',y}\delta_{u,w} \delta_{t,x}\delta_{u',w'} \delta_{t',x'} \;  \begin{aligned}

    \end{aligned} = \Phi([Z\otimes W])\Phi([Z'\otimes W']). 
\end{align*}
    Thus, $\Phi$ preserves the multiplication. 
    Also, it is easy to see that $\Phi$ preserves the identity:
    \begin{equation*}
    \Phi(1_{U\Join_pV})=\Phi\left(\sum_{s,v}\begin{aligned}
    %
    \end{aligned}\; = 1_{\mathbf{Tube}({_\EC}\EM_\ED)}.  
\end{equation*}
Second, for the coalgebra structure, as 
\begin{align*}
    \Delta([Z\otimes W]) & = \sum_{(Z),(W)}[Z^{(2)}\otimes W^{(1)}]\otimes [Z^{(1)}\otimes W^{(2)}] \\ 
    & = \sum_{i,k,\sigma} \sum_{j,l,\tau}\sqrt{\frac{d_k}{d_id_b}}\sqrt{\frac{d_l}{d_ad_j}}\left( \begin{aligned}
    %
    \end{aligned}   \\
    & = \Delta\comp \Phi([Z\otimes W]). 
\end{align*}
And $\Phi$ also preserves the counit: 
\begin{align*}
    \varepsilon_{U\Join_p V}([Z\otimes W]) & = p(Z,\varepsilon_R(S^{-1}(W))) \\
    & = p\left(\begin{aligned}
    %
    \end{aligned} \\
    & = \delta_{u,w}\delta_{t,x} \delta_{z,y}\delta_{t,u}\delta_{s,v}\delta_{\zeta,\nu}\delta_{\gamma,\mu}\sqrt{\frac{d_ad_zd_b}{d_s}} \\
    & = \varepsilon_{\mathbf{Tube}({_\EC}\EM_\ED)}\comp \Phi([Z\otimes W]). 
\end{align*}
    Finally, let us compute the antipode morphism. By definition of antipode of the quantum double, one obtains
\begin{align*}
    S([Z&\otimes W]) \\
    & = \sum_{(Z),(W)}[S(Z^{(2)})\otimes S(W^{(2)})] p(Z^{(1)},W^{(3)})p(Z^{(3)},S^{-1}(W^{(1)})) \\
    & = \sum_{i,k,\sigma,j,l,\tau} \sum_{m,n,\rho,c,e,\theta} \sqrt{\frac{d_k}{d_id_b}} \sqrt{\frac{d_l}{d_jd_b}} \sqrt{\frac{d_m}{d_ad_n}} \sqrt{\frac{d_c}{d_ad_e}} \\
    &\quad \quad   \frac{\delta_{l,e}\delta_{j,y}\delta_{t,x}\delta_{s,c}}{d_s} \;\begin{aligned}

\end{aligned} . 
\end{align*}
Applying F-move to the pairing diagrams, one gets 
\begin{align*}
    \begin{aligned}
        %
    \end{aligned} = [F_z^{\bar{a}v\bar{b}}]_{w\zeta\gamma}^{n\sigma\rho}\sqrt{d_ad_vd_zd_b}\;.  
\end{align*}
Plugging in these values and applying $\Phi$, we finally arrive at the following: 
\begin{align*}
    \Phi\comp & S([Z\otimes W]) \\ 
    & = \delta_{t,x}\delta_{u,w}\sum_{\sigma,\tau} \sum_{n,\rho,e,\theta}\frac{1}{d_ad_vd_b}\sqrt{\frac{d_z}{d_yd_sd_v}} \\
    &\quad \quad   \begin{aligned}
        %
    \end{aligned} \\ 
    & = \delta_{t,x}\delta_{u,w} \frac{d_z}{d_v} \sum_{e',\theta',\tau'}\underbrace{\sum_{e,\theta,\tau}[(F_{\bar{a}s\bar{b}}^y)^{-1}]_{e\theta\tau}^{e'\theta'\tau'} [(F_y^{\bar{a}s\bar{b}})^{-1}]_{e\theta\tau}^{t\mu\nu}}_{\delta_{e',t}\delta_{\theta',\mu}\delta_{\tau',\nu}} \\
    & \quad \quad \sum_{n',\sigma',\rho'} \underbrace{\sum_{n,\sigma,\rho}[F_z^{\bar{a}v\bar{b}}]_{n\sigma\rho}^{n'\sigma'\rho'}[(F_z^{\bar{a}v\bar{b}})^{-1}]_{w\zeta\gamma}^{n\sigma\rho}}_{\delta_{n',w}\delta_{\sigma',\zeta}\delta_{\rho',\gamma}}\;\begin{aligned}
    \end{aligned} \; = S\comp \Phi([Z\otimes W]). 
\end{align*}
In conclusion, we have shown that $\Phi$ is a morphisms of weak Hopf algebras. It remains to show that $\Phi$ is bijective. Clearly, $\Phi$ is surjective. To see it is injective, if $\Phi([Z\otimes W]) = 0$, then either $u\neq w$  or $t\neq x$. If $u\neq w$, say, choose $x'=w$ for $X\in V_L$ in \eqref{eq:left-counit}, then $Z\otimes W = Z\otimes XW - Z(X\rightharpoonup 1_U)\otimes W$ by Eq.~\eqref{eq:ideal-1}, which means $[Z\otimes W] = 0$ in the quantum double. Therefore, we finish the proof.  
\end{proof}

\begin{corollary}
    The quasitriangular structure of the the domain wall tube algebra $\mathbf{Tube}({_\EC}\EM_\ED)$ is given by 
\begin{align}
        R & =\sum_{a,b,s,t,u,v}\sum_{\mu,\nu,\gamma,\zeta} \frac{1}{[F_s^{aub}]_{t\mu\nu}^{v\zeta\gamma}}\sqrt{\frac{d_s}{d_ad_ud_b}} \; \begin{aligned}
    \end{aligned}\;.
\end{align}
\end{corollary}

\begin{proof}
    This follows from Theorem~\ref{thm:QD_tube} and Proposition~\ref{prop:R-matrix}. 
\end{proof}

%\subsection{Remark on Haar integral}

The Haar integral $\mathfrak{I}_{\rm bd}^L$ of  $\mathbf{Tube}_{\rm bd}({_{\EC}}\EM)$ is given in Eq.~\eqref{eq:bdHaarL}, namely
\begin{equation} 
   \mathfrak{I}_{\rm bd}^L =\frac{1}{\operatorname{rank} \EM} \sum_{a,x,z,\nu} \frac{1}{d_z}\sqrt{\frac{d_a}{d_xd_z}} \;\;
    \begin{aligned}%
    \end{aligned}\;.
\end{equation}
Similarly, the Haar integral of  $\mathbf{Tube}_{\rm bd}(\EM{_{\ED}})$ is given by
\begin{equation} \label{eq:bdHaarR}
   \mathfrak{I}_{\rm bd}^R =\frac{1}{\operatorname{rank} \EM} \sum_{b,s,t,\mu} \frac{1}{d_t}\sqrt{\frac{d_b}{d_sd_t}} \;\;
    \begin{aligned}%
    \end{aligned}\;.
\end{equation}
From the skew-pairing between the left and right boundary tube algebras, the relationship between their respective Haar integrals can be summarized in the following result:

\begin{proposition}
%1. The Haar integrals $\mathfrak{I}_{\mathrm{bd}}^L \in \mathbf{Tube}_{\mathrm{bd}}({_{\EC}}\EM)$ and $\mathfrak{I}_{\mathrm{bd}}^R \in \mathbf{Tube}_{\mathrm{bd}}(\EM{_{\ED}})$ form a dual integral pair under the skew-pairing introduced in Eq.~\eqref{eq:pairing_def}. 
%Namely, $\langle \bullet,\mathfrak{I}_{\mathrm{bd}}^L\rangle_p$ is non-degenerate functional over $\mathbf{Tube}_{\mathrm{bd}}(\EM{_{\ED}})$ and $\langle \mathfrak{I}_{\mathrm{bd}}^R, \bullet\rangle_p$ is non-degenerate functional over $\mathbf{Tube}_{\mathrm{bd}}({_{\EC}}\EM)$, and they satisfy
% \begin{equation} \label{eq:dual_haar_LR}
%     \mathfrak{I}_{\mathrm{bd}}^L \rightharpoonup  \mathfrak{I}_{\mathrm{bd}}^R = 1_R, \quad \mathfrak{I}_{\mathrm{bd}}^R \rightharpoonup  \mathfrak{I}_{\mathrm{bd}}^L =1_L,
% \end{equation}
% where $1_R$ and $1_L$ are unit of right and left boundary tube algebras.

The boundary tube Haar integrals are compatible with the Haar integral $\mathfrak{I}$ of the domain wall tube algebra $\Tube({_{\EC}}\EM_{\ED})$ in the sense that
$\Phi([\mathfrak{I}_{\mathrm{bd}}^R \otimes \mathfrak{I}_{\mathrm{bd}}^L]) = \mathfrak{I}$, where $\Phi$ is the isomorphism in Theorem~\ref{thm:QD_tube}.
\end{proposition}

\begin{proof}
    % 1. The non-degeneracy of $\langle \bullet,\mathfrak{I}_{\mathrm{bd}}^L\rangle_p$ and $\langle \mathfrak{I}_{\mathrm{bd}}^R, \bullet\rangle_p$ is clear, as the pairing $p$ is non-degenerate. It remains to show Eq.~\eqref{eq:dual_haar_LR}. 

Recall that the Haar integral $\mathfrak{I}$ of $\Tube({_{\EC}}\EM_{\ED})$ is given  by Eq.~\eqref{eq:haar_integral_2}. 
    It follows that 
\begin{align*}
    \Phi([\mathfrak{I}^R_{\rm bd}\otimes \mathfrak{I}^L_{\rm bd}]) & = \frac{1}{(\operatorname{rank} \EM)^2} \sum_{b,s,t,\mu} \frac{1}{d_t}\sqrt{\frac{d_b}{d_sd_t}} \sum_{a,x,z,\nu} \frac{1}{d_z}\sqrt{\frac{d_a}{d_xd_z}} \;\;\Phi\left(
    \begin{aligned}
    \end{aligned} \; = \mathfrak{I},
\end{align*} 
as expected. 
\end{proof}

\section{The representation category of domain wall tube algebra}
\label{sec:RepTube}

For a gapped domain wall ${_{\EC}}\EM_{\ED}$ separating two $2d$ topological orders, the excitations within the domain wall can undergo fusion and splitting, but cannot braid with one another, as the domain wall represents a $1d$ phase.
This means that the topological excitations in the domain wall form a fusion category $\EW$ \cite{Kitaev2012boundary}.

If we consider the $2d$ bulk excitations, they can be viewed as a domain wall ${_{\EC}}\EC_{\EC}$ of string-net model, where the two bulk phases connected by the domain wall are identical, and the domain wall shares the same input data as the bulk phase. In this case, the domain wall excitations correspond to the bulk excitations, and these excitations form a unitary modular tensor category \cite{Kitaev2003,KITAEV2006,Levin2005,simon2023topological,pachos2012introduction,wang2010topological}.

Several methods reveal that the algebraic structure underlying these domain wall excitations is a weak Hopf algebra \cite{jia2024weakTube}. When realizing the $2d$ topological phase using the quantum double model \cite{Kitaev2003,Buerschaper2013a,jia2023boundary,Jia2023weak}, the input data is a weak Hopf gauge symmetry $W$, and the topological excitations are characterized by representations of the quantum double $D(W)$, which can be interpreted as the charge symmetry.
Alternatively, in the string-net model realization of the topological phase, the topological excitations correspond to representations of the tube algebra \cite{lan2014topological,ocneanu1994chirality,ocneanu2001operator}. In Ref.~\cite{jia2024weakTube}, we demonstrate that the bulk tube algebra can be endowed with a weak Hopf algebra structure. For bulk excitations, the underlying weak Hopf charge symmetry possesses a quantum double structure, which implies the existence of an $R$-matrix. This $R$-matrix enables the derivation of the braiding structure of these excitations.

For a general domain wall, the domain wall phase is only a unitary fusion category, meaning there is no braiding between the domain wall excitations. However, the underlying algebraic structure remains a weak Hopf algebra. In the string-net model realization, this corresponds to the domain wall tube algebra we discussed in the previous sections.

In this section, we will demonstrate how the tube algebra can be used to characterize the domain wall excitations, including their charge types, fusion and splitting rules, quantum dimensions, and other properties. For simplicity, we assume that $d_m^L=d_m^R=d_m$ for all $m\in \EM$ throughout. The main result of this section is as follows:

\begin{theorem}\label{thm:equivalence}
The domain wall excitations can be equivalently characterized either by $\EC|\ED$ bimodule category endofunctors in $\Fun_{\EC|\ED}(\EM, \EM)$ or by representations of the domain wall tube algebra $\Tube(_{\EC}\EM_{\ED})$. More precisely, there exists a monoidal functor
\begin{equation}
F: \Fun_{\EC|\ED}(\EM, \EM) \to \Rep(\Tube(_{\EC}\EM_{\ED})),
\end{equation}
which preserves the topological charges, vacuum sector, and fusion rules.

% and the Frobenius-Perron dimensions of the two categories satisfy
% \begin{equation}
% \FPdim \Rep(\Tube(_{\EC}\EM_{\ED})) \geq \FPdim \Fun_{\EC|\ED}(\EM, \EM).
% \end{equation}
\end{theorem}

\subsection{Domain wall excitation as module functors}

The domain wall topological excitations of string-net models can be understood as $\EC|\ED$ bimodule category endofunctors of $\EM$ \cite{Kitaev2012boundary}. We will now provide a detailed discussion of this perspective as a preparation for constructing a proof of Theorem~\ref{thm:equivalence}.

Since domain wall phase is a $1d$ phase, its excitations form a unitary fusion category. To be more precise, since $\EM$ is an Abelian category, for which we can define direct sum $m\oplus n$, this induces an Abelian category structure over $\Fun_{\EC|\ED}(\EM,\EM)$. For two functors $\mathfrak{f},\mathfrak{g}$, their direct sum $\mathfrak{f}\oplus \mathfrak{g}$ characterizes the superposition of domain wall anyons.

It is convenient to use the Deligne tensor product $\EX \boxtimes \EY$ of fusion categories $\EX$ and $\EY$. The $\EC|\ED$ bimodule category can be equivalently regarded as a $\EC \boxtimes \ED^{\rm rev}$ left module category:
\begin{equation}
    {_{\EC}}\EM_{\ED} \leftrightarrow  {_{\EC \boxtimes \ED^{\rm rev}}}\EM.
\end{equation}
This is known as the folding trick for the string-net model of topological phases discussed in Section~\ref{sec:multi_tube}. The $\EC|\ED$ bimodule functors can also be interpreted as $\EC \boxtimes \ED^{\rm rev}$ left module functors. Thus, the domain wall excitations are characterized by $\EC \boxtimes \ED^{\rm rev}$ left module functors $\mathfrak{f} \in \Fun_{\EC \boxtimes \ED^{\rm rev}}(\EM,\EM)$.

For a fusion category $\EX$ and module categories $\EM$ and $\EN$ over $\EX$, an $\EX$-module functor consists of an (additive) functor $\mathfrak{f}: \EM \to \EN$ and natural isomorphisms
\begin{equation} \label{eq:bimoduleAsso}
    \alpha_{X,M,\mathfrak{f}}:  \mathfrak{f} (X \otimes M) \to X \otimes \mathfrak{f}(M).
\end{equation}
If we denote $\mathfrak{f}(M)$ as $M \otimes \mathfrak{f}$, then the map $\alpha_{X,M,\mathfrak{f}}:  (X \otimes M) \otimes \mathfrak{f}  \to X \otimes (M \otimes \mathfrak{f})$ resembles an associator. The axioms for these natural isomorphisms becomes: 

(i) Pentagon relation (the following diagram is commutative for all $X,Y\in \EX, M\in \EM$):
\begin{equation}
\begin{aligned}
      \resizebox{0.8\textwidth}{!}{ % Adjust width here (0.5\textwidth for half the column width)
\begin{tikzpicture}[node distance=2cm, every node/.style={align=center}]
    % Nodes
    \node (A) {$((X \otimes Y)\otimes M)\otimes \fk$};
    \node (B) [right=of A] {$(X \otimes (Y\otimes M))\otimes \fk$};
    \node (C) [below right=of B] {$X \otimes ((Y\otimes M)\otimes \fk)$};
    \node (D) [below left=of C] {$X \otimes( Y\otimes (M\otimes \fk))$};
    \node (E) [left=of D] {$(X \otimes Y)\otimes (M\otimes \fk)$};
    % Arrows
    \draw[->] (A) -- node[above] {$\fk(a_{X,Y,M})$} (B);
    \draw[->] (B) -- node[right] {$\alpha_{X,Y\otimes M,\fk}$} (C);
    \draw[->] (C) -- node[below] {$\alpha_{Y,M,\fk}$} (D);
    \draw[->] (E) -- node[below] {$a_{X,Y,\fk(M)}'$} (D);
    \draw[->] (A) -- node[right] {$\alpha_{X\otimes Y, M,\fk}$} (E);
\end{tikzpicture}}
\end{aligned}
\end{equation}
where $a_{\bullet,\bullet,\bullet}$ and $a_{\bullet,\bullet,\bullet}'$ are the associators for the left $\EX$-modules $\EM$ and $\EN$, respectively.

(ii) Triangle relation (the following diagram is commutative for all $M\in \EM$)
\begin{equation}
\begin{aligned}
      \resizebox{0.5\textwidth}{!}{ % Adjust width here (0.5\textwidth for half the column width)
\begin{tikzpicture}[node distance=1.3cm, every node/.style={align=center}]
    % Nodes
    \node (A) {$(\mathbb{1}\otimes M)\otimes \fk$};
    \node (B) [right=of A] {$\mathbb{1} \otimes (M\otimes \fk)$};
    \node (C) [below right=of B] {$M \otimes \fk$};
    % Arrows
    \draw[->] (A) -- node[above] {$\alpha_{\mathbb{1},M,\fk}$} (B);
    \draw[->] (B) -- node[above] {$l_{\fk(M)}'$} (C);
    \draw[->] (A) -- node[below] {$\fk(l_M)$} (C);
\end{tikzpicture}}
\end{aligned}
\end{equation}
where $l_\bullet$ and $l_\bullet'$ are the left unit isomorphisms for $\EM$ and $\EN$, respectively.

Note that for any left $\EX$-module category $\EM$, the endofunctor category $\Fun_{\EX}(\EM,\EM)$ forms a monoidal category, with the tensor product defined as follows:
\begin{equation}
\mathfrak{f} \otimes \mathfrak{g} := \mathfrak{g} \circ \mathfrak{f},
\end{equation}
and the tensor unit is given by $\mathbb{1} = \id_{\EM}$. The associator isomorphism and the left and right unit isomorphisms are all identities, since $(\fk \circ \gk) \circ \hk = \fk \circ (\gk \circ \hk)$ and $\fk \circ \id_{\EM} = \fk = \id_{\EM} \circ \fk$ for all $\fk, \gk, \hk$.
We equip $\EM$ with a right $\Fun_{\EX}(\EM,\EM)$-action as follows: $M \otimes \fk := \fk(M)$. Since $(M \otimes \fk) \otimes \gk = \gk \circ \fk(M)$ and $M \otimes (\fk \otimes \gk) = \gk \circ \fk(M)$, we define the associator as $\alpha_{M,\fk,\gk} = \id_{\gk \circ \fk(M)}$ and the right unit isomorphism as $r_{M} = \id_{M}$. This construction clearly establishes $\EM$ as a right $\Fun_{\EX}(\EM,\EM)$-module category.
The following result is well known in the mathematical literature:

\begin{proposition}[See, e.g., Ref.~\cite{etingof2016tensor}]\label{prop:bimodule}
    For any left $\EX$-module category $\EM$, the above construction makes $\EM$ a $\EX|\Fun_{\EX}(\EM,\EM)$ bimodule category.
\end{proposition}

\begin{proof}
Notice that $\alpha_{X,M,\fk}$ in Eq.~\eqref{eq:bimoduleAsso} provides a bimodule middle associator. The pentagon and triangle relations for it are precisely the middle associativity constraints of a bimodule category, related to the left $\EX$-module structure. For the right $\Fun_{\EX}(\EM,\EM)$-action, since the module associators are identities, the middle associativity constraint is automatically satisfied.
\end{proof}

When applied to the domain wall bimodule category ${_{\EC}}\EM_{\ED}$, we first use the folding trick to map it to a left module category ${_{\EC \boxtimes \ED^{\rm rev}}}\EM$. Then, applying Proposition~\ref{prop:bimodule}, we obtain an invertible bimodule category ${_{\EC \boxtimes \ED^{\rm rev}}}\EM_{\Fun_{\EC|\ED}(\EM,\EM)}$.

%\subsubsection{Excitation realized by Hamiltonian}

\subsection{Domain wall excitation as representation of tube algebra}

The domain wall excitation can be regarded as a disk region on the domain wall (see Figure \ref{fig:SNlattice}).
The corresponding disk space is 
\begin{equation}
    \mathbf{Disk}({{_\EC}\EM_{\ED}}) 
  %  = \bigoplus_{s,t\in\Irr(\EM)} \Hom_{\EM}(s,t) 
    = \mathrm{span}_{\mathbb{C}}\left\{\begin{aligned}

    \end{aligned} : s,t\in \Irr(\EM), \Psi\in \Hom_{\EM}(s,t)\right\}.  
\end{equation}
This disk region can be equipped with a left action of the domain wall tube algebra $\TubeM$ as follows
\begin{equation}\label{eq:TubeModule}
    \begin{aligned}%
    \end{aligned}\;. 
\end{equation}
It is straightforward to verify that the unit of $\TubeM$ acts as the identity, and that the action satisfies associativity.

Since $\TubeM$ is a finite-dimensional complex weak Hopf algebra, it is useful to first review some basics of the representation theory of weak Hopf algebras before discussing its representations.
A representation of a weak Hopf algebra $W$ is an algebra map $\Gamma: W \to \End(V)$, where $V$ is a complex vector space. The dimension of $V$ is referred to as the dimension of the representation. An irreducible representation is one that does not have any non-trivial sub-representations. A representation is called indecomposable if it is non-zero and cannot be expressed as a direct sum of two non-zero sub-representations.
Since we have established that the domain wall tube algebra is a finite-dimensional $C^*$ algebra, in particular semisimple, it follows that the notions of irreducible representations and indecomposable representations are equivalent.

For the category $\Rep(W)$ of finite-dimensional representations of $W$, the left counital subalgebra $W_L$ is the tensor unit with $W$-action given by
\begin{equation}\label{eq:tubeunitL}
    x \triangleright a = \varepsilon_L(xa) =\sum_{(x)}x^{\cone} a S(x^{\ctwo}), \;\forall~x \in W, \; \forall~a\in W_L.
\end{equation}
For two representations $V,U$, the action of $W$ on $V\otimes_{\Cbb} U$ is defined as
\begin{equation}
    h \triangleright (v\otimes w):= \sum_{(h)}(h^{\cone } \triangleright v) \otimes (h^{\ctwo}\triangleright w),
\end{equation}
and the tensor product of $V$ and $W$ is defined as
\begin{equation}
    V\otimes U = \{ x\in V\otimes_{\Cbb} U| \Delta(1)\triangleright x =x  \}.
\end{equation}
The dual representation of $V$ is $\hat{V}:=\Hom(V,\Cbb)$ equipped with the action $(h\triangleright f)(x):=f(S(h)\triangleright x)$.

%\subsubsection{Schur orthogonality relation}

For an endofunctor $\mathfrak{f}\in \Fun_{\EC|\ED}(\EM,\EM)$, we can introduce the following vector spaces 
\begin{equation}
    V_{\mathfrak{f}}:= \bigoplus_{x,y\in\Irr(\EM)}\Hom_{\EM}(\mathfrak{f}(x),y),\quad V^{\mathfrak{f}}:= \bigoplus_{x,y\in\Irr(\EM)}\Hom_{\EM}(x,\mathfrak{f}(y)). 
\end{equation}
Since we can denote $\mathfrak{f}(x)$ as $x\otimes \mathfrak{f}$, elements in $V_{\mathfrak{f}}$ and $V^{\mathfrak{f}}$ can be represented by diagrams as 
\begin{equation}
    \begin{aligned}
        \begin{tikzpicture}[scale=0.85]
            % \draw[white, fill=gray, fill opacity=0.2] (-0.5,-0.81) rectangle (0.81,0.81);
            \filldraw[black!60, fill=gray!15, dotted] (0,0) circle[radius=0.8];
            \draw[black, ->-=0.6, line width=.6pt] (0,0) -- (0,0.8); 
            \draw[black, ->-=0.6, line width=.6pt] (0,-0.8) -- (0,0); 
            \draw[black, ->-=0.6, line width=.6pt] (0.8,0) -- (0,0); 
            \node[ line width=0.6pt, dashed, draw opacity=0.5] (a) at (-0.2,0.5){$\scriptstyle y$};
            \node[ line width=0.6pt, dashed, draw opacity=0.5] (a) at (-0.2,-0.6){$\scriptstyle x$};
            \node[ line width=0.6pt, dashed, draw opacity=0.5] (a) at (0.6,-0.3){$\scriptstyle \mathfrak{f}$};
            \node[ line width=0.6pt, dashed, draw opacity=0.5] (a) at (-0.3,0){$\scriptstyle \alpha$};
            \node[ line width=0.6pt, dashed, draw opacity=0.5] (a) at (0,0){$\scriptstyle \bullet$};
        \end{tikzpicture}
    \end{aligned}\;\in V_{\mathfrak{f}},\;\;
    \begin{aligned}
        \begin{tikzpicture}[scale=0.85]
            % \draw[white, fill=gray, fill opacity=0.2] (-0.5,-0.81) rectangle (0.81,0.81);
            \filldraw[black!60, fill=gray!15, dotted] (0,0) circle[radius=0.8];
            \draw[black, ->-=0.6, line width=.6pt] (0,0) -- (0,0.8); 
            \draw[black, ->-=0.6, line width=.6pt] (0,-0.8) -- (0,0); 
            \draw[black, ->-=0.6, line width=.6pt] (0,0) -- (0.8,0); 
            \node[ line width=0.6pt, dashed, draw opacity=0.5] (a) at (-0.2,0.5){$\scriptstyle y$};
            \node[ line width=0.6pt, dashed, draw opacity=0.5] (a) at (-0.2,-0.6){$\scriptstyle x$};
            \node[ line width=0.6pt, dashed, draw opacity=0.5] (a) at (0.6,0.3){$\scriptstyle \mathfrak{f}$};
            \node[ line width=0.6pt, dashed, draw opacity=0.5] (a) at (-0.3,0){$\scriptstyle \alpha$};
            \node[ line width=0.6pt, dashed, draw opacity=0.5] (a) at (0,0){$\scriptstyle \bullet$};
        \end{tikzpicture}
    \end{aligned}\;\in V^{\mathfrak{f}}.
\end{equation}
In this subsection, we will proceed with $V^{\mathfrak{f}}$, and note that the arguments will apply to $V_{\mathfrak{f}}$ as well. As before, arrows of strings will be omitted sometimes for avoiding clustering of equation. 

To avoid further complications, we fix a gauge in which the Frobenius–Perron dimension of the domain wall string is identical under all actions.
Both $V_{\mathfrak{f}}$ and $V^{\mathfrak{f}}$ are modules over $\Tube(_{\EC}\EM_{\ED})$ by Eq.~\eqref{eq:TubeModule}. More precisely, the representation $\rho_{\mathfrak{f}}:\Tube(_{\EC}\EM_{\ED})\to \End(V^{\mathfrak{f}})$ is defined by 
    \begin{align}
        & \begin{aligned}
    \end{aligned}\;, \label{eq:tube_action}
    \end{align}
    where we have used the F-move and loop move twice.

The vector space $V^\mathfrak{f}$ has the inner product defined by 
\begin{equation} \label{eq:inner_prod}
    \Big{\langle} \begin{aligned}
        %
    \end{aligned} \!\!\!\!\!\! = \delta_{x,x'}\delta_{y,y'}\delta_{\alpha,\alpha'} \sqrt{d_xd_yd_\mathfrak{f}}.
\end{equation}
The labels $(\bullet)'$ after the first equality should be understood as the duals of the original $(\bullet)'$. With respect to this inner product, the space $V^\mathfrak{f}$ has an orthonormal basis $\{|x\xrightarrow{\alpha} y,\mathfrak{f}\rangle:x,y,\alpha\}$, where   
\begin{equation}
    |x\xrightarrow{\alpha} y,\mathfrak{f}\rangle := \left(d_xd_yd_\mathfrak{f}\right)^{-1/4}
    \begin{aligned}
        %
    \end{aligned}\;.
\end{equation}
The inner product \eqref{eq:inner_prod} can be generalized to arbitrary number of $\mathfrak{f}$. For example, we can define the inner product on $V^{\mathfrak{g}\otimes \mathfrak{f}}$ by  
\begin{equation} \label{eq:inner_prod_tensor}
    \Big{\langle} \,\begin{aligned}
        %
    \end{aligned}\!\!\!\!\!\!. 
    % \!\!\!\!\!\! = \delta_{x,x'}\delta_{y,y'}\delta_{\mathfrak{f},\mathfrak{f}'}\delta_{\alpha,\alpha'} \sqrt{d_xd_yd_\mathfrak{f}}. 
\end{equation}

Let us identify the general elements in $V^{\mathfrak{f}}\otimes V^{\mathfrak{g}}$ which is defined by $\Delta(1)\triangleright (V^{\mathfrak{f}}\otimes_{\mathbb{C}} V^{\mathfrak{g}})$. Note that $\Delta(1)$ is provided by Eq.~\eqref{eq:coprod_unit}, hence  
\begin{equation}
    \Delta(1) \triangleright \left(\begin{aligned}
        %
    \end{aligned}\;.
\end{equation}
Therefore, $V^{\mathfrak{f}}\otimes V^{\mathfrak{g}} \simeq V^{\mathfrak{g}\otimes \mathfrak{f}}$ by the following map
\begin{equation}\label{eq:Vasso}
    \begin{aligned}
        %
    \end{aligned}\;.
\end{equation}

\begin{theorem}
The functor $V^{\bullet} : \mathfrak{f} \mapsto V^{\mathfrak{f}}$ is a monoidal functor, with associator given by the inverse of Eq.~\eqref{eq:Vasso}:
\begin{equation}
    \eta_{\mathfrak{f},\mathfrak{g}} \colon V^{\mathfrak{f} \otimes \mathfrak{g}} \to V^{\mathfrak{f}} \otimes V^{\mathfrak{g}}.
\end{equation}
It satisfies the pentagon relation.
Note that on the right-hand side, the tensor product is taken over the tube algebra rather than over $\mathbb{C}$, and thus $\eta_{\mathfrak{f},\mathfrak{g}}$ is an isomorphism.
The tensor unit $\id_{\EM}$ of $\Fun_{\EC|\ED}(\EM,\EM)$ is mapped to the tensor unit $V^{\id_{\EM}}$ of $\Rep(\TubeM)$. From Eq.~\eqref{eq:left_counit_tube}, we have a basis $\mathbb{1} = \varepsilon_L(\TubeM) \in \Rep(\TubeM)$ given by
\begin{equation} \label{eq:tube_unit}
    e_s = \sum_{s' \in \Irr(\EM)} \;
    \begin{aligned}

    \end{aligned}
    \in \varepsilon_L(\TubeM).
\end{equation}
These elements are in one-to-one correspondence with the disk morphisms
\begin{equation}
    \Tilde{e}_{s} = 
    \begin{aligned}
        %
    \end{aligned}
    \in V^{\id_{\EM}}.
\end{equation}
This correspondence is a module map for $\TubeM$.
The unit morphism $l_M: V^{\id_{\EM}} \otimes V^{\mathfrak{f}} \to V^{\id_{\EM}\otimes \mathfrak{f}} = V^{\mathfrak{f}}$ can be defined as 
\begin{equation}
    \begin{aligned}
 \begin{aligned}
        %
    \end{aligned}
    \end{aligned}.
\end{equation}
In a similarly way, the functor $V_{\bullet}: \mathfrak{f} \mapsto V_{\mathfrak{f}}$ is also a monoidal functor.
\end{theorem}

\begin{proof}
The fact that the pentagon relation holds follows directly from the definition of the associator discussed previously. 
What remains is to establish the correspondence between the tensor units.

We will show that tube algebra basis elements acts in the same way for $e_s$ and $\Tilde{e}_s$.
Notice that $e_s \in \TubeM$, for tube algebra basis $X$,  using Eq.~\eqref{eq:tubeunitL} we have
\begin{equation}
\begin{aligned}
      X\triangleright e_s =&\varepsilon_L ( X \cdot e_s) = \varepsilon_L \left(
    \begin{aligned}
    \end{aligned}  \right) \\
   =& \delta_{u,s} \delta_{z,s} \delta_{x,w} \delta_{y,v} \delta_{\nu,\zeta} \delta_{\mu,\gamma}  \sqrt{\frac{d_ad_zd_b}{d_x}} e_{x}
\end{aligned}    \label{eq:Xes}
\end{equation}
where we have used Eq.~\eqref{eq:left_counit_tube}.

For $\Tilde{e}_s$, we have
\begin{equation}
    \begin{aligned}
    X \triangleright \Tilde{e}_s= &
        \begin{aligned}%
    \end{aligned}\\
    =&\delta_{s,u}\delta_{s,z} \delta_{v,y} \delta_{x,w} \delta_{\nu,\zeta} \delta_{\mu,\gamma} \sqrt{\frac{d_ad_z}{d_v}} \sqrt{\frac{d_vd_b}{d_x}} \Tilde{e}_x
    \end{aligned}
\end{equation}
which takes the same form as that of Eq.~\eqref{eq:Xes}.
\end{proof}

\begin{theorem}[Orthogonality of characters] \label{prop:ortho_char}
    Let $\mathfrak{f}, \mathfrak{g}\in\Fun_{\EC|\ED}(\EM,\EM)$ be simple. Denote by $\chi_{\mathfrak{f}}, \chi_{\mathfrak{g}}$ the characters of $\rho_\mathfrak{f}$ and $\rho_{\mathfrak{g}}$ respectively. Then \footnote{Recall that for a weak Hopf algebra $W$, its dual $\hat{W}=\Hom(W,\mathbb{C})$ is endowed with the inner product $ (\phi,\psi) = \langle\phi^*\psi,\mathfrak{I}\rangle$ where $\mathfrak{I}\in W$ is the Haar integral. The $*$-operation is given by $\langle \phi^*, w \rangle = \overline{\phi(S(w)^*)}$.  }
    \begin{equation}
        (\chi_{\mathfrak{f}},\chi_{\mathfrak{g}}) = \delta_{\mathfrak{f},\mathfrak{g}}. 
    \end{equation}
\end{theorem}

\begin{proof}
    First, we have 
    \begin{equation*}
        (\chi_{\mathfrak{f}},\chi_{\mathfrak{g}}) = \langle \chi_{\mathfrak{f}}^*\chi_{\mathfrak{g}},\mathfrak{I}\rangle = \langle \chi_{\bar{\mathfrak{f}}}\chi_{\mathfrak{g}},\mathfrak{I}\rangle = \langle \chi_{\bar{\mathfrak{f}}\otimes \mathfrak{g}},\mathfrak{I}\rangle = \Tr\rho_{\bar{\mathfrak{f}}\otimes \mathfrak{g}}(\mathfrak{I}),
    \end{equation*}
    where we used the facts that $\chi_{\mathfrak{f}}^* = \chi_{\bar{\mathfrak{f}}}$ and $\chi_\mathfrak{f}\chi_\mathfrak{g} = \chi_{\mathfrak{f}\otimes \mathfrak{g}}$ for simple $\mathfrak{f}$ and $\mathfrak{g}$ \cite{Nikshych2004semisimpleWHA}. Recall that the Haar integral $\mathfrak{I}$ of $\Tube(_{\EC}\EM_{\ED})$ is given by Eq.~\eqref{eq:haar_integral}; but it is more convenient to use the alternative form \eqref{eq:haar_integral_1} for the calculation here. To compute the trace, one chooses an orthonormal basis of $V^{\bar{\mathfrak{f}}\otimes \mathfrak{g}}$
    \begin{equation}
        | x,\bar{\mathfrak{f}}\xrightarrow{\alpha} y \xrightarrow{\beta} z,\mathfrak{g} \rangle:=  \left(d_xd_zd_\mathfrak{f}d_\mathfrak{g}\right)^{-1/4}\begin{aligned}

    \end{aligned}\;, x,y,z,\alpha,\beta,
    \end{equation}
    with respect to inner product \eqref{eq:inner_prod_tensor}. Therefore, we compute  
     
\begin{align*}
    \Tr\rho_{\bar{\mathfrak{f}}\otimes {\mathfrak{g}}}(\mathfrak{I}) & = \sum_{x,y,z,\alpha,\beta}  \frac{1}{\operatorname{rank} \EuScript{M} } \sum_{s,t} \frac{1}{d_t}\;\; 
    \Biggl \langle | x,\bar{\mathfrak{f}}\xrightarrow{\alpha} y \xrightarrow{\beta} z,{\mathfrak{g}} \rangle\; ,\;
    \begin{aligned}%
    \end{aligned} \\
    & = \sum_{x,y,z,\alpha,\beta}  \frac{1}{\operatorname{rank} \EM} \sum_{s,t} \frac{1}{d_t} \left(d_xd_zd_\mathfrak{f}d_\mathfrak{g}\right)^{-1/2} \delta_{x,t}\delta_{z,t}\delta_{z,s}\delta_{x,s} \delta_{\alpha,\beta}\delta_{\mathfrak{f},\mathfrak{g}}\sqrt{\frac{d_yd_s}{d_{\mathfrak{f}}}}\sqrt{\frac{d_td_y}{d_{\mathfrak{f}}}} d_{\mathfrak{f}} \\
    & = \sum_{x,y,\alpha}  \frac{1}{\operatorname{rank} \EM}  \frac{1}{d_x} \left(d_xd_xd_\mathfrak{f}d_\mathfrak{g}\right)^{-1/2}\delta_{\mathfrak{f},\mathfrak{g}}\sqrt{\frac{d_yd_x}{d_{\mathfrak{f}}}}\sqrt{\frac{d_xd_y}{d_{\mathfrak{f}}}}  d_{\mathfrak{f}}\\
    & = \delta_{\mathfrak{f},\mathfrak{g}}\sum_{x}  \frac{1}{\operatorname{rank} \EM}  \frac{1}{d_xd_{\mathfrak{f}}}  \left(\sum_{y}N_{x\bar{\mathfrak{f}}}^yd_y\right)\\
    & = \delta_{\mathfrak{f},\mathfrak{g}},  
\end{align*}
as expected. This completes the proof. 
\end{proof}

\begin{corollary}
    If the endofunctor $\mathfrak{f}\in \Fun_{\EC|\ED}(\EM,\EM)$ is simple, then $V^{\mathfrak{f}}$ is an irreducible representation over $\Tube(_{\EC}\EM_{\ED})$. 
\end{corollary}

\begin{proof}
    By a standard result from representation theory, the representation $V^\mathfrak{f}$ is irreducible if and only if its character $\chi_\mathfrak{f}$ satisfies $(\chi_\mathfrak{f},\chi_\mathfrak{f})=1$. Hence the claim follows from Proposition~\ref{prop:ortho_char}. 
\end{proof}

The above results imply that the domain wall topological excitations can be regarded as representations of the domain wall tube algebra. 
Previously, it was conjectured that $\Fun_{\EC|\ED}(\EM,\EM)$ is equivalent to $\Rep(\TubeM)$~\cite{jia2024weakTube}; for the boundary tube algebra, a similar conjecture has also been proposed~\cite{Kitaev2012boundary,bridgeman2023invertible}. 
Here, we demonstrate that this is not the case in general. However, $\Fun_{\EC|\ED}(\EM,\EM)$ can be embedded into $\Rep(\TubeM)$.
Notice that boundary tube algebra is a special case of the domain wall tube algebra, thus this will also give the negative answer to the conjecture for boundary tube algebra.

To see this, recall that an additive functor $F: \EC \to \ED$ is called \textit{injective} if and only if it is fully faithful, i.e., bijective on the sets of morphisms. It is called \textit{surjective} if every simple object of $\ED$ is a subquotient of some object $F(X)$ for some $X \in \EC$.

\begin{proposition}[{\cite[Section 6.3]{etingof2016tensor}}] A tensor functor \( F: \EC \to \ED \) between fusion categories \(\EC\) and \(\ED\) is an equivalence if and only if it is fully faithful and \(\operatorname{FPdim} \EC = \operatorname{FPdim} \ED\).
\end{proposition}

% For a $C^*$ weak Hopf algebra $W$, it is known that $\Rep(W)$ is pseudo-unitary~\cite{Nikshych2004semisimpleWHA, etingof2005fusion, etingof2016tensor}, and we have
% \begin{equation}
%     \dim_{\mathbb{C}} W = \operatorname{FPdim} \, \Rep(W),
% \end{equation}
% namely, the dimension of the weak Hopf algebra as a complex vector space coincides with the Frobenius–Perron dimension of $\Rep(W)$.

We have show that the functor
\begin{equation}
    F: \Fun_{\EC|\ED}(\EM,\EM) \to \Rep(\Tube({_{\EC}}\EM_{\ED}))
\end{equation}
defines an embedding of $\Fun_{\EC|\ED}(\EM,\EM)$ into $\Rep(\Tube({_{\EC}}\EM_{\ED}))$. 
To establish an equivalence between $\Fun_{\EC|\ED}(\EM,\EM)$ and $\Rep(\Tube({_{\EC}}\EM_{\ED}))$ (which is generally believed to be true), their Frobenius–Perron dimensions must coincide.
It follows from Eq.~\eqref{eq:tube_basis} that
\begin{equation}
 \dim_{\mathbb{C}} \mathbf{Tube}({_{\EC}}\EM_{\ED}) = \sum_{a,b,x,y,z,u,v,w} N_{x}^{yb} N_{y}^{az} N_{au}^v N_{vb}^w.
\end{equation}
But the Frobenius–Perron dimension  is not easy to calculate.
On the other hand, observe that
\begin{equation}
    \Fun_{\EC|\ED}(\EM,\EM) \simeq_{\otimes} \Fun_{\EC \boxtimes \ED^{\mathrm{rev}}}(\EM,\EM) =: (\EC \boxtimes \ED^{\mathrm{rev}})_{\EM}^{\vee}.
\end{equation}
By the fact that $\mathcal{Z}((\EC \otimes \ED^{\mathrm{rev}})_{\EM}^{\vee}) \cong \mathcal{Z}(\EC \otimes \ED^{\mathrm{rev}})$ and applying~\cite[Corollary 8.14]{etingof2005fusion}, we obtain
\begin{equation}
    \FPdim \Fun_{\EC|\ED}(\EM,\EM) = \FPdim(\EC \boxtimes \ED^{\mathrm{rev}}) = \FPdim \EC \cdot \FPdim \ED,
\end{equation}
where we have used the property of Frobenius–Perron dimension for Deligne tensor product.
We left the proof of equivalence for our future work.

% This embedding offers a meaningful way to characterize domain wall excitations via the domain wall tube algebra, as it preserves the topological charges, fusion structures, and other essential features.

\begin{remark}
    The discussion in this section can be extended to the case of multimodule domain walls via the folding trick. The topological excitations on a multimodule domain wall can  be characterized as representations of the corresponding multimodule domain wall tube algebra.
\end{remark}

\section{Domain wall defect tube algebra}
\label{sec:DefectTube}

When considering the defects between two domain walls ${_{\EC}\EM_{\ED}}$ and ${_{\EC}\EN_{\ED}}$, we can construct a tube algebra $\mathbf{Tube}_{\rm defect}[{_{\EC}\EM_{\ED}};{_{\EC}\EN_{\ED}}]$ by labeling the top and bottom of the tube with objects and morphisms from $\EN$ and $\EM$, respectively:

\begin{definition}[Domain wall defect tube algebra]
    For gapped domain walls ${_{\EC}}\EM_{\ED}$ and $_{\EC}\EN_{\ED}$,  the domain wall defect tube algebra  $\mathbf{Tube}_{\rm defect}[_{\EC}\EM_{\ED};{_{\EC}\EN_{\ED}}]$ is the vector space spanned by the following basis:
\begin{equation} \label{eq:defect_tube_basis}
 \left\{  
   \begin{aligned}
        \begin{tikzpicture}[scale=0.65]
        \filldraw[black!60, fill=gray!15, dotted, even odd rule] (0,0) circle[radius=0.5] (0,0) circle[radius=1.5];
             \draw[line width=.6pt,black] (0,0.5)--(0,1.5);
             \draw[line width=.6pt,black] (0,-0.5)--(0,-1.5);
             \draw[red] (0,0.8) arc[start angle=90, end angle=270, radius=0.8];
             \draw[blue] (0,1.3) arc[start angle=90, end angle=-90, radius=1.3];
            \node[ line width=0.6pt, dashed, draw opacity=0.5] (a) at (0,1.7){$\scriptstyle w$};
             \node[ line width=0.6pt, dashed, draw opacity=0.5] (a) at (0,-1.7){$\scriptstyle x$};
            \node[ line width=0.6pt, dashed, draw opacity=0.5] (a) at (-1,0){$\scriptstyle a$};
            \node[ line width=0.6pt, dashed, draw opacity=0.5] (a) at (1.1,0){$\scriptstyle b$};
            \node[ line width=0.6pt, dashed, draw opacity=0.5] (a) at (-0.2,-1){$\scriptstyle y$};
            \node[ line width=0.6pt, dashed, draw opacity=0.5] (a) at (-0.2,-1.35){$\scriptstyle \mu$};
            \node[ line width=0.6pt, dashed, draw opacity=0.5] (a) at (0.2,-0.8){$\scriptstyle \nu$};
            \node[ line width=0.6pt, dashed, draw opacity=0.5] (a) at (0,-0.3){$\scriptstyle z$};
            \node[ line width=0.6pt, dashed, draw opacity=0.5] (a) at (0,0.3){$\scriptstyle u$};
            \node[ line width=0.6pt, dashed, draw opacity=0.5] (a) at (-0.2,1){$\scriptstyle v$};
            \node[ line width=0.6pt, dashed, draw opacity=0.5] (a) at (-0.2,1.3){$\scriptstyle \gamma$};
            \node[ line width=0.6pt, dashed, draw opacity=0.5] (a) at (0.2,0.8){$\scriptstyle \zeta$};
        \end{tikzpicture}
    \end{aligned}
   \;:\; 
   \begin{aligned}
     & a\in \Irr(\EC),b\in \Irr(\ED), x,y,z\in \Irr(\EM),u,v,w\in \Irr(\EN), \\
     & \mu\in\Hom_{\EM}(x,y\otimes b)\neq 0,\nu\in\Hom_{\EM}(y,a\otimes z)\neq 0,\\
     & \zeta\in\Hom_{\EN}(a\otimes u,v)\neq 0,\gamma\in \Hom_{\EN}(v\otimes b,w)\neq 0  
   \end{aligned}
    \right\}.
\end{equation}
It is clear that this construction defines an algebra via gluing of basis elements, in a manner analogous to the domain wall tube algebra. However, it no longer carries a natural coalgebra structure.
\end{definition}

For defects between domain walls defined by multimodule categories, we similarly define the \emph{multimodule domain wall defect tube algebra}
\begin{equation}
\mathbf{Tube}_{\rm defect}[{_{\{\EC_{\alpha}\}_{\alpha\in I}}} \EM_{\{\ED_{\beta}\}_{\beta\in J}};{_{\{\EC_{\alpha}\}_{\alpha\in I}}} \EN_{\{\ED_{\beta}\}_{\beta\in J}}]
\end{equation}
by replacing the top and bottom labels of the tube with simple objects and morphisms from $\EN$ and $\EM$, respectively.

Defects between domain walls can be characterized by modules over the defect tube algebra $\mathbf{Tube}_{\rm defect}[_{\EC}\EM_{\ED};{_{\EC}\EN_{\ED}}]$. Given a bimodule functor $\mathfrak{f} \in \Fun_{\EC|\ED}(\EM, \EN)$, we define the associated vector space on a disk region by
\begin{equation}
    V_{\mathfrak{f}} := \bigoplus_{x \in \Irr(\EM), y\in \Irr(\EN)} \Hom_{\EN}(\mathfrak{f}(x), y).
\end{equation}
We can defined $x\otimes \mathfrak{f}:=\mathfrak{f}(x)$; a basis element of $V_{\mathfrak{f}}$ can be diagrammatically represented as
\begin{equation} \label{eq:defectBasis}
    \begin{aligned}
        \begin{tikzpicture}[scale=0.85]
            \filldraw[black!60, fill=gray!15, dotted] (0,0) circle[radius=0.8];
            \draw[black, ->-=0.6, line width=.6pt] (0,0) -- (0,0.8); 
            \draw[black, ->-=0.6, line width=.6pt] (0,-0.8) -- (0,0); 
            \draw[black, ->-=0.6, line width=.6pt] (0.8,0) -- (0,0); 
            \node[ line width=0.6pt, dashed, draw opacity=0.5] at (-0.2,0.5){$\scriptstyle y$};
            \node[ line width=0.6pt, dashed, draw opacity=0.5] at (-0.2,-0.6){$\scriptstyle x$};
            \node[ line width=0.6pt, dashed, draw opacity=0.5] at (0.6,-0.3){$\scriptstyle \mathfrak{f}$};
            \node[ line width=0.6pt, dashed, draw opacity=0.5] at (-0.3,0){$\scriptstyle \alpha$};
            \node[ line width=0.6pt, dashed, draw opacity=0.5] at (0,0){$\scriptstyle \bullet$};
        \end{tikzpicture}
    \end{aligned} \;\in V_{\mathfrak{f}}.
\end{equation}
We emphasize that, unlike the domain wall excitations, for domain wall defects, the direction of the functor $\mathfrak{f}$ is fixed: it must go from $\EM$ to $\EN$. Reversing its direction does not make sense, as $\mathfrak{f}$ is defined as a functor from $\EM$ to $\EN$ and not vice versa.

The disk space $ V_{\mathfrak{f}}$ is a module over $\mathbf{Tube}_{\rm defect}[_{\EC}\EM_{\ED};{_{\EC}\EN_{\ED}}]$ via the action 
\begin{equation}
    \rho_{\mathfrak{f}}:\mathbf{Tube}_{\rm defect}[_{\EC}\EM_{\ED};{_{\EC}\EN_{\ED}}] \to \End(V_{\mathfrak{f}})
\end{equation}
which is defined by 
    \begin{align}
        & \begin{aligned}
    \end{aligned}\;. 
    \end{align}
Note that evaluating the right-hand side to obtain a linear combination of the basis elements in Eq.~\eqref{eq:defectBasis} is more subtle than in the case of domain wall excitations. For instance, defining the $F$-move requires handling objects such as $\mathfrak{f}\in \Fun_{\EC|\ED}(\EM,\EN)$, $t, v \in \EN$, $s \in \EM$, and $a \in \EC$, which lie in four different categories.
To address this issue, we introduce a multifusion category arising from the left $\EC$-module structure (since in this case, the labels in $\ED$ do not appear). We define $\EK_0 = \EC$, $\EK_1 = \EM$, and $\EK_2 = \EN$, and set $\EP_{i,j} = \Fun_{\EC}(\EK_i, \EK_j)$. The multifusion category is then given by
\begin{equation}
    \EP = \bigoplus_{i,j=0}^2 \EP_{i,j}.
\end{equation}
We emphasize that $\EP_{0,0} \simeq \EC$, $\EP_{0,1} \simeq \EM$, and $\EP_{0,2} \simeq \EN$. Within this category, we can define the $F$-move
$[F_{a,s,\mathfrak{f}}^v]_{t,\zeta,\alpha}^{t',\zeta',\alpha'}$.
Similarly, when dealing with the local moves involving right $\ED$-action and $\mathfrak{f}$, one needs to introduce a corresponding multifusion category $\EQ = \bigoplus_{i,j=0}^2 \EQ_{i,j}$ built from the right module structure.

The associativity and unit axioms are easy to verify.
For multimodule categories, the domain wall defects are characterized by a multimodule functor $\mathfrak{f}$, we similarly obtain a module $V_{\mathfrak{f}}$ over $\mathbf{Tube}_{\rm defect}[{_{\{\EC_{\alpha}\}_{\alpha\in I}}} \EM_{\{\ED_{\beta}\}_{\beta\in J}};{_{\{\EC_{\alpha}\}_{\alpha\in I}}} \EN_{\{\ED_{\beta}\}_{\beta\in J}}]$.

%\section{Fermionization of tube algebra}

%\section{G-graded tube algebra}

%\section{Application I: weak Hopf quantum double model from weak Hopf tube algebra}

%A crucial application of weak Hopf tube algebra is that we can use it as the input data for a weak Hopf quantum double model.

%\section{Application II: Extendended string-net model with an action of weak Hopf tube algebra}

\section{Conclusion and discussion}
\label{sec:conclusion}

In this work, we establish the theory of weak Hopf tube algebras for describing $2$d gapped phases arising in Turaev-Viro topological quantum field theories (TQFTs). While substantial progress has been made, several important questions remain open and deserve further exploration.

(1) \emph{Anyon Condensation Theory via Tube Algebras.}  
Anyon condensation is a fundamental mechanism in topological phases, describing phase transitions induced by the condensation of a set of anyons. Within the framework of weak Hopf tube algebras, such condensation processes can be interpreted as instances of weak Hopf symmetry breaking. The topological phases before and after condensation are then characterized by distinct weak Hopf symmetries, encoded respectively in different weak Hopf tube algebras. A systematic study of anyon condensation from this perspective offers a promising and insightful direction for future research.

(2) \emph{$G$-graded Extension of Tube Algebra.}  
In the study of symmetry-enriched topological (SET) phases, it is natural to consider a $G$-graded fusion category as the input for the string-net model. In this setting, one obtains $G$-graded versions of the boundary and domain wall tube algebras. These graded algebras serve as essential tools for understanding SET phases from the perspective of weak Hopf symmetries.

(3) \emph{Systematic Investigation of Quantum $N$-Tuple Algebras.}  
We have briefly introduced the notion of quantum $N$-tuple algebras, which exhibit a rich hierarchical structure and merit a systematic and detailed investigation. These algebras are also expected to play a crucial role in the study of higher-dimensional topological phases.

(4) \emph{Applications in Boundary and Domain Wall SymTFT.}  
The boundary tube algebra has been employed to study boundary SymTFTs. It is natural to extend this framework to domain wall SymTFTs. Given that the domain wall tube algebra possesses a richer weak Hopf algebra structure—encompassing the boundary tube algebras as subalgebras and forming a Drinfeld quantum double—it is compelling to systematically establish the connection between domain wall SymTFTs and domain wall tube algebras.

All of the above directions are part of our ongoing research and will be explored in detail in future work.

%\vspace{1em}
\subsection*{Acknowledgements}
We express our gratitude to Liang Kong for valuable discussions on tube algebra and symmetry-enriched topological phases. We extend our sincere thanks to Song Cheng for the insightful conversations during ICBS 2024 and Z.J. also acknowledges support by Song Cheng and Jinsong Wu during his visiting at BIMSA and Dagomir Kaszlikowski for his support; additionally, he thanks Ansi Bai for sharing his work on boundary tube algebra and appreciate the discussions with him and Zhi-Hao Zhang on this topic.
S.T. extends his thanks to Zhengwei Liu and Jinsong Wu for their constant support, and to Ansi Bai, Yilong Wang and Zhi-Hao Zhang for their helpful discussions.
We also acknowledge Kantaro Ohmori for discussion.
Z.J. is supported by the National Research Foundation in Singapore, the A*STAR under its CQT Bridging Grant, CQT-Return of PIs EOM YR1-10 Funding and CQT Young Researcher Career Development Grant.

%\vspace{1em}

\appendix

\section{Basics of fusion categories and weak Hopf algebras} \label{sec:app_fusion}

Throughout the paper, all algebraic data are over the field $\mathbb{C}$ of complex numbers. We follow the convention adopted in our previous work \cite{jia2024weakTube}. Nevertheless, let us recall some basic definitions so as to fix the notations. More details can be found in Ref.~\cite{etingof2016tensor}. 

\emph{Multifusion categories.} ---  Recall that a multifusion category $\EC$ is a finite semisimple $\mathbb{C}$-linear abelian rigid monoidal category. More precisely, $\EC$ is equipped with a tensor product bifunctor $\otimes:\EC\times\EC\to \EC$, together with a natural isomorphism $\{a_{X,Y,Z}:(X\otimes Y)\otimes Z \to X\otimes (Y\otimes Z)\}_{X,Y,Z\in\EC}$, satisfying the pentagon axiom which connects different ways of tensoring four objects. It has a distinguished object $\one$, called the tensor unit, and two natural isomorphisms $\{l_X:\one\otimes X\to X\}_{X\in\EC}$ and $\{r_X: X\otimes \one\to X\}_{X\in \EC}$ fulfilling the triangle axiom. These data endow $\EC$ a monoidal category structure. If the tensor product is bilinear on morphism, it is a multitensor category. Semisimplicity means that every object can be written as a finite direct sum of simple objects, where an object $X$ is simple if and only if $\End_\EC(X)\simeq\mathbb{C}$. By finiteness assumption, the set, denoted as $\Irr(\EC)$, of inequivalent simple objects in $\EC$ is a finite set, and each Hom-space is finite dimensional. $\EC$ is rigid if every object $X\in \EC$ has a left dual $({^\vee}\!X,b_X':\one\to {^\vee}\!X\otimes X,d_X':X\otimes {^\vee}\!X\to \one)$ and a right dual $(X^\vee, b_X:\one\to X\otimes X^\vee,d_X:X^\vee\otimes X\to \one)$. If in addition the tensor unit $\one$ is simple,  $\EC$ is called a fusion category. A braiding on a monoidal category is a natural isomorphism $c_{X,Y}:X\otimes Y\xrightarrow{\sim} Y\otimes X$ such that the hexagon axiom holds. A braided monoidal category is a monoidal category with a braiding.

\emph{Module categories.} --- A left module category $\EM$ over a multitensor category $\EC$ is a locally finite abelian category equipped with a left $\EC$-action $\otimes:\EC\times \EM\to \EM$, together with a natural isomorphism $\{m_{X,Y,M}:(X\otimes Y)\otimes M\to X\otimes (Y\otimes M)\}_{X,Y\in \EC, M\in \EM}$ obeying the pentagon axiom. The action $\otimes$ is bilinear on morphisms and exact in the first variable. It is convenient to use the notation ${_\EC}\EM$ to indicate a left $\EC$-module $\EM$. Similarly one can define a right $\EC$-module $\EN$ with natural isomorphism $\{n_{N,Y,X}:N\otimes (Y\otimes X)\to (N\otimes Y)\otimes X\}_{N\in\EN, Y,X\in\EC}$ and use the notation $\EN_\EC$. A ${\EC}|\ED$-bimodule category $\EM$ is both a left $\EC$-module category and a right $\ED$-module category such that these two structures are compatible in a natural way. Namely, there is a natural isomorphism $\{b_{X,M,Z}:(X\otimes M)\otimes Z \xrightarrow{\sim} X\otimes (M\otimes Z)\}_{X\in\EC, M\in \EM, Z\in \ED}$ compatible with the left $\EC$-action and the right $\ED$-action. We use ${_\EC}\EM_{\ED}$ to denote a bimodule category over $\EC|\ED$. 

By $\Fun_\EC(\EM,\EN)$ and $\Fun_{\EC|\ED}(\EM,\EN)$ we denote the respective categories of module functors. For instance, a module functor $F\in \Fun_\EC(\EM,\EN)$ satisfies $F(X\otimes M) \xrightarrow{\sim} X\otimes F(M)$ for $X\in \EC$, $M\in\EM$. If $\EC$ is a multifusion category, then $\Fun_\EC(\EM,\EN)$ is semisimple for any module categories $\EM, \EN$ \cite{etingof2005fusion}. In particular, the dual category $\EC^\vee_\EM:=\Fun_\EC(\EM,\EM)$ is a rigid semisimple monoidal category, with the tensor product given by composition of functors, the right and left duals given by the right and left adjoint functors.

\emph{Drinfeld centers.} --- For a monoidal category $\EC$, its Drinfeld center $\mathcal{Z}(\EC)$ is a category whose objects are the half-braidings, i.e., pairs $(X,\gamma_{X,\bullet})$ where $X\in\EC$ and $\gamma_{X,\bullet}:X\otimes (\bullet) \xrightarrow{\sim} (\bullet)\otimes X$ is a natural isomorphism satisfying certain coherence condition. A morphism $f:(X,\gamma_{X,\bullet})\to (Z,\gamma_{Z,\bullet})$ is a morphism $f:X\to Z$ such that $\gamma_{Z,Y}\comp(f\otimes\id_Y) = (\id_Y\otimes f)\comp\gamma_{X,Y}$ holds for all $Y\in \EC$. This is  a braided monoidal category with the braiding $c_{(X,\gamma_{X,\bullet}),(Z,\gamma_{Z,\bullet})}=\gamma_{X,Z}$. Indeed, $\mathcal{Z}(\EC)$ inherits many properties from $\EC$ if the later possesses them, such as rigidity, semisimplicity, etc.  There is a canonical equivalence $\mathcal{Z}(\EC)\simeq \Fun_{\EC|\EC}(\EC,\EC)$. 

\emph{Unitarity and pivotal structures.} --- A $\dagger$-category is a $\mathbb{C}$-linear category $\EC$ equipped with a non-degenerate star functor $\dagger:\EC\to \EC^{\rm op}$, provided by $X^\dagger = X$ for $X\in \EC$ and $f^\dagger:Y\to X$ for $f:X\to Y$. It obeys the properties $f^{\dagger\dagger} = f$, $(\lambda f)^\dagger=\bar{\lambda}f^\dagger$ and $(g\comp f) = f^\dagger\comp g^\dagger$. A unitary $\dagger$-category is the one satisfies the non-degeneracy condition: $f\comp f^\dagger=0$ implies $f=0$. A monoidal $\dagger$-category imposes additional compatibility conditions: $(g\otimes h)^\dagger= g^\dagger\otimes h^\dagger$, $a_{X,Y,Z}^\dagger = a_{X,Y,Z}^{-1}$, $l_X^\dagger=l_X^{-1}$, and $r_X^\dagger=r_X^{-1}$. 

A pivotal structure on a rigid tensor category $\EC$ is a natural isomorphism $j:\id \to (\bullet)^{\vee\vee}$. It consists of a collection of isomorphisms $j_X:X\xrightarrow{\sim} X^{\vee\vee}$ and satisfies $j_{X\otimes Y} = j_X\otimes j_Y$. A category endowed with such structure is called a pivotal category. With a pivotal structure, one can define the left and right trace $\Tr_L(f)$, $\Tr_R(f)\in \End(\one)$ for $f\in \End(X)$. $j$ is called spherical if $\Tr_L(f)=\Tr_R(f)$ holds for all $f$. It is shown that any unitary fusion category admits a unique pivotal structure which is spherical \cite{etingof2005fusion}. 

\emph{Quantum dimension.} --- Given $X, Y\in \EC$, their tensor $X\otimes Y$ can be written as a direct sum of simple objects, $X\otimes Y = \oplus_{Z\in\Irr(\EC)} N_{X,Y}^ZZ$, where $N_{X,Y}^Z = \dim \Hom(X\otimes Y,Z)\in \mathbb{Z}_{\geq 0}$ is the multiplicity. Fixing $X\in \EC$, consider the non-negative integer square matrix $N_X:=\{N_{X,Y}^Z\}_{Y,Z\in \Irr(\EC)}$. By Frobenius-Perron theorem, it has non-negative real eigenvalues, among which the largest one dominates the absolute values of all other eigenvalues. Denote by $d_X$ the largest eigenvalue, called the quantum dimension of $X$; it satisfies $d_Xd_Y=\sum_{Z\in\Irr(\EC)}N^Z_{X,Y}d_Z$. The total dimension of $\EC$ is $d_\EC:=\sum_{X\in\Irr(\EC)}d_X^2$. 
More generally, one can define the quantum dimension $d_M$ for $M\in {_\EC}\EM$; it is completely determined by the properties: (i) $d_M >0$ ($M\in \Irr(\EM)$), (ii) $d_{X\otimes M} = d_Xd_M$ ($X\in \EC$, $M\in \EM$) and (iii) $\sum_{M\in\Irr(\EM)}d_M^2=d_\EC$ \cite{etingof2010fusion}.  Note that this definition depends on the left action of $\EC$. Similarly, one can define the quantum dimension of objects in a right module. When considering a bimodule, it is sometimes necessary to distinguish between the quantum dimensions defined by the left and right module structures; hence the notations $d_M^L$ and $d_M^R$ will be used in this case. In most cases we consider, these two notions are assumed to coincide.

\emph{Graphical calculus.} --- A monoidal category is strict if all natural isomorphisms $a, l, r$ are identities. It is well know that any monoidal category is equivalent to a strict one. Graphical calculus is a useful tool for representing morphisms, especially when dealing with sequential compositions of morphisms. Moreover, it becomes more concise for graphical calculus of strict categories. 
% We will assume the strictness in the following. 
In diagrams, we read morphisms from bottom to top:
\begin{equation} \label{eq:arrow-convention}
\vcenter{\hbox{
\begin{tikzpicture}
\draw [black, ->-=0.5, line width=0.6pt] (0,0.4) -- (0, 1);
\draw [black, line width=0.6pt] (-0.2,0) rectangle (0.2,0.4); 
\draw [black, ->-=0.6, line width=0.6pt] (0,-0.6) -- (0, 0);
\node [above] at (0.3, 0.5) {\tiny $Y$};
\node [above] at (0.3, -0.5) {\tiny $X$};
\node [ line width=0.2pt, dashed, draw opacity=0.5] at (0,0.2) {\small $f$};
\end{tikzpicture}
}}
\in\Hom(X,Y)\,,\quad
\id_X = \vcenter{\hbox{
\begin{tikzpicture}
\draw [black, ->-=0.5, line width=0.6pt] (0,-1) -- (0, 0);
\node [above] at (0.3, -0.5) {\tiny $X$};
\end{tikzpicture}
}} = \vcenter{\hbox{
\begin{tikzpicture}
\draw [black, ->-=0.5, line width=0.6pt] (0,0) -- (0,-1);
\node [above] at (0.3, -0.5) {\tiny $X^*$};
\end{tikzpicture}
}}.
\end{equation} 
The arrow labeled by the tensor unit $\one$ can be omitted, and isotopic diagrams represent the same morphism. For example, the ``birth'' and ``death'' morphisms are 
\begin{equation}
    b_X = \vcenter{\hbox{
\begin{tikzpicture}
\draw [black, ->-=0.52, line width=0.6pt] (0,0) [partial ellipse=-180:0:0.5 and 0.6];
\node [above] at (-0.5, 0) {\tiny $X$};
\node [above] at (0.6, 0) {\tiny $X^*$};
\end{tikzpicture}
}},\quad d_X = \vcenter{\hbox{
\begin{tikzpicture}
\draw [black, ->-=0.52, line width=0.6pt] (0,0) [partial ellipse=180:0:0.5 and 0.6];
\node [below] at (-0.5, 0) {\tiny $X$};
\node [below] at (0.6, 0) {\tiny $X^*$};
\end{tikzpicture}
}}. 
\end{equation}
The arrows can be omitted for simplicity. In a braided monoidal category, the braiding $c_{X,Y}$ is represented by 
\begin{equation}
c_{X,Y}=\begin{aligned}\begin{tikzpicture}
\braid[
			width=0.6cm,
			height=0.4cm,
			line width=0.6pt,
			style strands={1}{black},
			style strands={2}{black}] (Kevin)
			s_1^{-1} ;	\end{tikzpicture}\end{aligned} \;\;. 
\end{equation}
A good source for introducing the graphical calculus is Ref.~\cite{turaev2016quantum}.

\emph{F-symbols.} --- The associator $a_{i,j,k}:(i\otimes j)\otimes k\to i\otimes (j\otimes k)$ of $\EC$ induces isomorphisms of vector spaces
\begin{equation}
\begin{aligned}
     F_l^{ijk}:\Hom(l,(i\otimes j)\otimes k) \to \Hom(l,i\otimes(j\otimes k)), \\
     F^l_{ijk}: \Hom((i\otimes j)\otimes k,l) \to \Hom(i\otimes(j\otimes k),l).
\end{aligned} 
\end{equation}
If $\EC$ is unitary, they satisfy $(F_l^{ijk})^\dagger = (F_l^{ijk})^{-1}$, $(F^l_{ijk})^\dagger = (F^l_{ijk})^{-1}$.
Fixing basis for respective spaces, we obtain the F-symbols by the following
\begin{align}
    \begin{aligned}

    \end{aligned}. 
\end{align}
One important property is that 
\begin{equation} \label{eq:F-symbol-delta}
    \sum_{m,\alpha,\beta} [F_l^{ijk}]^{m\alpha\beta}_{n\mu\nu}[F^l_{ijk}]^{m\alpha\beta}_{n'\mu'\nu'}=\delta_{n,n'}\delta_{\mu,\mu'}\delta_{\nu,\nu'}.
\end{equation}
Similar identity holds for the F-symbols introduced below.

Now consider a bimodule category ${_\EC}\EM_{\ED}$. The left $\EC$-action and $m_{i,j,x}:(i\otimes j)\otimes x\to i\otimes (j\otimes x)$ induces the following F-symbols:  
% For $i, j, u \in\Irr(\EC)$, $x,z\in\Irr(\EM)$, $\alpha\in V_{ij}^u$, and $\beta\in V_{ux}^z$, 
\begin{align}
    \begin{aligned}

    \end{aligned}.   
\end{align}
Similarly, the right $\ED$-action and $n_{x,j,i}:x\otimes(j\otimes i) \to x\otimes (j\otimes i)$ induces the following F-symbols: 
\begin{align}
    \begin{aligned}
        %
    \end{aligned}.
\end{align}
Finally, the associator $b_{i,x,j}:(i\otimes x)\otimes j\to i\otimes (x\otimes j)$ induces the following F-symbols: 
\begin{align}
    \begin{aligned}
        %
    \end{aligned}. 
\end{align}
All operations from the left diagram to the right diagram (or the inverse) are referred to F-moves. 
Here are some useful properties of the F-symbols.

\begin{proposition} \label{prop:F-symbols_property}
    For a $\EC|\ED$ bimodule $\EM$, its F-symbols satisfy the identities 
    \begin{gather}
        \sum_{u,\alpha,\beta} [F_{ijx}^z]_{v\mu\nu}^{u\alpha\beta}[F_z^{ijx}]_{v'\mu'\nu'}^{u\alpha\beta} = \delta_{v,v'}\delta_{\mu,\mu'}\delta_{\nu,\nu'}, \\ 
        \sum_{u,\alpha,\beta} [F_{xji}^z]_{v\mu\nu}^{u\alpha\beta}[F_z^{xji}]_{v'\mu'\nu'}^{u\alpha\beta} = \delta_{v,v'}\delta_{\mu,\mu'}\delta_{\nu,\nu'}, \\
        \sum_{u,\alpha,\beta} [F_{ixj}^z]_{v\mu\nu}^{u\alpha\beta}[F_z^{ixj}]_{v'\mu'\nu'}^{u\alpha\beta}\sqrt{\frac{d_u^R}{d_u^L}} = \delta_{v,v'}\delta_{\mu,\mu'}\delta_{\nu,\nu'}\sqrt{\frac{d_x^Rd_v^Ld_z^R}{d_x^Ld_v^Rd_z^L}}. 
    \end{gather}
\end{proposition}

\begin{proof}
    The proof of the three identities above corresponds to computing the value of the following three diagrams:
    \begin{equation*}
    \begin{aligned}

    \end{aligned}. 
\end{equation*}
For example, on the one hand, using loop move the value of the third diagram is equal to 
\begin{equation}
    \begin{aligned}
        %
    \end{aligned} . \label{eq:F-symbols_value1}
\end{equation}
On the other hand, applying F-move to the third diagram results in 
\begin{align}
    \begin{aligned}
        %
    \end{aligned} . \label{eq:F-symbols_value2}
\end{align}
Then the third identity follows by comparing Eq.~\eqref{eq:F-symbols_value1} and Eq.~\eqref{eq:F-symbols_value2}. 
\end{proof}

\emph{Weak Hopf algebras.} --- A weak bialgebra in a braided multifusion category $\EC$ is an object $W\in\EC$ with an associative algebra structure $(W,\mu,\eta)$ and a coassociative coalgebra structure $(W,\Delta,\varepsilon)$, where 
\begin{equation}
    \mu=\begin{aligned}
    %
\end{aligned}\,\,\, , 
\end{equation}
such that the following compatibility holds
\begin{gather} \label{eq:weak_bialg}
    \small 
    \begin{aligned}
    %
			\end{aligned}\,\,\,.
\end{gather}
A weak Hopf algebra is a weak bialgebra  $(W,\mu,\eta,\Delta,\varepsilon)$ equipped with a morphism $S: W\to W$, called an antipode, which satisfies the antipode axioms:
\begin{gather} \label{eq:weak_antipode}
            \begin{aligned}	
				%
	   \end{aligned}\,\,\,.
\end{gather}
The map on the right-hand side of the first (resp.~second) identity is denote by $\varepsilon_L$ (resp.~$\varepsilon_R$), called the left (resp.~right) counital morphism. 

All Hopf algebras, including group algebras, are example of weak Hopf algebras; there are also examples that do not come from Hopf algebras. In this paper, we mainly consider weak Hopf algebras in $\EC=\Vect_\mathbb{C}$, the category of finite dimensional complex vector spaces. In this case, it is convenient to adopt Sweedler's notation for comultiplication $\Delta(x) = \sum_{(x)} x^{(1)}\otimes x^{(2)}$, $(\id\otimes\Delta)\comp\Delta(x) = \sum_{(x)} x^{(1)}\otimes x^{(2)}\otimes x^{(3)}$, etc. In particular, the left and right counit morphisms are $\varepsilon_L(x) = \sum_{(1_W)}\varepsilon(1_W^{(1)}\cdot x)1_W^{(2)}$ and $\varepsilon_R(x) = \sum_{(1_W)}1_W^{(1)}\varepsilon(x\cdot 1_W^{(2)})$ respectively. Denote $W_L=\varepsilon_L(W)$, $W_R = \varepsilon_R(W)$, which are weak Hopf subalgebras of $W$. 

For a weak Hopf algebra $W\in \Vect_\mathbb{C}$, there is a canonical equivalence $\Rep(D(W))\simeq \mathcal{Z}(\Rep(W))$ between braided tensor categories. Here $D(W)$ is the Drinfeld quantum double of $W$, whose representation category $\Rep(D(W))$ has the braiding provided by the $R$-matrix. The concrete construction of $D(W)$ is provided by  Refs.~\cite{nikshych2003invariants,Jia2023weak}; an important thing is that $D(W)$ is again a weak Hopf algebra. One motivation to tackle with weak Hopf algebras is that any multifusion category $\EC$ is equivalent to the category $\Rep(W)$ of finite dimensional representations of some weak Hopf algebra $W$. This algebra is generally not unique, but they are Morita equivalent to each other. The category $\EC$ is a fusion category if and only if the weak Hopf algebra $W$ is connected, which, tautologically, means that the trivial representation of $W$ is irreducible. 

\section{Proof of Theorem~\ref{thm:QD-algebra}} \label{sec:app_QD}

In this section, we adopt a convention analogous to the Einstein summation convention: all summation symbols for comultiplication are omitted. The appearance of superscripts such as $\cone$, $\ctwo$, etc., implicitly indicates that comultiplication is being applied.
For example, 
\begin{gather*}
    \Delta(x) = x^{(1)}\otimes x^{(2)},\quad (\id\otimes \Delta)\comp \Delta(x) = x^{(1)}\otimes x^{(2)}\otimes x^{(3)}, \\
    \varepsilon_L(x) = \varepsilon(1_W^{(1)}\cdot x)1_W^{(2)},\quad \varepsilon_R(x) = 1_W^{(1)}\varepsilon(x\cdot 1_W^{(2)}),\quad \text{etc}. 
\end{gather*}

Suppose $p:U\times V\to \mathbb{C}$ is a non-degenerate skew-pairing in the sense of Definition~\ref{def:skew-pairing}. 
Recall that $U^{\rm cop}$ has the following left and right $V$-module structures 
\begin{equation}
    v \rightharpoonup u = u^{(1)}p(u^{(2)},v),\quad u \leftharpoonup v = p(u^{(1)},v)u^{(2)},\quad u\in U^{\rm cop}, v\in V. 
\end{equation}
Here the comultiplication $\Delta^{\rm op}(u) = u^{(1)}\otimes u^{(2)}$ is taken in $U^{\rm cop}$. 
Sometimes, we write $\Delta^{\rm op}(u) = u^{(1)}_{\rm op}\otimes u^{(2)}_{\rm op} = u^{(2)}\otimes u^{(1)}$ to distinguish the comultiplication in $U^{\rm cop}$ from that in $U$, where expressions without subscript $(-)_{\rm op}$ are understood to be taken in the original $U$. 
Since we adopt this convention, it is more convenient to treat the skew-pairing $p:U\times V\to \mathbb{C}$ as a canonical pairing $p:U^{\rm cop}\times V \to \mathbb{C}$ in what follows. Always keep in mind that the antipode of $U^{\rm cop}$ is $S_{U^{\rm cop}}=S^{-1}_U$. 

It is immediate that 
\begin{equation}
    p(v\rightharpoonup u,v') = p(u,v'v),\quad p(u\leftharpoonup v, v') = p(u,vv'). 
\end{equation}
In particular, if $x \in V_L$, and $y\in V_R$
\begin{align*}
    p(x\rightharpoonup u,v) &= p(u,vx) = p(1_U,v^{(1)}x)p(u,v^{(2)}) = p((x\rightharpoonup 1_U)u,v), \\
    p(y\rightharpoonup u,v) &= p(u,vy) = p(u,v^{(1)})p(1_U,v^{(2)}y) = p(u(y\rightharpoonup 1_U), v). 
\end{align*}
Hence if follows from the non-degeneracy of $p$ that 
\begin{equation}
    x\rightharpoonup u = (x\rightharpoonup 1_U)u, \quad y\rightharpoonup u = u(y\rightharpoonup 1_U). 
\end{equation}
Similarly, one has 
\begin{equation}
    u\leftharpoonup x = (1_U\leftharpoonup x)u,\quad u \leftharpoonup y = u(1_U\leftharpoonup y). 
\end{equation}
The following relations are useful: for any $x\in V_L$, $y\in V_R$, and any $v\in V$, 
\begin{gather}
    1_U\leftharpoonup x = 1_U\leftharpoonup S(x),\quad 1_U\leftharpoonup y = 1_U\leftharpoonup S^{-1}(y), \\
    x\rightharpoonup 1_U = S^{-1}(x) \rightharpoonup 1_U,\quad y\rightharpoonup 1_U = S(y)\rightharpoonup 1_U, \\
    v\rightharpoonup 1_U = \varepsilon_L(v) \rightharpoonup1_U,\quad 1_U\leftharpoonup v = 1_U\leftharpoonup \varepsilon_R(v). 
\end{gather}
Their proofs are straightforward by using the non-degeneracy of $p$. Note that the last line actually comes from $\varepsilon(vv') = \varepsilon(v\varepsilon_L(v'))=\varepsilon(\varepsilon_R(v)v')$. 

We divide the proof of Theorem~\ref{thm:QD-algebra} into several steps. 

\emph{Step 1}. --- The set $J$ of $U\otimes V$ generated by 
\begin{equation}
        u\otimes xv - u(x\rightharpoonup 1_U)\otimes v,\quad u\otimes yv - u(1_U \leftharpoonup y)\otimes v,\quad \forall~ x\in V_L,\; \forall~y\in V_R, 
\end{equation} 
is a two-sided ideal under the multiplication defined by 
\begin{equation}
    (u\otimes v)(u'\otimes v') = uu'^{(2)}\otimes v^{(2)}v'p(u'^{(3)},v^{(1)})p(u'^{(1)},S^{-1}(v^{(3)})).
\end{equation}
In fact, note that $(v\rightharpoonup 1_U)\in (U^{\rm cop})_R$, $(1_U\leftharpoonup v) \in (U^{\rm cop})_L$, and for any weak Hopf algebra $W$, we have $\Delta(ux) = u^{(1)}x\otimes u^{(2)}$, and $\Delta(zy) = z^{(1)}\otimes z^{(2)}y$, for $x\in W_L$, $y\in W_R$; thus for any $w\otimes z\in U\otimes V$, one has
\begin{align*}
    &(w\otimes z)(u(x\rightharpoonup 1_U)\otimes v)  \\
    & \quad \quad \quad = wu^{(2)}\otimes z^{(2)} v p(u^{(3)}(x\rightharpoonup 1_U),z^{(1)})p(u^{(1)},S^{-1}(z^{(3)})) \\
    & \quad \quad \quad = wu^{(2)}\otimes z^{(3)} v p(u^{(3)},z^{(1)})p(x\rightharpoonup 1_U,z^{(2)})p(u^{(1)},S^{-1}(z^{(4)})) \\
    & \quad \quad \quad = wu^{(2)}\otimes z^{(3)} v p(u^{(3)},z^{(1)})p(1_U,z^{(2)}x)p(u^{(1)},S^{-1}(z^{(4)})) \\
    & \quad \quad \quad = wu^{(2)}\otimes z^{(2)}x v p(u^{(3)},z^{(1)})p(u^{(1)},S^{-1}(z^{(3)})) = (w\otimes z)(u\otimes xv), \\
    &(w\otimes z)(u(1_U\leftharpoonup y)\otimes v)  \\
    & \quad \quad \quad = wu^{(2)} \otimes z^{(2)} v p(u^{(3)},z^{(1)})p(u^{(1)}(1_U\leftharpoonup y),S^{-1}(z^{(3)})) \\
    & \quad \quad \quad = wu^{(2)} \otimes z^{(2)} v p(u^{(3)},z^{(1)})p(u^{(1)},S^{-1}(z^{(4)}))p(1_U\leftharpoonup y,S^{-1}(z^{(3)})) \\
    & \quad \quad \quad = wu^{(2)} \otimes z^{(2)} v p(u^{(3)},z^{(1)})p(u^{(1)},S^{-1}(z^{(4)}))p(1_U,z^{(3)}y) \\
    & \quad \quad \quad = wu^{(2)} \otimes z^{(2)}y v p(u^{(3)},z^{(1)})p(u^{(1)},S^{-1}(z^{(3)})) = (w\otimes z)(u\otimes yv). 
\end{align*}
    Here the sixth equality follows from 
    \begin{align*}
        p(1_U,yS^{-1}(z^{(3)})) &= p(1_U,S^{-1}(z^{(3)}S(y))) = p(S(1_U),z^{(3)}S(y)) \\
        & = p(S(y)\rightharpoonup 1_U,z^{(3)}) = p(y\rightharpoonup 1_U,z^{(3)}) = p(1_U,z^{(3)}y). 
    \end{align*}
    Similarly, one has 
    \begin{gather*}
        (u(x\rightharpoonup 1_U)\otimes v)(w\otimes z) = (u\otimes xv)(w\otimes z), \\
        (u(1_U\leftharpoonup y)\otimes v)(w\otimes z) = (u\otimes yv)(w\otimes z). 
    \end{gather*}
Thus $J$ is a two-sided ideal, as $(w\otimes z)J = J(w\otimes z)=0$ for any $w\otimes z\in U\otimes V$. 
Denote by $[u\otimes v]$ the equivalence class of $u\otimes v\in U\otimes V$ in the quotient space $U\otimes V/J$. 

\emph{Step 2}. --- The algebra structure. First we verify the unit property: 
    \begin{align*}
        [1_U\otimes 1_V][u\otimes v] & = [u^{(2)}\otimes 1_V^{(2)}v] p(u^{(3)},1_V^{(1)})p(u^{(1)},S^{-1}(1_V^{(3)})) \\
        & = [u^{(2)}\otimes 1_V^{(2)}1_V'^{(1)}v] p(u^{(3)},1_V^{(1)})p(u^{(1)},S^{-1}(1_V'^{(2)})) \\
         & = [u^{(2)}(1_V^{(2)} \rightharpoonup 1_U)p(u^{(3)},1_V^{(1)})(1_U\leftharpoonup 1_V'^{(1)})p(S(u^{(1)}),1_V'^{(2)})\otimes v]  \\
         & = [u^{(2)}\varepsilon_R(u^{(3)})\varepsilon_L(S(u^{(1)}))\otimes v]  = [u\otimes v], \\
         [u\otimes v][1_U\otimes 1_V] & = [u1_U^{(2)}\otimes v^{(2)}]p(1_U^{(3)},v^{(1)})p(1_U^{(1)},S^{-1}(v^{(3)})) \\ 
        & = [u1_U'^{(2)}1_U^{(1)}\otimes v^{(2)}]p(1_U^{(2)},v^{(1)})p(1_U'^{(1)},S^{-1}(v^{(3)})) \\
        & = [u(1_U'\leftharpoonup S^{-1}(v^{(3)}))(v^{(1)}\rightharpoonup 1_U)\otimes v^{(2)}] \\
        & = [u(1_U'\leftharpoonup \varepsilon_R(S^{-1}(v^{(3)})))(\varepsilon_L(v^{(1)})\rightharpoonup 1_U)\otimes v^{(2)}] \\
        & = [u\otimes \varepsilon_R(S^{-1}(v^{(3)}))\varepsilon_L(v^{(1)})v^{(2)}] = [u\otimes v]. 
    \end{align*}
    Here, we used the fact that $\Delta(1_V) \in V_R\otimes V_L$. The fifth equality holds because 
    \begin{equation*}
        u^{(2)}_{\rm op}\varepsilon_R^{\rm cop}(u^{(3)}_{\rm op})\varepsilon_L^{\rm cop}(S(u^{(1)}_{\rm op}))   = u^{(2)}(S^{-1}\comp\varepsilon_R)(u^{(1)})(S^{-1}\comp\varepsilon_L)(S(u^{(3)})) = u,
    \end{equation*}
    as $\varepsilon_L^{\rm cop} = S^{-1}\comp \varepsilon_L$, and $\varepsilon_R^{\rm cop} = S^{-1}\comp \varepsilon_R$. 
   Then we verify the associativity of the multiplication: 
    \begin{align*}
        & ([u \otimes  v][u'\otimes v'])[u''\otimes v''] \\
         =\; & [(uu'^{(2)})u''^{(2)}\otimes (v^{(3)}v'^{(2)})v'']  p(u''^{(1)}, S^{-1}(v^{(4)}v'^{(3)})) \\
        &  \times p(u''^{(3)},v^{(2)}v'^{(1)}) p(u'^{(1)},S^{-1}(v^{(5)})) p(u'^{(3)},v^{(1)}) \\ 
        =\; & [u(u'^{(2)}u''^{(3)})\otimes v^{(3)}(v'^{(2)}v'')]  p(u''^{(1)}, S^{-1}(v'^{(3)})) p(u''^{(2)}, S^{-1}(v^{(4)})) \\
        &  \times p(u''^{(4)},v^{(2)}) p(u''^{(5)},v'^{(1)}) p(u'^{(1)},S^{-1}(v^{(5)})) p(u'^{(3)},v^{(1)}) \\ 
        =\; & [u(u'^{(2)}u''^{(3)})\otimes v^{(2)}(v'^{(2)}v'')]   p(u'^{(1)}u''^{(2)}, S^{-1}(v^{(3)})) p(u'^{(3)}u''^{(4)},v^{(1)})\\
        &   \times p(u''^{(1)}, S^{-1}(v'^{(3)})) p(u''^{(5)},v'^{(1)}) \\ 
        = \; & [u \otimes  v]([u'\otimes v'][u''\otimes v'']).
    \end{align*}

    \emph{Step 3}. --- The morphisms 
        \begin{gather*}
         \Delta([u\otimes v])  = [u^{(2)}\otimes v^{(1)}] \otimes [u^{(1)}\otimes v^{(2)}], \quad \varepsilon([u\otimes v]) = p(u,\varepsilon_R(S^{-1}(v))), \\
         S([u\otimes v]) = [S(u^{(2)})\otimes S(v^{(2)})] p(u^{(1)},v^{(3)})p(u^{(3)},S^{-1}(v^{(1)})), 
    \end{gather*}
    are well-defined. First, the assertion for $\Delta$ is easy to verify. 
    Second, for the counit, it follows that 
    \begin{align*}
        \varepsilon([u(x\rightharpoonup 1_U)\otimes v]) & = p(u,1_V^{(1)}) p(1_U,\varepsilon_R(S^{-1}(v))1_V^{(2)}S^{-1}(x))  \\ 
        & = p(u,1_V^{(1)}) p(1_U,1_V^{(2)}S^{-1}(x\varepsilon_L(v)))  \\ 
        & = p(u,\varepsilon_R(S^{-1}(xv))) = \varepsilon([u\otimes xv]), \\
        \varepsilon([u(1_U\leftharpoonup y)\otimes v]) & = p(u,1_V^{(1)}) p(1_U,y\varepsilon_R(S^{-1}(v))1_V^{(2)})  \\ 
        & = p(u,1_V^{(1)}) p(1_U,S^{-1}(\varepsilon_L(v)S(y))1_V^{(2)})  \\ 
        & = p(u,\varepsilon_R(S^{-1}(yv))) = \varepsilon([u\otimes yv]). 
    \end{align*}
    The third equality holds because $x\varepsilon_L(v) = xv^{(1)}S(v^{(2)}) = \varepsilon_L(xv)$, and similarly, the seventh equality holds because $\varepsilon_L(v)S(y) = v^{(1)}S(yv^{(2)}) = \varepsilon_L(yv)$.  Finally, the antipode is well-defined: 
    \begin{align*}
        S([u\otimes xv]) &  = [S(u^{(2)})\otimes S(v^{(2)})] p(u^{(1)},v^{(3)})p(u^{(3)},S^{-1}(xv^{(1)})) \\
        & = [S(u^{(2)})\otimes S(v^{(2)})] p(u^{(1)},v^{(3)})p(S^{-1}(x)\rightharpoonup u^{(3)},S^{-1}(v^{(1)})) \\
          & = [S(u^{(2)})\otimes S(v^{(2)})] p(u^{(1)},v^{(3)})p(u^{(3)}(x\rightharpoonup 1_U),S^{-1}(v^{(1)})) \\
          & = S([u(x\rightharpoonup 1_U)\otimes v]), \\
          S([u\otimes yv]) &  = [S(u^{(2)})\otimes S(v^{(2)})] p(u^{(1)},yv^{(3)})p(u^{(3)},S^{-1}(v^{(1)})) \\
        & = [S(u^{(2)})\otimes S(v^{(2)})] p(u^{(1)} \leftharpoonup y,v^{(3)})p(u^{(3)},S^{-1}(v^{(1)})) \\
          & = [S(u^{(2)})\otimes S(v^{(2)})] p(u^{(1)}(1_U\leftharpoonup y),v^{(3)})p(u^{(3)},S^{-1}(v^{(1)})) \\
          & = S([u(1_U \leftharpoonup y)\otimes v]). 
    \end{align*}

    \emph{Step 4}. --- The coalgebra structure. The coassociativity of the comultiplication is easy to establish. The counit property is verified as follows: 
    \begin{align*}
        (\varepsilon\otimes \id)\comp \Delta([u\otimes v]) & = p(u^{(2)},\varepsilon_R(S^{-1}(v^{(1)}))) [u^{(1)}\otimes v^{(2)}] \\
        & =  [(S^{-1}(\varepsilon_L(v^{(1)})) \rightharpoonup u)\otimes v^{(2)}] \\
        & =  [u(\varepsilon_L(v^{(1)}) \rightharpoonup 1_U)\otimes v^{(2)}] = [u\otimes v], \\
        (\id\otimes \varepsilon)\comp \Delta([u\otimes v]) & = [u^{(2)}\otimes v^{(1)}]p(u^{(1)},\varepsilon_R(S^{-1}(v^{(2)}))) \\ 
        & = [(u \leftharpoonup \varepsilon_R(S^{-1}(v^{(2)})))\otimes v^{(1)}]\\ 
        & = [u (1_U\leftharpoonup \varepsilon_R(S^{-1}(v^{(2)})))\otimes v^{(1)}] = [u\otimes v].
    \end{align*}

    \emph{Step 5}. --- The comultiplication $\Delta$ is multiplicative: 
    \begin{align*}
        \Delta([u&\otimes v])\Delta([u'\otimes v']) \\
        & = [u^{(2)}\otimes v^{(1)}][u'^{(2)}\otimes v'^{(1)}] \otimes [u^{(1)}\otimes v^{(2)}][u'^{(1)}\otimes v'^{(2)}] \\
        & = [u^{(2)}u'^{(5)}\otimes v^{(2)}v'^{(1)}]p(u'^{(6)},v^{(1)})p(u'^{(4)},S^{-1}(v^{(3)})) \\
        & \quad \otimes [u^{(1)}u'^{(2)}\otimes v^{(5)}v'^{(2)}]p(u'^{(3)},v^{(4)})p(u'^{(1)},S^{-1}(v^{(6)})) \\
        & = [u^{(2)}u'^{(4)}\otimes v^{(2)}v'^{(1)}] \otimes [u^{(1)}u'^{(2)}p(u'^{(3)},S^{-1}(\varepsilon_L(v^{(3)})))\otimes v^{(4)}v'^{(2)}]\\
        &\quad \times p(u'^{(5)},v^{(1)})p(u'^{(1)},S^{-1}(v^{(5)})) \\
        & = [u^{(2)}u'^{(3)}\otimes v^{(2)}v'^{(1)}] \otimes [u^{(1)}(S^{-1}(\varepsilon_L(v^{(3)}))\rightharpoonup u'^{(2)})\otimes v^{(4)}v'^{(2)}]\\
        &\quad \times p(u'^{(4)},v^{(1)})p(u'^{(1)},S^{-1}(v^{(5)})) \\
        & = [u^{(2)}u'^{(3)}\otimes v^{(2)}v'^{(1)}] \otimes [u^{(1)}u'^{(2)}(\varepsilon_L(v^{(3)}) \rightharpoonup 1_U)\otimes v^{(4)}v'^{(2)}]\\
        &\quad \times p(u'^{(4)},v^{(1)})p(u'^{(1)},S^{-1}(v^{(5)})) \\ 
        & = [u^{(2)}u'^{(3)}\otimes v^{(2)}v'^{(1)}] \otimes [u^{(1)}u'^{(2)}\otimes \varepsilon_L(v^{(3)})v^{(4)}v'^{(2)}]\\
        &\quad \times p(u'^{(4)},v^{(1)})p(u'^{(1)},S^{-1}(v^{(5)})) \\
        & = [u^{(2)}u'^{(3)}\otimes v^{(2)}v'^{(1)}] \otimes [u^{(1)}u'^{(2)}\otimes v^{(3)}v'^{(2)}]\\
        &\quad \times p(u'^{(4)},v^{(1)})p(u'^{(1)},S^{-1}(v^{(4)})) \\
        & = \Delta([u\otimes v][u'\otimes v']). 
    \end{align*}

    \emph{Step 6}. --- The counit $\varepsilon$ is weak multiplicative:
    \begin{align*}
        \varepsilon(&[u\otimes v][u'\otimes v'][u''\otimes v'']) \\
        & = p(uu'^{(2)}u''^{(2)},\varepsilon_R(S^{-1}(v^{(3)}v'^{(2)}v'')))p(u'^{(3)},v^{(1)})\\
        & \quad \times p(u'^{(1)},S^{-1}(v^{(5)})) p(u''^{(3)},v^{(2)}v'^{(1)})p(u''^{(1)},S^{-1}(v^{(4)}v'^{(3)})) \\
        & = p(S(\varepsilon_R(uu'^{(2)}u''^{(2)})),v^{(3)}v'^{(2)}v'')p(u'^{(3)},v^{(1)})\\
        & \quad \times p(u'^{(1)},S^{-1}(v^{(5)})) p(u''^{(3)},v^{(2)}v'^{(1)})p(u''^{(1)},S^{-1}(v^{(4)}v'^{(3)})) \\
        & = p(1_U^{(1)}S(\varepsilon_R(uu'^{(2)}u''^{(2)})),v^{(3)}v'^{(2)})p(1_U^{(2)},v'')p(u'^{(3)},v^{(1)})\\
        & \quad \times p(u'^{(1)},S^{-1}(v^{(5)})) p(u''^{(3)},v^{(2)}v'^{(1)})p(S(u''^{(1)}),v^{(4)}v'^{(3)}) \\
        & = p(u''^{(3)}1_U^{(1)}S(\varepsilon_R(uu'^{(2)}u''^{(2)}))S(u''^{(1)}),v^{(2)}v')\\
        & \quad \times p(1_U^{(2)},v'')p(u'^{(3)},v^{(1)})p(u'^{(1)},S^{-1}(v^{(3)})) \\
        & = p(S(\varepsilon_R(uu'^{(2)}))S(\varepsilon_L(u''1_U^{(1)})),v^{(2)}v')\\
        & \quad \times p(1_U^{(2)},v'')p(u'^{(3)},v^{(1)})p(u'^{(1)},S^{-1}(v^{(3)})) \\
        & = p(S(\varepsilon_R(uu'^{(2)})),v^{(2)}v'^{(1)})p(S(\varepsilon_L(u''1_U^{(1)})),v^{(3)}v'^{(2)})\\
        & \quad \times p(1_U^{(2)},v'')p(u'^{(3)},v^{(1)})p(u'^{(1)},S^{-1}(v^{(4)})) \\
        & = p(S(\varepsilon_R(uu'^{(2)})),v^{(2)}v'^{(1)})p(1_U'^{(1)},v^{(3)})
        p(S(\varepsilon_L(u''1_U^{(1)}))1_U'^{(2)},v'^{(2)})\\
        & \quad \times p(1_U^{(2)},v'')p(u'^{(3)},v^{(1)})p(S(u'^{(1)}),v^{(4)}) \\
        & = p(S(\varepsilon_R(uu'^{(2)})),v^{(2)}v'^{(1)})
        p(S(\varepsilon_L(u''1_U^{(1)}))1_U'^{(2)},v'^{(2)})\\
        & \quad \times p(1_U^{(2)},v'')p(u'^{(3)},v^{(1)})p(1_U'^{(1)}S(u'^{(1)}),v^{(3)}) \\
        & = p(S(\varepsilon_R(uu'^{(3)})),v^{(2)}v'^{(1)})
        p(S(\varepsilon_L(u''1_U^{(1)}))\varepsilon_L(S(u'^{(1)})),v'^{(2)})\\
        & \quad \times p(1_U^{(2)},v'')p(u'^{(4)},v^{(1)})p(S(u'^{(2)}),v^{(3)}) \\
        & = p(S(\varepsilon_R(uu'^{(3)})),v^{(2)}v'^{(1)})
        p(u''^{(3)}1_U^{(1)}S(\varepsilon_R(u'^{(1)}u''^{(2)}))S(u''^{(1)}),v'^{(2)})\\
        & \quad \times p(1_U^{(2)},v'')p(u'^{(4)},v^{(1)})p(u'^{(2)},S^{-1}(v^{(3)})) \\
        & = p(S(\varepsilon_R(uu'^{(3)})),v^{(2)}v'^{(1)})
        p(u''^{(3)},v'^{(2)})p(1_U^{(1)}S^{-1}(\varepsilon_R(u'^{(1)}u''^{(2)})),v'^{(3)})\\
        & \quad \times p(S(u''^{(1)}),v'^{(4)}) p(1_U^{(2)},v'')p(u'^{(4)},v^{(1)})p(u'^{(2)},S^{-1}(v^{(3)})) \\
        & = p(S(\varepsilon_R(uu'^{(3)})),v^{(2)}v'^{(1)})
        p(u''^{(3)},v'^{(2)})p(S(\varepsilon_R(u'^{(1)}u''^{(2)})),v'^{(3)}v'')\\
        & \quad \times p(u''^{(1)},S^{-1}(v'^{(4)})) p(u'^{(4)},v^{(1)})p(u'^{(2)},S^{-1}(v^{(3)})) \\
        & = p(uu'^{(3)},\varepsilon_R(S^{-1}(v^{(2)}v'^{(1)}))) p(u'^{(4)},v^{(1)})p(u'^{(2)},S^{-1}(v^{(3)})) \\
        & \quad \times p(u'^{(1)}u''^{(2)},\varepsilon_R(S^{-1}(v'^{(3)}v'')))p(u''^{(3)},v'^{(2)})p(u''^{(1)},S^{-1}(v'^{(4)})) \\
        & = \varepsilon([u\otimes v][u'^{(2)}\otimes v'^{(1)}])\varepsilon([u'^{(1)}\otimes v'^{(2)}][u''\otimes v'']). 
    \end{align*}
    The ninth equality is from $u^{(1)}\otimes \varepsilon_L(u^{(2)}) = 1_U^{(1)}u\otimes 1_U^{(2)}$. The identity \[
    \varepsilon([u\otimes v][u'\otimes v'][u''\otimes v'']) = \varepsilon([u\otimes v][u'^{(1)}\otimes v'^{(2)}])\varepsilon([u'^{(2)}\otimes v'^{(1)}][u''\otimes v''])
    \]
    is shown similarly.

    \emph{Step 7}. --- The unit $1=[1_U\otimes 1_V]$ is weak comultiplicative:
    \begin{align*}
        (\Delta&(1)\otimes 1)(1\otimes \Delta(1)) \\
        & = [1_U^{(2)}\otimes 1_V^{(1)}] \otimes [1_U^{(1)}1_U'^{(3)} \otimes 1_V^{(3)}1_V'^{(1)}]\otimes [1_U'^{(1)}\otimes 1_V'^{(2)}]p(1_U'^{(4)},1_V^{(2)})p(1_U'^{(2)},S^{-1}(1_V^{(4)})) \\
        & = [1_U^{(2)}\otimes 1_V^{(1)}] \otimes [1_U^{(1)}(1_U'^{(2)}\leftharpoonup S^{-1}(1_V^{(4)})) \otimes 1_V^{(3)}1_V'^{(1)}]\otimes [1_U'^{(1)}\otimes 1_V'^{(2)}]p(1_U'^{(3)},1_V^{(2)}) \\
        & = [1_U^{(2)}\otimes 1_V^{(1)}] \otimes [1_U^{(1)}1_U'^{(2)} (1_U\leftharpoonup S^{-1}(1_V^{(4)})) \otimes 1_V^{(3)}1_V'^{(1)}]\otimes [1_U'^{(1)}\otimes 1_V'^{(2)}]p(1_U'^{(3)},1_V^{(2)}) \\
        & = [1_U^{(2)}\otimes 1_V^{(1)}] \otimes [1_U^{(1)}1_U'^{(2)}  \otimes S^{-1}(1_V^{(4)})1_V^{(3)}1_V'^{(1)}]\otimes [1_U'^{(1)}\otimes 1_V'^{(2)}]p(1_U'^{(3)},1_V^{(2)}) \\
        & = [1_U^{(2)}\otimes 1_V^{(1)}] \otimes [1_U^{(1)}1_U'^{(2)}(S^{-1}(\varepsilon_R(1_V^{(3)}))\rightharpoonup 1_U)  \otimes 1_V'^{(1)}]\otimes [1_U'^{(1)}\otimes 1_V'^{(2)}]p(1_U'^{(3)},1_V^{(2)}) \\
        & = [1_U^{(2)}\otimes 1_V^{(1)}] \otimes [1_U^{(1)}1_U'^{(2)}p(1_U'^{(3)},1_V^{(2)})p(1_U''^{(2)},S^{-1}(\varepsilon_R(1_V^{(3)}))) 1_U''^{(1)}   \otimes 1_V'^{(1)}]\otimes [1_U'^{(1)}\otimes 1_V'^{(2)}] \\
        & = [1_U^{(2)}\otimes 1_V^{(1)}] \otimes [1_U^{(1)}1_U'^{(2)}p(1_U'^{(3)}\varepsilon_R(S(1_U''^{(2)})),1_V^{(2)}) 1_U''^{(1)}   \otimes 1_V'^{(1)}]\otimes [1_U'^{(1)}\otimes 1_V'^{(2)}] \\
        & = [1_U^{(2)}\otimes 1_V^{(1)}] \otimes [1_U^{(1)}(1_V^{(2)} \rightharpoonup 1_U'^{(2)}\varepsilon_R(S^{-1}( 1_U''^{(2)})))  1_U''^{(1)}  \otimes 1_V'^{(1)}]\otimes [1_U'^{(1)}\otimes 1_V'^{(2)}] \\
        & = [1_U^{(2)}\otimes 1_V^{(1)}] \otimes [1_U^{(1)}(1_V^{(2)} \rightharpoonup 1_U) 1_U'^{(2)}\varepsilon_R(S^{-1}( 1_U''^{(2)}))   1_U''^{(1)} \otimes 1_V'^{(1)}]\otimes [1_U'^{(1)}\otimes 1_V'^{(2)}] \\
        & = [1_U^{(2)}\otimes 1_V^{(1)}] \otimes [1_U^{(1)}1_U'^{(2)} \otimes  1_V^{(2)}1_V'^{(1)}]\otimes [1_U'^{(1)}\otimes 1_V'^{(2)}] \\
        & = [1_U^{(3)}\otimes 1_V^{(1)}] \otimes [1_U^{(2)}\otimes 1_V^{(2)}]\otimes [1_U^{(1)}\otimes 1_V^{(3)}] = (\id\otimes \Delta)\comp \Delta(1). 
    \end{align*}
    Similarly, $(1\otimes \Delta(1))(\Delta(1)\otimes 1) = (\id\otimes \Delta)\comp \Delta(1)$ is proven.

    \emph{Step 8}. --- The antipode axioms hold. 
    First we compute that left counit:
    \begin{align*}
        \varepsilon_L([u\otimes v]) &= p(1_U^{(2)}u^{(2)},\varepsilon_R(S^{-1}(1_V^{(2)}v)))p(u^{(3)},1_V^{(1)})\\
        &\quad\times p(u^{(1)},S^{-1}(1_V^{(3)}))[1_U^{(1)}\otimes 1_V^{(4)}] \\
         & = p(1_U^{(2)},1_V'^{(1)})p(u^{(2)},\varepsilon_R(S^{-1}(1_V^{(2)}v))1_V'^{(2)})\\
        &\quad\times p(u^{(3)},1_V^{(1)})p(u^{(1)},S^{-1}(1_V^{(3)}))[1_U^{(1)}\otimes 1_V^{(4)}] \\
        & = p(u,S^{-1}(1_V^{(3)})\varepsilon_R(S^{-1}(1_V^{(2)}v))1_V'^{(2)}1_V^{(1)})\\
        &\quad\times [(1_V'^{(1)}\rightharpoonup 1_U)\otimes 1_V^{(4)}] \\
        % & = p(u,S^{-1}(\varepsilon_R(1_V^{(3)}))S^{-1}(\varepsilon_L(v))S^{-1}(1_V^{(2)})1_V'^{(2)}1_V^{(1)})\\
        % &\quad\times [(S(1_V'^{(1)})\rightharpoonup 1_U)\otimes 1_V^{(4)}] \\
        & = p(u,S^{-1}(\varepsilon_L(v))S^{-1}(\varepsilon_R(1_V^{(3)}))S^{-1}(1_V^{(2)})1_V'^{(2)}1_V^{(1)})\\
        &\quad\times [1_U\otimes S(1_V'^{(1)})1_V^{(4)}] \\
        & = p(u,S^{-1}(\varepsilon_L(v))S^{-1}(S(1_V'^{(2)})1_V^{(2)})1_V^{(1)})\\
        &\quad\times [1_U\otimes S(1_V'^{(1)})1_V^{(3)}] \\
        & = p(u,S^{-1}(\varepsilon_L(v))S^{-1}(1_V^{(1)})) [1_U\otimes 1_V^{(2)}]. 
    \end{align*}
    On the other hand, we have
    \begin{align*}
        [u^{(2)}&\otimes v^{(1)}]S([u^{(1)}\otimes v^{(2)}]) \\
        & = p(S(u^{(2)}),v^{(1)})p(S(u^{(4)}),S^{-1}(v^{(3)}))p(u^{(1)},v^{(6)})\\
        & \quad \times  p(u^{(5)},S^{-1}(v^{(4)}))[u^{(6)}S(u^{(3)})\otimes v^{(2)}S(v^{(5)})] \\
        & = p(S(u^{(3)}),S^{-1}(\varepsilon_L(v^{(3)})))p(u^{(1)},v^{(5)}S^{-1}(v^{(1)}))\\
        & \quad \times  [u^{(4)}S(u^{(2)})\otimes v^{(2)}S(v^{(4)})] \\ 
        & = p(S(u^{(3)}),S^{-1}(1_V^{(2)}))p(u^{(1)},v^{(3)}S^{-1}(v^{(1)}))\\
        & \quad \times  [u^{(4)}S(u^{(2)})\otimes 1_V^{(1)}\varepsilon_L(v^{(2)})] \\ 
        & = p(u^{(1)},S^{-1}(1_V'^{(1)}\varepsilon_L(v)))   [u^{(3)}(S(u^{(2)}) \leftharpoonup S^{-1}(1_V^{(2)}))\otimes 1_V^{(1)}1_V'^{(2)}] \\ 
        & = p(u^{(1)},S^{-1}(1_V'^{(1)}\varepsilon_L(v)))   [S(\varepsilon_L(u^{(2)}))(1_U \leftharpoonup S^{-1}(1_V^{(2)}))\otimes 1_V^{(1)}1_V'^{(2)}] \\ 
        & = p(1_U^{(1)}u,S^{-1}(1_V'^{(1)}\varepsilon_L(v)))   [S(1_U^{(2)})\otimes S^{-1}(1_V^{(2)})1_V^{(1)}1_V'^{(2)}] \\
        & = p(1_U^{(1)},S^{-1}(1_V'^{(2)}))p(u,S^{-1}(\varepsilon_L(v))S^{-1}(1_V'^{(1)}))   [S(1_U^{(2)})\otimes 1_V'^{(3)}] \\
        & = p(1_U^{(2)},1_V'^{(2)})p(u,S^{-1}(\varepsilon_L(v))S^{-1}(1_V'^{(1)}))   [1_U^{(1)}\otimes 1_V'^{(3)}] \\
        & = p(u,S^{-1}(\varepsilon_L(v))S^{-1}(1_V'^{(1)}))   [(1_V'^{(2)}\rightharpoonup 1_U)\otimes 1_V'^{(3)}] \\
        & = p(u,S^{-1}(\varepsilon_L(v))S^{-1}(1_V'^{(1)}))   [(\varepsilon_L(1_V'^{(2)})\rightharpoonup 1_U)\otimes 1_V'^{(3)}] \\
        & =\varepsilon_L([u\otimes v]),
    \end{align*}
    where we have used the relation $w^{(1)}\otimes \varepsilon_L(w^{(2)}) = 1^{(1)}w\otimes 1^{(2)}$. Similarly, one can show that 
    \begin{equation*}
        \varepsilon_R([u\otimes v]) = p(1_U^{(1)}S^{-1}(\varepsilon_R(u)),v)[1_U^{(2)}\otimes 1_V] = S([u^{(2)}\otimes v^{(1)}])[u^{(1)}\otimes v^{(2)}]. 
    \end{equation*}
    Finally, one computes 
    \begin{align*}
        S(&[u^{(3)}\otimes v^{(1)}])[u^{(2)}\otimes v^{(2)}]S([u^{(1)}\otimes v^{(3)}]) \\
        & = S([u^{(2)}\otimes v^{(1)}])\varepsilon_L([u^{(1)}\otimes v^{(2)}])  \\
        & = [S(u^{(3)})\otimes S(v^{(2)})][1_U\otimes 1_V^{(2)}]p(u^{(2)},v^{(3)})\\
        & \quad \times p(u^{(4)},S^{-1}(v^{(1)}))p(u^{(1)},S^{-1}(\varepsilon_L(v^{(4)}))S^{-1}(1_V^{(1)})) \\ 
        & = [S(u^{(2)})\otimes S(v^{(2)})1_V^{(2)}]p(u^{(3)},S^{-1}(v^{(1)}))p(u^{(1)},S^{-1}(S(v^{(3)})1_V^{(1)})) \\
        & = S([u\otimes v]).
    \end{align*}
    This completes the proof of Theorem~\ref{thm:QD-algebra}.

\section{Skew-pairing of boundary tube algebra for general bimodule category} \label{app:general_pairing}

In this appendix, we clarify the claim in Remark~\ref{rmk:general_pairing}. In this part, $\EM$ does not necessarily satisfy $d^L_m=d^R_m$. The pairing  $p: \mathbf{Tube}_{\rm bd}(\EM_{\ED}) \times \mathbf{Tube}_{\rm bd}({_{\EC}}\EM)    \to \mathbb{C}$ is defined by 
\begin{equation} 
\begin{aligned}
    p\left(\begin{aligned}

    \end{aligned}\;,
\end{aligned}
\end{equation}
where $\bar{s}$ is the dual object defined by the right action of $\ED$. Let us show that $p$ satisfies the conditions in Definition~\ref{def:skew-pairing}.     The identities 
\[
p(1,Y) = \varepsilon(Y),\quad p(X,1)=\varepsilon(X),
\]
can be straightforwardly verified, mirroring the pseudo-unitary module category argument: 
    \begin{align*}
\sum_{s,v}\, 
    p\left(\begin{aligned}
    %
\end{aligned}\right). 
\end{align*}
Let us verify the identity 
\[
p(X,YY') = \sum_{(X)} p(X^{(1)},Y')p(X^{(2)},Y).
\]
On the one hand, the computation of the left-hand side is proceed as follows:   
\begin{align*}
    p(X,YY')=&\quad p\left(\begin{aligned}
%
    \end{aligned} \\
    & = \delta_{z,w'}\delta_{y,x'} \delta_{s,w}\delta_{t,x}\delta_{u,y'}\delta_{v,z'} \frac{1}{d_s^R} \sum_{k,\sigma} \sqrt{\frac{d_k^R}{d_y^Rd_b}} \sum_{c,\alpha,\beta} [F^{a'ub}_k]_{y\sigma\nu'}^{c\alpha\beta} \sum_{e,\tau,\lambda}[F_s^{ayb}]_{t\mu\nu}^{e\tau\lambda} \\
    & \quad \times \delta_{k,e}\delta_{\sigma,\lambda}\sqrt{\frac{d_y^Rd_b}{d_k^R}}\delta_{c,v}\delta_{\beta,\gamma}\sqrt{\frac{d_u^Rd_b}{d_v^R}}\delta_{k,z}\delta_{\alpha,\zeta'}\sqrt{\frac{d_{a'}d_v^L}{d_z^L}}\delta_{\tau,\zeta} \sqrt{\frac{d_ad_z^L}{d_s^L}}d_s^R \\ 
    & = \delta_{z,w'}\delta_{y,x'} \delta_{s,w}\delta_{t,x}\delta_{u,y'}\delta_{v,z'}  \sum_{\sigma}  [F^{a'ub}_z]_{y\sigma\nu'}^{v\zeta'\gamma} [F_s^{ayb}]_{t\mu\nu}^{z\zeta\sigma} \sqrt{\frac{d_bd_u^Rd_{a'}d_v^Ld_a}{d_v^Rd_s^L}}\;. 
\end{align*}
On the other hand, the right-hand side is equal to the following  
\begin{align*}
& \quad \sum_{(X)} p(X^{(1)},Y')p(X^{(2)},Y) \\
& = \sum_{i,k,\sigma}\sqrt{\frac{d_k^R}{d_i^Rd_b}}\;
p\left(\begin{aligned}

    \end{aligned}\\
    & \quad \times \delta_{z,w'}\delta_{y,x'}\delta_{u,y'}\delta_{v,z'}\delta_{s,w}\delta_{t,x}\sqrt{\frac{d_z^R}{d_y^Rd_b}}\frac{1}{d_z^Rd_s^R} \\
    & = \delta_{z,w'}\delta_{y,x'}\delta_{u,y'}\delta_{v,z'}\delta_{s,w}\delta_{t,x}\sum_{\sigma} \sum_{c,\alpha,\beta} [F_z^{a'ub}]_{y\sigma\nu'}^{c\alpha\beta} \delta_{\beta,\gamma} \delta_{c,v}\sqrt{\frac{d_u^Rd_b}{d_v^R}}\delta_{\alpha,\zeta'}\sqrt{\frac{d_{a'}d_v^L}{d_z^L}}d_z^R  \\ 
    & \quad \times \sum_{e,\tau,\lambda}[F_s^{ayb}]_{t\mu\nu}^{e\tau\lambda} \delta_{\lambda,\sigma} \delta_{e,z}\sqrt{\frac{d_y^Rd_b}{d_z^R}}\delta_{\tau,\zeta}\sqrt{\frac{d_ad_z^L}{d_s^L}}d_s^Rg \sqrt{\frac{d_z^R}{d_y^Rd_b}}\frac{1}{d_z^Rd_s^R} \\
    & = \delta_{z,w'}\delta_{y,x'}\delta_{u,y'}\delta_{v,z'}\delta_{s,w}\delta_{t,x}\sum_{\sigma}[F_z^{a'ub}]_{y\sigma\nu'}^{v\zeta'\gamma}[F_s^{ayb}]_{t\mu\nu}^{z\zeta\sigma}\sqrt{\frac{d_u^Rd_bd_{a'}d_v^Ld_a}{d_v^Rd_s^L}}\;, 
    \end{align*}
    which is equal to the expression above, as expected. One uses a similar technique to show the identity 
    \[
        p(XX',Y) = \sum_{(Y)} p(X,Y^{(1)})p(X',Y^{(2)}). 
    \]
    In remains to verify the antipode identity 
    \[
        p(S(X),Y) = p(X,S^{-1}(Y)). 
    \]
    First, the left-hand side is equal to 
\begin{align*}
\begin{aligned}
    & \quad p(S(X),Y) \\
    & = p\left(S\left(\begin{aligned}

    \end{aligned} \\
    & = \frac{d_u^R}{d_v^R}\frac{\delta_{u,w}\delta_{v,x}\delta_{s,y}\delta_{t,z}}{d_u^R} \sum_{t,\mu,\zeta}[F_s^{\bar{a}ub}]_{t\mu\zeta}^{t'\mu'\zeta'} \delta_{t',v}\delta_{\mu',\nu}\sqrt{\frac{d_ad_s^L}{d_v^L}}\delta_{\zeta',\gamma}\sqrt{\frac{d_v^Rd_b}{d_u^R}}d_u^R \\ 
    & = \delta_{u,w}\delta_{v,x}\delta_{s,y}\delta_{t,z}[F_s^{\bar{a}ub}]_{t\mu\zeta}^{v\nu\gamma}  \sqrt{\frac{d_u^Rd_ad_s^Ld_b}{d_v^Rd_v^L}}\;.
\end{aligned}
\end{align*}
Second, the right-hand side is equal to 
\begin{align*}
\begin{aligned}
    & \quad p(X,S^{-1}(Y)) \\
    & = p\left(\begin{aligned}

    \end{aligned} \\
    & =  \frac{d_y^L}{d_x^L}\frac{\delta_{s,y}\delta_{t,z}\delta_{u,w}\delta_{v,x}}{d_s^R} \sum_{t',\mu',\zeta'} [F_s^{\bar{a}ub}]_{t\mu\zeta}^{t'\mu'\zeta'} \delta_{t',v}\delta_{\zeta',\gamma}\sqrt{\frac{d_u^Rd_b}{d_v^R}}\delta_{\mu',\nu}\sqrt{\frac{d_ad_v^L}{d_s^L}}d_s^R \\
    & = \delta_{s,y}\delta_{t,z}\delta_{u,w}\delta_{v,x} [F_s^{\bar{a}ub}]_{t\mu\zeta}^{v\nu\gamma} \sqrt{\frac{d_s^Ld_u^Rd_bd_a}{d_v^Ld_v^R}}\;.
\end{aligned}
\end{align*}
Both expressions are identical. Hence we clarify the claim in Remark~\ref{rmk:general_pairing}.

\begin{remark}
    With this general skew-pairing $p$, one can demonstrate that the morphism 
    \begin{align*}
        \Phi:\mathbf{Tube}_{\rm bd}(\EM_{\ED}) & \Join_p \mathbf{Tube}_{\rm bd}({_{\EC}}\EM)  \to \mathbf{Tube}({_\EC}\EM_\ED), \\
        \begin{aligned}

    \end{aligned}
    \end{align*}
    is an isomorphism of weak bialgebras. 
    The proof follows a similar approach to the corresponding part of Theorem~\ref{thm:QD_tube}, with careful handling of F-symbols. 
\end{remark}

\bibliographystyle{apsrev4-1-title}
\bibliography{Jiabib}

\end{document}